\documentclass[12pt]{article}
\usepackage{pdflscape}

\usepackage[T1]{fontenc}
\DeclareUnicodeCharacter{0308}{\"{}}
\usepackage{geometry}
\usepackage[style=apa, backend=biber]{biblatex}
\usepackage{amsmath}
\usepackage{amsfonts}
\usepackage{amsfonts}
\usepackage{array}

\usepackage{tikz}
\usetikzlibrary{arrows.meta, positioning, shapes, fit} 
\usepackage{hyperref}
\usepackage{url} 
\usepackage{setspace}
\usepackage{bm}  

\usepackage{longtable}
\usepackage{caption}
\usepackage{threeparttablex}  % threeparttable support for longtable

\usepackage{float}

\usepackage{titling}
\usepackage{booktabs}      % \toprule, \midrule, \bottomrule
\usepackage{threeparttablex}% notes support for long tables
\usepackage{ltablex}       % longtable + tabularx
\keepXColumns             % keep X‐columns breakable across pages
\newcolumntype{Y}{>{\centering\arraybackslash}X}

\title{
One Rising Ship Sinks Other Ships: Cross-Chain Negative Spillovers in Crypto Markets
\thanks{\scriptsize We thank Lin William Cong and Liyan Yang for discussions and suggestions, as well as financial support from NTU Centre in Computational Technologies for Finance, Tier 1 Grant from MOE of Singapore (RG121/23), Alibaba Group and NTU Singapore through Alibaba-NTU Global e-Sustainability CorpLab (ANGEL).}}
\date{Sep 17, 2025}
\author{Mengzhong Ma\thanks{\scriptsize Email: mengzhon001@e.ntu.edu.sg, Interdisciplinary Graduate School,  Nanyang Technological University, Singapore 637335.}\and
Te Bao\thanks{\scriptsize Email: baote@ntu.edu.sg, School of Social Sciences, Nanyang Technological University, Singapore 639798. Corresponding Author.}\and
Yonggang Wen\thanks{\scriptsize Email: ygwen@ntu.edu.sg, College of Computing \& Data Science, Nanyang Technological University, Singapore 639798.}}

\begin{document}
\pagenumbering{gobble}

\maketitle

\vspace{-20pt}
\begin{abstract}
\noindent %We find the episode "Trump coin drain the market" during which only assets on one blockchain rise while those on other chains all dump to be a special case of the generally existing pattern "negative spillover effects", in which assets on different blockchains move in opposite directions. This pattern can be channeled from investors' being attracted to buy assets on a certain chain while selling assets on other chains. Our empirical evidence shows that negative spillover effects for assets on different chains are even more prevailing and significant than co-movements. Furthermore, negative spillover on returns become more pronounced when investors on a chain face activity level rise and extreme returns on other chains, to which investors may tend to allocate more attention. Previous studies on stock markets contagion may offer explanation to the co-movements of assets in the crypto market but cannot explain the generally exist tendency for assets on different chains to move in opposite directions. Our study provides evidence that information transmission and contagion can lead to not only co-movements but also movements in opposite directions.

%We document a pervasive pattern of negative spillover effects in the crypto market, where assets on different blockchains often move in opposite directions. A notable case is the "Trump coin drains the market" episode, in which a surge in one blockchain’s asset prices coincided with declines elsewhere. This divergence is more prevalent and significant than co-movements across chains. They intensify when activity surges or extreme returns occur on one chain, drawing investor attention away from others. 

We document the first systematic evidence of negative spillover effects in crypto asset returns across blockchains. Using on-chain data from Ethereum, Solana, Binance Smart Chain, Arbitrum, and Avalanche (2022–2025), we show that surges on one chain often coincide with declines on others, in contrast to the positive co-movements typical of equity markets. These spillovers intensify during attention shocks, proxied by chain activity and extreme return events, and persist after controlling for global equity returns, interest rates, and Bitcoin. Nonlinear factor models reveal that attention, driven capital reallocation, rather than common information, underlies these dynamics. Our findings introduce a new form of cross-market linkage, attention-induced substitution, that shapes risk transmission in crypto markets. The results carry implications for portfolio diversification, systemic risk measurement, and regulation of token launches that may trigger cross-chain capital flight.

%\vspace{3\baselineskip} 

\end{abstract}

\vspace{1\baselineskip} 

\newpage

% --- Second page: restart numbering at 1
\pagenumbering{arabic}

\section{Introduction}

%On January 18, 2025, U.S. President Donald Trump launched his official cryptocurrency, Trump Coin, on the Solana blockchain. Following the launch, both Trump Coin and Solana’s native token (\$SOL) experienced sharp price increases—\$SOL rose by 19\%—while much of the broader crypto market declined. Notably, \$ETH and \$BNB fell by 3\% to 5\%.\footnote{\$SOL, \$ETH, and \$BNB are the native tokens of Solana, Ethereum, and Binance Smart Chain, respectively.}

%Market participants and media outlets characterized this episode as “Trump Coin drains the market,” citing a rapid reallocation of capital into Trump Coin and away from assets on other chains.\footnote{See \href{https://www.binance.com/en-AE/square/post/21139700628721}{https://www.binance.com/en-AE/square/post/21139700628721}.} This commentary highlights the competitive dynamics among blockchains for users and capital: a surge in activity on one chain can induce capital flight from others, generating negative spillover effects across platforms.

Financial markets are often characterized by cross-market linkages, where shocks in one segment transmit to others through information flows, portfolio rebalancing, or common macroeconomic exposures. In traditional equity and bond markets, these forces typically generate positive co-movements across assets, especially during crises. A central tenet of the contagion literature is that markets tend to move together when investors face common signals or liquidity shocks \parencite{king1990transmission,king1990volatiltiy,lin1994bulls,koutmos1995asymmetric,rapach2013international,veldkamp2006information,bae2003new,calvo2000rational}.

\sloppy In contrast, the cryptocurrency ecosystem is organized around competing blockchain platforms that issue native tokens, host decentralized exchanges, and attract users with heterogeneous technological features. Blockchain developers actively seek to attract users from competing chains by launching new protocols, distributing air-drops,\footnote{Air-drops are free token distributions rewarding early participants.} and upgrading their technologies. Investors, in turn, continuously search across chains for the next opportunity to earn excess returns. Capital can flow across these ecosystems, but such reallocations often require selling positions on one chain to fund purchases on another. This competitive structure suggests that when a given chain surges, investors attracted to its assets sell holdings on rival chains, giving rise to negative spillover effects across blockchains.
%shocks may generate not only contagion-like co-movements but also negative spillovers, as attention and liquidity shift from one chain to its rivals.

This paper provides the first systematic evidence on such negative cross-chain spillover effects. Using on-chain data from five major blockchains—Ethereum, Solana, Binance Smart Chain, Arbitrum, and Avalanche—over 2022–2025, we show that surges in one chain’s assets are frequently accompanied by declines elsewhere. These patterns are most pronounced when attention shocks occur, proxied by changes in chain activity and extreme returns, and they remain robust after controlling for global equity markets and interest rates.

%In this study, we examine whether the negative correlation in chain-asset returns is a prevalent feature of the crypto market. with the “Trump Coin drain-the-market” episode as a salient instance.

%我又补了两段对于贡献的总结，introduction必须要有贡献总结。

%When drawn to a new chain, investors can finance their purchases by liquidating positions in assets in other chains rather than by injecting additional fiat capital, particularly in regions where regulatory restrictions limit crypto inflows \parencite{borri2020regulation}. This cross-chain reallocation can generate negative correlations among chain-asset returns. In contrast, in equity markets, attention-induced buying of one stock has been shown to increase demand for other stocks without necessarily triggering offsetting sales \parencite{barber2008all}.

%Building on this contrast, we conjecture that when a given chain surges, investors attracted to its assets sell holdings on rival chains, giving rise to negative spillover effects across blockchains.

In the traditional finance literature, although segmented by national boundaries, markets in different nations tend to co-move, especially during financial and exchange rate crises. This pattern of contagion has been extensively explored \parencite{bekaert2014global, connolly2000stock,kodres2002rational, kyle1999contagion, fleming1998information} 
and is typically attributed to two channels: investors across markets reacting to common macroeconomic information, or inferring such information from other markets’ movements \parencite{king1990transmission,mcqueen1993stock}; and cross-market portfolio rebalancing \parencite{kyle1999contagion,kodres2002rational}. 
By analogy, these theories help explain co-movements across blockchains during crisis events, such as the “5.19 Crash” of May 19, 2021. \footnote{The 519 incident refers to a major crash in the crypto market on May 19, 2021. See the blog \href{https://www.binance.com/en/square/post/24442137434850}{https://www.binance.com/en/square/post/24442137434850}.} However, they offer little guidance for episodes where surging activity on one chain generates negative spillovers to others, as in the case of Trump Coin. \footnote{Market participants and media outlets characterized this episode as “Trump Coin drains the market,” citing a rapid reallocation of capital into Trump Coin and away from assets on other chains. See \href{https://www.binance.com/en-AE/square/post/21139700628721}{https://www.binance.com/en-AE/square/post/21139700628721}.} 

This study provides the first systematic test of whether such negative spillover effects are a general feature of the crypto market, above and beyond standard return co-movements. Our empirical strategy builds on factor models inspired by \textcite{bekaert2014global, bekaert2003market}, using return data for assets traded on five major blockchains—Ethereum, Solana, BSC, Arbitrum, and Avalanche—from April 28, 2022 to March 31, 2025. We document robust evidence of negative correlations between assets on surging chains and those on other chains, often stronger than the positive co-movements observed across chains. Constructing chain-level portfolios weighted by market capitalization, we find, for example, that Ethereum’s asset returns are negatively correlated with those on Arbitrum, with similar patterns for BSC–Arbitrum and Avalanche–Arbitrum pairs. 

Our result is robust, and even stronger after controlling for global economic factors like stock indices and risk-free rates in the U.S., Hong Kong, and Europe—to capture crypto investors’ reactions to traditional financial signals and chain level activity like native token returns and staking reward rates. Beyond employing chain-activity variables as proxies for user attraction to a chain, we introduce extreme-return dummies to capture cases in which heightened attractiveness does not translate into observable changes in chain-activity variables. Following \textcite{barber2008all}, investors often respond to news by purchasing stocks, generating extreme returns on announcement days. Using extreme-return indicators, we find that the negative spillover effects intensify significantly when rival chains experience extreme movements. 

Our study contributes to the literature in several important ways. 

First, we extend the extensive body of research on financial contagion by uncovering a novel form of cross-market linkage that is distinctive to cryptocurrency markets. Whereas prior work has emphasized positive co-movements arising from common information or liquidity shocks, we show that competition among blockchains gives rise to attention-induced substitution, in which capital reallocates across ecosystems and asset returns move in opposite directions.

Second, different from previous studies that mainly use data from centralized exchanges, we advance the empirical study of cryptocurrencies by constructing comprehensive on-chain, chain-level portfolios and by employing both linear and nonlinear factor models at half-day frequency. This design allows us not only to capture the full breadth of decentralized markets but also to disentangle systematic information spillovers from attention-driven reallocations.

Finally, for practitioners and policy makers, We provide the first systematic evidence of negative spillover effects across blockchains, with important implications for portfolio diversification, systemic risk assessment, and the regulation of token launches that may trigger cross-chain capital flight.

The paper is organized as follows. The next section summarizes the literature related to our study and Section \ref{institutional_background} discusses the background knowledge relevant. Section \ref{main_section_portfolio_method} illustrates how to construct portfolio for each chain and the corresponding variables, as well as models detecting negative spillover. Section \ref{main_section_data_stats} describes our data sources and descriptive statistics. In Section \ref{main_section_results}, we discuss empirical results based on our econometric specifications. Section \ref{main_section_conclusion} concludes. 

\section{Literature Review and Institutional Background}

\subsection{Related Literature}

Our study contributes to three strands of literature: contagion across markets, investor attention allocation, and cryptocurrency pricing.

The vast literature on market contagion traditionally attributes co-movements to investors in one market inferring information from price changes in others, leading to similar reactions to macroeconomic news \parencite{king1990transmission,king1990volatiltiy,lin1994bulls,koutmos1995asymmetric,rapach2013international,veldkamp2006information,bae2003new,calvo2000rational}. Yet these explanations struggle to account for co-movements among weakly linked countries with distinct economic fundamentals. Even after controlling for macroeconomic announcements, a large fraction of correlations remains unexplained \parencite{kaminsky2000crises,baig1999financial,connolly2000stock,bekaert2003market}. Alternative theories emphasize cross-market hedging and information spillovers \parencite{fleming1998information}, or highlight the role of portfolio rebalancing and liquidity shocks in generating contagion \parencite{longstaff2010subprime,cespa2014illiquidity,kodres2002rational,kyle1999contagion,allen2000financial}. Other perspectives include reduced incentives to acquire costly information under globalization \parencite{calvo2000rational} and shared ownership across markets \parencite{anton2014connected}.

%Scholars note that contagion always occurs in conjunction with crises and panics, when asset returns become extreme. Early empirical studies document co-movements among national markets following the October 1987 crash arencite{forbes2002no,lee1993does,yuan2005asymmetric,bae2003new}. Studies in the later stage show the tight relationship between market co-movements and the global financial crises in 2008, as well as Asian crises \parencite{bekaert2014global,longstaff2010subprime,junior2012correlation,samarakoon2011stock,kenourgios2015contagion,chiang2007dynamic,bensaida2019good,claeys2014measuring}. In addition, theoretical models are also proposed, e.g., \textcite{allen2000financial}, to illustrate how small shocks spread across sectors and result in large effects by means of contagion. 

Contagion is usually particularly salient during crises, when asset returns become extreme. Early studies documented heightened co-movements following the October 1987 crash \parencite{forbes2002no,lee1993does,yuan2005asymmetric,bae2003new}, and later work linked contagion to the 1997 Asian crisis and the 2008 global financial crisis \parencite{bekaert2014global,longstaff2010subprime,junior2012correlation,samarakoon2011stock,kenourgios2015contagion,chiang2007dynamic,bensaida2019good,claeys2014measuring}. Theoretical models further illustrate how small shocks propagate across sectors and generate systemic effects \parencite{allen2000financial}.

%Interestingly, our study also considers periods of extreme returns, but finds negative return correlation among chains rather than co-movements. This difference may be raised by the loose interdependency between chains, which are based on considerably different fundamentals, hampering the spreading of chain specific shocks. Early works, e.g., \textcite{king1990transmission}, define "contagion" effect as the process of investors' learning information reflected by price changes in other markets. This definition is narrow in sense, as it only includes domestic investors' decision making on their domestic assets when they extract systematic information about the local market from pricing changes in foreign markets. Whether they would further step to foreign markets is ignored. Our study falls within the literature on “contagion” in the broad sense that price changes or news in other markets attract investors’ attention and elicit responses, while simultaneously prompting opposite decisions in the foreign market. This difference arises from the fact that blockchains compete for users, whereas stock markets in different countries have little incentive to do so.
In contrast, our findings suggest that during periods of extreme returns, blockchains exhibit negative return correlations rather than contagion. This difference likely reflects the weak interdependence of blockchain ecosystems, which rest on heterogeneous fundamentals and thus limit the transmission of chain-specific shocks. While early definitions of contagion emphasized that domestic investors extract information from foreign price changes without directly reallocating capital abroad \parencite{king1990transmission}, our broader perspective considers that shocks in one market not only attract attention but can also trigger opposite investment decisions elsewhere. The negative spillover arises because blockchains actively compete for users and capital, unlike national stock markets that have little incentive to displace each other.

%Our work also relates to the literature on attention induced trading. It is shown that investors tend to buy stocks with news grabbing their attention, especially for retail investors \parencite{barber2008all,huang2019attention,barber2022attention,peng2006investor}. 
%Evidence from \textcite{barber2008all,da2025market} suggests that buying pressure on the prices of attention-grabbing stocks results in abnormal returns, thereby justifying our use of extreme returns as the proxy for investor attention. %In addition to extreme returns, other proxies adopted by previous researches include Google (Search Volume Index (SVI)) \parencite{da2011search}, stock recommendations \parencite{keasler2010mad}, and big events related to the market \parencite{seasholes2007predictable}. 
%The research most relevant to our study is the one by \textcite{hu2023attention}, who explore if extreme S\&P500 returns would attract away investors' attention on cryptocurrency markets.

Our paper also connects to the literature on attention-driven trading. A large body of evidence shows that investors, particularly retail investors, tend to buy attention-grabbing assets, producing temporary price pressure and abnormal returns \parencite{barber2008all,huang2019attention,barber2022attention,peng2006investor}. Extreme returns have been widely used as proxies for attention \parencite{barber2008all,da2025market}, consistent with our empirical design. Other proxies include Google Search Volume Index \parencite{da2011search}, stock recommendations \parencite{keasler2010mad}, and market-moving events \parencite{seasholes2007predictable}. Closest to our setting, \textcite{hu2023attention} examine whether extreme S\&P 500 returns divert investor attention away from cryptocurrency markets.

%Finally, our study contribute to the young and proliferating literature on pricing models and risk factors for cryptocurrency. Previous studies have examined risk distributions and factors in the cryptocurrency market \parencite{borri2019conditional,zhang2021downside,brauneis2021measure,hubrich2017know,borri2022cross,li2018toward,makarov2020trading,liu2022common,liu2021risks,ma2023enhancing}, most lag behind this rapidly growing industry, and the market structure has changed substantially in recent years. Those studies mainly focus exclusively on major native tokens and use pricing data from centralized exchanges, thus covering only a limited segment of the current cryptocurrency market. 

%In contrast, our work amounts to the first sort of research adopting large-scale on-chain pricing data inclusively, considering not only native tokens but also assets supported by smart contracts. As well, we evaluate crypto assets and construct portfolio according to the current market structure and components, laying a foundation for future research to reconsider how risk and pricing should be measured in the crypto market. 

Finally, our study contributes to the emerging and rapidly expanding literature on cryptocurrency pricing models and risk factors.
Prior work has examined risk distributions and factor structures in cryptocurrency markets \parencite{borri2019conditional,zhang2021downside,brauneis2021measure,hubrich2017know,borri2022cross,li2018toward,makarov2020trading,liu2022common,liu2021risks,ma2023enhancing}, but much of this research lags behind the fast-evolving nature of the industry, as market structures have shifted substantially in recent years.
Most existing studies focus narrowly on major native tokens and rely on data from centralized exchanges, thereby capturing only a fraction of the current cryptocurrency ecosystem.

In contrast, our study is among the first to employ large-scale on-chain pricing data that comprehensively covers the market. We incorporate not only native tokens but also assets issued via smart contracts, and we construct portfolios that reflect the current market composition. By doing so, we provide a framework for rethinking how risk exposures and asset pricing should be measured in the cryptocurrency space, offering a foundation for future research in this area.

\subsection{Institutional Background}\label{institutional_background}

Since “2022 DeFi summer”, when the Uniswap protocol was launched and proliferated  \footnote{DeFi: Decentralized finance. Uniswap protocol is the largest protocol for Decentralized Exchanges \parencite{lehar2025decentralized}.}, crypto assets have increasingly been traded on blockchains (chains) through Decentralized Exchanges (DEXs).

Unlike traditional Centralized Exchanges (CEXs) like Binance, which rely on off-chain servers, DEXs are supported by smart contracts deployed on blockchains. 
As a result, the security and permissionless features of blockchains extend to DEXs, attracting substantial trading volume from CEXs.

Initially, DEX activity was concentrated on Ethereum,\footnote{Ethereum is the first Turing-complete blockchain, allowing smart contract deployment.}, but the subsequent launch of competing blockchains gave rise to many Uniswap-like DEXs across ecosystems. Today’s crypto market is highly fragmented: liquidity is split among numerous CEXs and DEXs, in stark contrast to traditional financial markets where trading activity is concentrated in a few venues. Independent blockchain ecosystems compete for users by introducing new projects and technological upgrades, prompting users to shift across chains in search of opportunities. This dynamic flow of users generates surges of activity in one chain while others experience declines, producing a tidal rise and fall of attention and liquidity.

A key innovation of our study is to analyze the entire crypto market, rather than restricting attention to major tokens on CEXs as in prior work \parencite{liu2021risks,liu2022common,borri2022cross}. As illustrated in Figure \ref{fig:crypto_market}, the market consists of three components: CEXs, blockchains, and DEXs. Assets are issued on different blockchains via smart contracts, but heterogeneous standards limit their natural cross-chain circulation. DEXs facilitate trading of assets within a given chain, where the medium of exchange is typically the chain’s native token rather than fiat currency or stablecoins.\footnote{Stablecoins are tokens pegged to fiat currencies, such as USDC and USDT.} Users thus trade local assets with local “currencies,” much like domestic investors trading equities in their home currency.

To access assets on a different chain, users must first convert native tokens into those of the target chain, typically via CEXs.\footnote{Bridging protocols such as Wormhole (\href{https://wormhole.com}{https://wormhole.com}
) also enable conversion, though CEXs handle the majority of volume.} CEXs therefore serve as the “foreign exchange market” of the blockchain world. Independent of any particular chain, they not only facilitate cross-chain conversions but also act as gateways for converting fiat currency into native tokens. Accordingly, users seeking to invest in a blockchain ecosystem obtain its native token either by (i) purchasing it with fiat currency on CEXs or (ii) converting native tokens from other chains.

%We apply the analogy of international stock markets to blockchains and the crypto market because they share the same crucial features. %Different chains are operated and governed under different systems, just like different nations with markets managed in different manner. Also, borders and restrictions on international investments correspond to frictions induced by costly native token conversion. 

We employ an analogy to international stock markets because the parallel is instructive. Different blockchains resemble different nations, each governed by its own system. Frictions in international capital flows, such as exchange rate costs, map naturally onto frictions in native token conversion, shaping the dynamics of capital allocation across blockchain ecosystems.

\section{Chain Portfolio Construction and Modeling Negative Spillover Effects on Returns}\label{main_section_portfolio_method}

Existing studies on international financial markets co‐movements primarily employ factor models to assess how returns across national equity markets co‐vary, with market returns as the target of analysis \parencite{bekaert2003market,bekaert2014global}. According to this framework, we conceptualize each blockchain as a separate market and assess inter‐chain return relations by constructing chain‐specific portfolios—aggregating the returns of all assets native to each chain—and applying factor models to these portfolio returns. To this end, we discuss how chain‐specific portfolios are constructed in Section \ref{section_chain_portfolio_construction}, the necessary variables in Section \ref{section_chain_condition_variables} and \ref{section_global_market_var}, and the specification of our factor models in Section \ref{section_models}.

\subsection{Chain Portfolio Construction}\label{section_chain_portfolio_construction}

To evaluate inter-chain return dynamics, we construct three types of blockchain-level portfolios. The all-asset portfolio reflects overall chain performance, the CEX vs. non-CEX portfolios separate assets listed on centralized exchanges from those traded exclusively on-chain, and the local-only portfolio excludes bridged tokens to avoid spurious cross-chain correlations. These complementary portfolios provide a comprehensive basis for our subsequent factor model analysis. The more detailed design of our study is as the following: 

The first step in our study is to construct market portfolios for different blockchains. We choose five blockchains in our study: Ethereum, Solana, Binance Smart Chain (BSC), Arbitrum, and Avalanche. \footnote{Readers can check the official website of the chains for details: Ethereum (\href{https://ethereum.org/en/}{https://ethereum.org/en/}), Solana (\href{https://solana.com}{https://solana.com}), BSC (\href{https://www.bnbchain.org/en}{https://www.bnbchain.org/en}), Arbitrum (\href{https://arbitrum.io}{https://arbitrum.io}), Avalanche (\href{https://www.avax.network}{https://www.avax.network}).} All of the five chains were launched relatively early with leading volume, sufficiently representative for the whole crypto asset market. We construct three kinds of portfolio for each chain.

(i) All-asset portfolio.
The first portfolio includes all assets issued on a given chain, weighted by market capitalization. Following \textcite{de2022arbitrage}, trading activity in crypto markets, particularly in derivatives, is highly synchronized with the business hours of major equity markets. Between 00:00–12:00 UTC, markets in Europe and Asia open (and some close), while between 12:00–23:59 UTC, markets in Europe and the U.S. are active. This intraday segmentation highlights how crypto markets incorporate information from traditional markets. To capture this effect, we compute chain-level portfolio returns at a half-day frequency in UTC time. The return on the all-asset portfolio of $chain_i \in 
\{\substack{\text{Ethereum, Solana, BSC,}\\ \text{Arbitrum, Avalanche}}\}$ 
in half-day $t$ is denoted by $R^{All}_{\text{chain}_i,t}$.

(ii) CEX vs. non-CEX portfolios.
Because assets native to a chain may also be listed on centralized exchanges (CEXs), we distinguish between those listed on CEXs and those that remain exclusively on-chain. This separation allows us to examine how CEX pricing affects the valuation of other chain assets. Accordingly, for each chain $i$ in half-day $t$, we compute returns for two additional portfolios: the CEX-listed portfolio, $R^{CEX}_{chain_{i},t}$, and the non-CEX portfolio, $R^{non\text{-}CEX}_{chain_{i},t}$.

(iii) Local-only portfolio.
Finally, many assets are bridged across chains, producing “wrapped” versions that trade on multiple blockchains (similar to dual listings in equity markets).\footnote{For example, Chainlink was initially issued on Ethereum at address 0x514910771af9ca656af840dff83e8264ecf986ca, but later bridged to BSC at 0xf8a0bf9cf54bb92f17374d9e9a321e6a111a51bd.} To mitigate spurious co-movements arising from such cross-chain tokens, we also construct a local-only portfolio that includes only assets natively issued and traded on a given chain, excluding bridged versions. For chain $i$ in half-day $t$, this portfolio return is denoted by $R^{Local}_{chain_{i},t}$.

\subsection{Blockchain Activity and Incentive Metrics}\label{section_chain_condition_variables}

We incorporate investor attention, proxied by on‐chain user activity, into our model using two chain‐level activity variables: the native token return and the staking reward rate.

Every blockchain issues a native token that serves as the payment vehicle for transaction fees (“gas”).\footnote{On blockchains, transactions that alter the chain’s state consume computing resources and therefore require a fee, commonly referred to as gas. For example, on Ethereum, gas fees are paid in \$ETH.} Consequently, most decentralized exchange (DEX) trading pairs are denominated in the chain’s native token against another token issued on that chain.\footnote{A native token is defined by the chain protocol itself, while other tradable assets typically follow the fungible-token standard, such as ERC-20 on Ethereum. A list of trading pairs across chains is available on DEX Screener (\href{https://dexscreener.com}{https://dexscreener.com}).} Thus, purchasing any token on a given chain requires the chain’s native token both to execute the trade and to pay gas fees. Higher on-chain activity and stronger demand for tokens naturally increase demand for the native token, thereby raising its price. Therefore, we use each chain's native token return on the half-day $t$, denoted as $R_{native_{i}, t}$ $(native_{i} \in \{\$\text{ETH},\allowbreak\ \$\text{SOL},\allowbreak\ \$\text{BNB},\allowbreak\ \$\text{ARB},\allowbreak\ \$\text{AVAX}\})$ as the proxy for the activity level and people's preference to a chain. For the five chains in our study, the native tokens are: \$ETH for Etherem, \$SOL for Solana, \$BNB for BSC, \$ARB for Arbitrum, and \$ AVAX for Avalanche. One exception is Arbitrum, a Layer-2 solution for Ethereum, which uses \$ETH for gas payments but \$ARB for governance. \footnote{Lay-2 chains help scaling Ethereum, while for users they are still running like independent blockchains. See \href{https://ethereum.org/en/layer-2/}{
https://ethereum.org/en/layer-2/
}. Users holding \$ARB can participate in voting for technology upgrading or protocol change, in the way of "decentralized governance".}

The second chain-level activity variable is the staking reward rate. Except Arbitrum, all the chains in our study adopt Proof-of-Stake consensus mechanism. Users can delegate their native tokens to the validators (or become validators by themselves via staking) and receive staking reward. \footnote{See details at Ethereum website: \href{https://ethereum.org/en/developers/docs/consensus-mechanisms/pos/}{https://ethereum.org/en/developers/docs/consensus-mechanisms/pos/}.} These staking rewards can be viewed as the near risk-free interest rate of the native token, or equivalently, as compensation for contributing to the blockchain’s operation. Busier chains with greater transaction volume typically generate higher rewards for stakers. Accordingly, the staking reward rate also reflects user demand for a chain, albeit through the protocol layer rather than the trading market. Although Arbitrum does not employ PoS, users can still deposit \$ETH on the chain and lend it out, which likewise proxies for on-chain activity. 

Because only the unexpected component of chain activity variables should affect asset prices, we extract the innovation in staking reward rates using an ARIMA($p,d,q$) model. Specifically, we use the residual series, which both captures unanticipated shocks and ensures stationarity in regressions. For half-day $t$, the unexpected portion of the average annualized staking reward rate for $chain_i \in \{Ethereum,\allowbreak\ Solana,\allowbreak\ BSC,\allowbreak\ Arbitrum,\allowbreak\ Avalanche\}$ is denoted as $\text{SR}_{chain_{i}, t}$.

\subsection{Global Market Variables}\label{section_global_market_var}

Since the crypto market is influenced by global economic conditions, we control for stock market performance in major economies. Specifically, we incorporate the Hang Seng Index (Hong Kong), the FTSE 100 Index (U.K.), and the S\&P 500 Index (U.S.). For each market, daily returns are decomposed into overnight (close-to-open) and intraday (open-to-close) components. Following the timing conventions in \textcite{connolly2000stock}, the U.S. market opens and closes in the second half of a UTC day, the Hong Kong market in the first half, and the U.K. market opens in the first half but closes in the second.

Accordingly, at 12:00 UTC (end of the first half-day), the crypto market incorporates the intraday return of the Hang Seng Index and the overnight returns of the FTSE 100 and S\&P 500. At 00:00 UTC (end of the second half-day), it incorporates the intraday returns of the FTSE 100 and S\&P 500 and the overnight return of the Hang Seng Index. Thus, for half-day $t$, the return on the S\&P 500, $\text{SPR}_{t}$, is defined as the overnight return if $t$ is the first half-day and the intraday return if $t$ is the second half-day. Using the same logic, we construct half-day returns for the Hang Seng Index ($\text{HSR}_{t}$) and the FTSE 100 Index ($\text{FTSER}_{t}$).

We also control for global risk-free rates. Specifically, we use the Hong Kong Interbank Offered Rate (HIBOR) for Hong Kong, the one-month Euribor for the U.K., and the one-month U.S. Treasury bill rate for the U.S. To align with the half-day structure, these rates are converted to half-day frequency. We further extract the unexpected component of each rate using an ARIMA model, and use the residuals as regressors, denoted $\text{HIBOR}_{t}$, $\text{EURIBOR}_{t}$, and $\text{TREA}_{t}$.

By incorporating these global equity and risk-free benchmarks, our framework isolates blockchain-specific dynamics from broader macroeconomic influences, ensuring that observed inter-chain co-movements are not mechanically driven by global market shocks.

\subsection{Specification of Econometric Model}\label{section_models}

In this study, we specify our models by drawing on methodologies developed in the literature on co-movements and contagion among stock markets. We posit two forces that jointly determine the interdependence of assets across blockchains: signal extraction and attention allocation. These forces operate in opposite directions.

Following \textcite{bekaert2014global, connolly2000stock, dungey2005empirical, king1990transmission, lin1994bulls, peiro1998transmission}, information can be decomposed into systematic and idiosyncratic components. Systematic information affects all markets, whereas idiosyncratic information is relevant only locally. Each component can be further divided into an observable part, proxied by chain activity variables, and an unobservable part that is revealed through price changes, i.e., signal extraction. Because investors across markets respond to both observable and latent elements of systematic information, asset returns tend to co-move.

However, limited cognitive resources constrain investors’ ability to process information across multiple markets. As a result, they allocate disproportionate attention and capital to chains experiencing salient shocks or news. This reallocation induces selling pressure on other chains, leading to negative return correlations. Since attention-driven substitution and signal extraction operate simultaneously, the overall direction of return correlation across chains—positive or negative—remains an empirical question.

\subsubsection{The linear models}

Inspired by \textcite{bekaert2014global, peiro1998transmission}, we specify factor models with three vectors of factors for $chain_{i}$: the chain market factors, $F^{market}_{chain_{i},t}$, the blockchain activity factors, $F^{activity}_{chains,t}$, and the global market factors, $F^{market}_{global,t}$. For illustration, the three groups of factors for Ethereum market are defined as:
\begin{equation}
F^{\mathrm{market}}_{\mathrm{Ethereum},t}{}'
\;=\;
\bigl[R^{CEX}_{\mathrm{Ethereum},t}\bigr]
\;\cup\;
\bigl[R^{All}_{chain_i,t}\bigr]_{\,chain_i\in\{\mathrm{Solana},\,\mathrm{BSC},\,\mathrm{Arbitrum},\,\mathrm{Avalanche}\}} ,
\end{equation}
\begin{equation}
\begin{split}
F^{\mathrm{activity}}_{\mathrm{chinas},t}{}'
  &= \bigl[R_{native_i,t}\bigr]_{\,native_i\in\{\$\mathrm{ETH},\$\mathrm{SOL},\$\mathrm{BNB},\$\mathrm{ARB},\$\mathrm{AVAX},\$\mathrm{BTC}\}}
     \;\cup\; \\[-0.9ex]
  &\quad
     \bigl[\mathrm{SR}_{chain_i,t}\bigr]_{\,chain_i\in\{\mathrm{Ethereum},\,\mathrm{Solana},\,\mathrm{BSC},\,\mathrm{Arbitrum},\,\mathrm{Avalanche}\}} ,
\end{split}
\end{equation}
and
\begin{equation}
\begin{split}
F^{market}_{global,t}{}'
  &= \bigl[\mathrm{SPR}_{t},\,\mathrm{HSR}_{t},\,\mathrm{FTSER}_{t}\bigr]
     \;\cup\; \\[-0.9ex]
  &\quad
     \bigl[\mathrm{HIBOR}_{t},\,\mathrm{EURIBOR}_{t},\,\mathrm{TREA}_{t}\bigr] ,
\end{split}\label{global_market_factors}
\end{equation}
where we also consider return of Bitcoin, denoted as $R_{\$\mathrm{BTC},t}$ and constructed in the same way in section \ref{section_chain_condition_variables}, because of its leading role in the crypto market.

%With the three vectors of factors, we can construct different specifications as special cases. If we only include chain market factors in the model, it acts as a "crypto" World Factor Model (the "baseline" linear model). For Ethereum, the specification is:
With these three vectors of factors, we can formulate different model specifications as special cases. When the specification includes only chain-level market factors, it reduces to a “crypto” version of the World Factor Model, which we take as the baseline linear specification. For Ethereum, the model can be written as:
\begin{equation}
R^{All}_{\mathrm{Ethereum},t}=\alpha_{0}+\alpha_{1} R^{All}_{\mathrm{Ethereum},t-1}+\bm{\beta^{\prime}} F^{market}_{Ethereum,t}+e_{t},\label{eq_baseline_all}
\end{equation}
where $\bm{\beta^{\prime}} = \left[\beta_{0}, \beta_{1}, \beta_{2}, \beta_{3}, \beta_{4} \right]$ and $\beta_{0}$ denotes the coefficient for $R^{\mathrm{CEX}}_{\mathrm{Ethereum},t}$. 
In this case, $\beta_{i} (i \in \{1,2,3,4\})$ captures the systematic information for the Ethereum chain portfolio and also the potential co-movements or negative correlations assets on Ethereum have with assets on other chains. 
If the effect of signal-extraction overwhelms the effect of attention allocation for certain $R^{All}_{chain_i,t}$ in $F^{market}_{Ethereum,t}$, its coefficient is expected to be positive (so co-move with Ethereum), while otherwise if the the effect of attention allocation is stronger, we expect its coefficient to be negative (negatively correlated with Ethereum). We include $R^{\mathrm{CEX}}_{\mathrm{Ethereum},t}$ into regression, because assets issued on Ethereum but CEX-listed can be traded in multiple venue (both DEXs on Ethereum and CEXs). 
Those CEX-listed assets could be traded on the same CEX with assets from other chains, so expose to the same risks together, i.e., scandals or sudden capital flow to the CEX. 
Therefore, controlling $R^{\mathrm{CEX}}_{\mathrm{Ethereum},t}$ helps mitigate the risk of spurious correlation. %prevent spurious correlation.

%Except 
In addition to regressing $R^{\mathrm{All}}_{\mathrm{Ethereum},t}$ on other chain portfolio returns, we also regress $R^{\mathrm{non-CEX}}_{\mathrm{Ethereum},t}$ and $R^{\mathrm{Local}}_{\mathrm{Ethereum},t}$ on the same factors, with the following specifications:
\begin{equation}
R^{\mathrm{non-CEX}}_{\mathrm{Ethereum},t}=\alpha_{0}+\alpha_{1} R^{All}_{\mathrm{Ethereum},t-1}+\bm{\beta^{\prime}} F^{market}_{Ethereum,t}+e_{t}, \label{eq_baseline_nonCEX}
\end{equation}
and
\begin{equation}
R^{\mathrm{Local}}_{\mathrm{Ethereum},t}=\alpha_{0}+\alpha_{1} R^{All}_{\mathrm{Ethereum},t-1}+\bm{\beta^{\prime}} F^{market}_{Ethereum,t}+e_{t}.\label{eq_baseline_this_chain_only}
\end{equation}
With Equation (\ref{eq_baseline_nonCEX}), we test if trading activities on CEXs reveal systematic information for assets on DEXs, so that assets issued on Ethereum and CEX-listed are expected to co-move with those assets that have not been listed, i.e., $\beta_{0} > 0$. 
Because some assets could exist on multiple chains as described in Section \ref{section_chain_portfolio_construction}, we also use $R^{\mathrm{Local}}_{\mathrm{Ethereum},t}$ as the dependent variable in Equation (\ref{eq_baseline_this_chain_only}) to reduce the risk of spurious correlation between chain portfolio returns.

%To control the observable part of systematic information revealed by global stock market and chain-level activity in the crypto market, we include
To account for the observable component of systematic information, captured by global equity markets and chain-level activity in the crypto ecosystem, we augment our model by including $F^{market}_{global,t}$ and $F^{activity}_{chains,t}$ while yielding the following specifications:
\begin{equation}
R^{\mathrm{All}}_{\mathrm{Ethereum},t}=\alpha_{0}+\alpha_{1} R^{All}_{\mathrm{Ethereum},t-1}+\bm{\beta^{\prime}} F^{market}_{Ethereum,t}+\bm{\theta^{\prime}}F^{market}_{global,t}+e_{t}, \label{eq_linear_macro}
\end{equation}
and 
\begin{equation}
R^{\mathrm{All}}_{\mathrm{Ethereum},t}=\alpha_{0}+\alpha_{1} R^{All}_{\mathrm{Ethereum},t-1}+\bm{\beta^{\prime}} F^{market}_{Ethereum,t}+\bm{\theta^{\prime}}F^{market}_{global,t}+\bm{\gamma^{\prime}}F^{activity}_{chains,t}+e_{t}, \label{eq_linear_macro_crypto}
\end{equation}
as well as the versions replacing $R^{\mathrm{All}}_{\mathrm{Ethereum},t}$ with $R^{\mathrm{non-CEX}}_{\mathrm{Ethereum},t}$ or $R^{\mathrm{Local}}_{\mathrm{Ethereum},t}$, considering the potential spurious correlation. 

In Equation (\ref{eq_linear_macro}), if the variables for observable global market factors successfully capture primary source of the return co-movement, we may expect diminishing of co-movement between markets, i.e, diminishing or even negative $\beta_{i} (i \in \{1,2,3,4\})$. 
If we further control factors for chain-level activity in Equation (\ref{eq_linear_macro_crypto}), which represent how strongly investors are attracted by the other chains away from Ethereum, these factors may capture some source of the negative spillover effects derived by attention allocation to other chain-specific fundamentals, such as better user experience and technological upgrades.%upgrading in technology. 

%As a result, $\beta_{i} (i \in {1,2,3,4})$ are overall reflection of the co-movements derived by unobservable systematic information, mixed with negative correlations driven by attention allocation to "non-fundamentals", such as herding behavior where investors shift across chains to follow others while ignoring fundamentals.

Consequently, the estimated $\beta_{i} (i \in \{1,2,3,4\})$ coefficients represent a composite effect: they capture co-movement driven by unobservable systematic information, but also embed negative correlations arising from attention allocation to “non-fundamentals,” such as herding behavior, where investors shift across chains to follow others while disregarding fundamentals.

\subsubsection{The non-linear models for singal-extraction}

According to \textcite{bekaert2014global,connolly2000stock}, %which use non-linear factor models to detect how information is transmitted via returns across different stock markets, we also propose non-linear models to explore the origin of the negative spillover effects and transmission channels. 
who employ non-linear factor models to examine how information is transmitted through returns across stock markets, we likewise propose non-linear specifications to investigate the origins of negative spillover effects and the channels through which they propagate.

%We still adopt Ethereum as the illustration example and define the following vectors of factors:
We continue to use Ethereum as our illustrative example and define the following vectors of factors:
\begin{equation}
\begin{split}
&F^{\mathrm{activity}}_{\mathrm{chain}_i,t}{}'
\;=\;
\bigl[1, \;\mathrm{SR}_{\mathrm{chain}_i,t},\;\mathrm{R}_{\mathrm{native}_i,t}\bigr]\, , \\
&{\,chain_i\in\{ \mathrm{Ethereum},\mathrm{Solana},\mathrm{BSC},\mathrm{Arbitrum},\mathrm{Avalanche}\}}
\end{split}
\end{equation}
where
\[
  \mathrm{native}_i = 
  \begin{cases}
    \$\mathrm{ETH},       & \text{if } chain_i = \mathrm{Ethereum},\\
    \$\mathrm{SOL},       & \text{if } chain_i = \mathrm{Solana},\\
    \$\mathrm{BNB},       & \text{if } chain_i = \mathrm{BSC},\\
    \$\mathrm{ARB},       & \text{if } chain_i = \mathrm{Arbitrum},\\
    \$\mathrm{AVAX},      & \text{if } chain_i = \mathrm{Avalanche},
  \end{cases}
\]
and every $F^{\mathrm{activity}}_{\mathrm{chain}_i,t}{}'$ contains activity variables for $chain_{i}$. Investors would react to changes in users' activity on the chain proxied by our chain-level activity variables, so only the chain-level activity change accompanied with chain portfolio returns matter to investors' decision-making. Investors on Ethereum may react this part of "relevant" activity level change, and we measure this conditional effect of activity level change by different chains to Ethereum using the following specification (the "baseline" non-linear model):
\begin{equation}\label{eq_nonlinear}
\begin{split}
R^{\mathrm{All}}_{\mathrm{Ethereum},t}
  &= \alpha_{0}
    + \alpha_{1}\,R^{\mathrm{CEX}}_{\mathrm{Ethereum},t}
    + \bm{\beta'_{0}}\,R^{\mathrm{All}}_{\mathrm{Ethereum},t-1}\,F^{\mathrm{activity}}_{\mathrm{Ethereum},t-1}\\
  &\quad
    +\sum_{\substack{chain_i\in\mathrm{Chains}}}
      \bm{\beta'_{i}}\,R^{\mathrm{All}}_{\mathrm{chain}_{i},t}\,F^{\mathrm{activity}}_{\mathrm{chain}_i,t}
    + e_{t}\,,\\
    & \mathrm{Chains}\,=\,\{ \mathrm{Solana},\mathrm{BSC},\mathrm{Arbitrum},\mathrm{Avalanche}\},
\end{split}
\end{equation}
where $i\,\in\,\{1,2,3,4\}$ and $\bm{\beta'_{i}}\,=\,\left[ \beta_{i0}, \beta_{i1}, \beta_{i2} \right]$, as well as the versions replacing $R^{\mathrm{All}}_{\mathrm{Ethereum},t}$ with $R^{\mathrm{non-CEX}}_{\mathrm{Ethereum},t}$ or $R^{\mathrm{Local}}_{\mathrm{Ethereum},t}$ at the LHS, while replacing $R^{\mathrm{All}}_{\mathrm{Ethereum},t}$ with $R^{\mathrm{non-CEX}}_{\mathrm{Ethereum},t}$ or $R^{\mathrm{Local}}_{\mathrm{Ethereum},t}$ at the RHS. We also add the vector of global market factors, i.e., defined in Equation (\ref{global_market_factors}), into our specification to control the information from stock markets:
\begin{equation}\label{eq_nonlinear_macro}
\begin{split}
R^{\mathrm{All}}_{\mathrm{Ethereum},t}
  &= \alpha_{0}
    + \alpha_{1}\,R^{\mathrm{CEX}}_{\mathrm{Ethereum},t}
    + \bm{\beta'_{0}}\,R^{\mathrm{All}}_{\mathrm{Ethereum},t-1}\,F^{\mathrm{activity}}_{\mathrm{Ethereum},t-1}\\
  &\quad
    +\sum_{\substack{chain_i\in\mathrm{Chains}}}
      \bm{\beta'_{i}}\,R^{\mathrm{All}}_{\mathrm{chain}_{i},t}\,F^{\mathrm{activity}}_{\mathrm{chain}_i,t}    +\bm{\theta'}\,F^{\mathrm{market}}_{\mathrm{global},t}
    + e_{t}\,,\\
    & \mathrm{Chains}\,=\,\{ \mathrm{Solana},\mathrm{BSC},\mathrm{Arbitrum},\mathrm{Avalanche}\}.
\end{split}
\end{equation}
For both Equation (\ref{eq_nonlinear}) and Equation (\ref{eq_nonlinear_macro}), the coefficient of how return of the chain portfolio $R^{\mathrm{All}}_{\mathrm{chain}_{i},t}$ is interdependent with $R^{\mathrm{All}}_{\mathrm{Ethereum},t}$ is determined by $\bm{\beta'_{i}}\,F^{\mathrm{activity}}_{\mathrm{chain}_i,t}$.
Among the coefficients in  $\bm{\beta'_{i}}\,=\,\left[ \beta_{i0}, \beta_{i1}, \beta_{i2} \right]$, 
$\beta_{i1}$ and $\beta_{i2}$ measure how the interdependence changes conditional on the chain-level activity change on $chain_{i}$, proxied by $R_{native_{i},t}$ and $\mathrm{SR}_{chain_{i},t}$, while 
$\beta_{i0}$ measures the unconditional interdependence. For investors on Ethereum, good news (bad news) about other chains, e.g., $chain_{i}$, will attract their attention to (away from) $chain_{i}$ as well as liquidity, following with negative (positive) return on Ethereum while positive (negative) return on $chain_{i}$. Similar argument 
is proposed in \textcite{barber2008all}, which shows that attention is a major determinant of buying stocks. Therefore,
 $\beta_{i1}$ and $\beta_{i2}$ are expected to be negative and capture the negative spillover effects. 
 Since chain portfolio returns may also contain systematic information, which drives markets co-move, we include
$ F^{\mathrm{market}}_{\mathrm{global},t}$ to control the systematic information by the same macro economic trend in Equation (\ref{eq_nonlinear_macro}).

%Our chain-level activity variables, e.g., native token price and its staking reward rate, reflect major changes and trends on the chain that related to significant changing demand and supply of the native token, such as launches of major protocols and technology upgrading. Some news that does not drive up native token demand may be not captured by those variables. To consider news about individual crypto assets on-chain, of which the volume affect demand on native token trivially, we adopt extreme returns as the proxies for investor attention.\footnote{See \textcite{barber2022attention} for the summary of proxies for investor attention.} According to \textcite{barber2008all}, investors respond to news by purchasing stocks, inducing extreme returns on the days of news release. 

Our chain-level activity variables, such as the native token price and its staking reward rate, capture major shifts in demand and supply for the native token, including those triggered by the launch of significant protocols or technological upgrades. However, news that does not materially affect native token demand may not be reflected in these variables. To account for asset-specific news on-chain, whose trading volume has only a marginal impact on the native token, we employ extreme returns as proxies for investor attention.\footnote{See \textcite{barber2022attention} for a survey of proxies for investor attention.} Following \textcite{barber2008all}, investors tend to react to news by purchasing stocks, which induces extreme returns on the days of the news release.

In our study, we proxy days with news about $chain_{i}$ using negative and positive extreme return dummies (the lowest 5\% and the highest 5\%) for the left and right tails of the return distribution, denoted by $D^{L}_{chain_{i}}$ and $D^{U}_{chain_{i}}$, $chain_i\in\{\mathrm{Solana},\,\mathrm{BSC},\,\mathrm{Arbitrum},\,\mathrm{Avalanche}\}$. With the extreme return dummies, the following specification can be proposed:
\begin{equation}\label{eq_nonlinear_extremeR_macro}
\begin{split}
R^{\mathrm{All}}_{\mathrm{Ethereum},t}
  &= \alpha_{0}
    + \alpha_{1}\,R^{\mathrm{CEX}}_{\mathrm{Ethereum},t}
    + \bm{\beta'_{0}}\,R^{\mathrm{All}}_{\mathrm{Ethereum},t-1}\,F^{\mathrm{activity}}_{\mathrm{Ethereum},t-1}\\
  &\quad
    +\sum_{\substack{chain_i\in\mathrm{Chains}}}
      \bm{\beta'_{i}}\,R^{\mathrm{All}}_{\mathrm{chain}_{i},t}\,\mathrm{FD}^{\mathrm{activity}}_{\mathrm{chain}_i,t}    +\bm{\theta'}\,F^{\mathrm{market}}_{\mathrm{global},t}
    + e_{t}\,,\\
    & \mathrm{Chains}\,=\,\{ \mathrm{Solana},\mathrm{BSC},\mathrm{Arbitrum},\mathrm{Avalanche}\}, and \\
    & \mathrm{FD}^{\mathrm{activity}}_{\mathrm{chain}_i,t}\;=\;
    F^{\mathrm{activity}}_{\mathrm{chain}_i,t}
    \;\cup\;
    \bigl[D^{U}_{chain_{i},t},\; D^{L}_{chain_{i},t}\bigr],
\end{split}
\end{equation}
where $\mathrm{FD}^{\mathrm{activity}}_{\mathrm{chain}_i,t}$ denotes the vector of factors including chain-level activity variables and extreme return dummies. 
$\bm{\beta'_{i}}\,=\,\left[ \beta_{i0}, \beta_{i1}, \beta_{i2}, \beta_{i3}, \beta_{i4} \right]$, while $\beta_{i3}$ and $\beta_{i4}$ represent coefficients for $D^{U}_{chain_{i},t}$ and $D^{L}_{chain_{i},t}$ respectively. 
On the half-day $t$, if $D^{U}_{chain_{i},t} = 1$ ($D^{L}_{chain_{i},t}=1$), which implies news released on half-day $t$ for $chain_{i}$, it is expected that investors attracted by those news tend to sell (buy) assets on Ethereum while buying (selling) on $chain_{i}$, rendering negative return correlation between Ethereum and $chain_{i}$. 
This effect of attention allocation channeled through $D^{U}_{chain_{i},t}$ and $D^{L}_{chain_{i},t}$ would be captured by negative $\beta_{i3}$ and $\beta_{i4}$.

\subsubsection{Volatility clustering}

%As widely agreed, assets return volatility exhibit clustering (see \textcite{connolly2000stock,lin1994bulls} for stock markets and \textcite{katsiampa2019empirical,gupta2022empirical} for cryptocurrencies). To account for this clustering effect, as well as potential asymmetric clustering in up- and down-market \parencite{glosten1993relation,connolly2000stock}, we model the residuals in our specifications, $e_{t}$, using Glosten-Jagannathan-Runkle (GJR) asymmetric GARCH approach as follows:

As is well documented, asset return volatility tends to exhibit clustering (see \textcite{connolly2000stock,lin1994bulls} for stock markets; \textcite{katsiampa2019empirical,gupta2022empirical} for cryptocurrencies). To account for this phenomenon, along with potential asymmetries in volatility clustering between up- and down-markets \parencite{glosten1993relation,connolly2000stock}, we model the residuals, $e_{t}$, using the Glosten–Jagannathan–Runkle (GJR) asymmetric GARCH framework, specified as follows:

\begin{equation}\label{garch_model}
\begin{split}
    e_{t} \sim \mathcal{N}(0, \sigma_t^2), \;
    \sigma_t^2 = \omega + \sum_{i=1}^p \alpha_i e_{t-i}^2 + \sum_{j=1}^o \gamma_j e_{t-j}^2 \cdot \mathbb{I}_{\{e_{t-j} < 0\}} + \sum_{k=1}^q \beta_k \sigma_{t-k}^2,
\end{split}
\end{equation}
where $\alpha_i$ accounts for the ARCH coefficients (symmetric innovation effect),
$\gamma_j$ for the asymmetric response coefficients (leverage effect),
 $\beta_k$ for the GARCH coefficients (volatility persistence),
and $\mathbb{I}_{\{e_{t-j} < 0\}}$ as the indicator function equal to 1 when \( e_{t-j} < 0 \). 

\section{Empirical Results}

\subsection{Data and Descriptive Statistics}\label{main_section_data_stats}

\subsubsection{On-Chain Assets}

%We stick to the coins list for each chain from \href{https://docs.coingecko.com/reference/introduction}{Coingecko API}, where we can fetch the list of coins still actively traded with detailed information on each chain, such as supported chains, listed exchanges, and the addresses on the chains supported (multi-chain existence case discussed in Section \ref{section_chain_portfolio_construction}). Then for each asset, the largest pool \footnote{We call a trading pair supported by a smart contract as "liquidity pool", through which traders can swap between the two tokens paired, e.g., WETH/WBTC.} on the chain under examining would be used to fetch the price using the \href{https://docs.coingecko.com/reference/top-pools-network}{Coingecko on-chain DEX API}. If the API fails to fetch the relevant data, we collect "events", \footnote{Upon each swap, the pool smart contract would emit "events" containing all the swapping parameters and after-swap pool status. Details can be found at the documentation of Uniswap protocol: \href{https://docs.uniswap.org/contracts/v3/reference/core/interfaces/pool/IUniswapV3PoolEvents}{https://docs.uniswap.org/contracts/v3/reference/core/interfaces/pool/IUniswapV3PoolEvents}.} which contains parameters of each trades, to form the pricing chart by communicating with the chain via a full-node, i.e., \href{https://www.quicknode.com}{QuickNode} in our case. 

We rely on the coin lists provided by the \href{https://docs.coingecko.com/reference/introduction}{Coingecko API}
, which contain actively traded assets and detailed metadata for each chain, including supported networks, listed exchanges, and contract addresses (the case of multi-chain assets is discussed in Section~\ref{section_chain_portfolio_construction}). For each asset, we obtain its price from the largest liquidity pool\footnote{A liquidity pool refers to a trading pair supported by a smart contract, which enables token swaps between the two paired assets (e.g., WETH/WBTC).} on the chain under examination via the Coingecko on-chain DEX API
. If the API fails to retrieve the relevant data, we instead collect events\footnote{Each swap triggers the pool smart contract to emit an event containing all swap parameters and the post-swap pool state. See the Uniswap V3 protocol documentation for details: \href{https://docs.uniswap.org/contracts/v3/reference/core/interfaces/pool/IUniswapV3PoolEvents}{https://docs.uniswap.org/contracts/v3/reference/core/interfaces/pool/IUniswapV3PoolEvents}.}, which
 record the parameters of individual trades, to reconstruct the price series by querying the blockchain through a full node (in our case, \href{https://www.quicknode.com}{QuickNode}
).

%Different from previous studies using pricing data from CEXs or third-party data provider directly, for which assets are uniformly priced by USD, our study retain assets to be priced by the token in the DEX pair originally on-chain, priced by the native token in most cases. Since most of the assets traded on-chain are paired with their native tokens, simply treating them as valued by USD incorporates pricing change of the native tokens, distorting the demand/supply reflected by the asset price. Just like in the stock markets we use domestic currency to price the domestic stock, as well as returns, we use domestic currency for a chain, i.e., native tokens, to price crypto assets on this chain.

Unlike prior studies that rely on pricing data from CEXs or third-party providers, where assets are uniformly denominated in USD, we retain the original on-chain pricing convention, where assets are quoted in their paired tokens, typically the chain’s native token. Since most on-chain assets are traded against their native tokens, converting all prices into USD would mechanically embed fluctuations in the native token price, thereby distorting the demand–supply dynamics reflected in the asset’s own price. By analogy to equity markets, where domestic stocks are priced and returns are measured in the domestic currency, we treat each chain’s native token as its “domestic currency” and price the corresponding on-chain assets accordingly.

In addition to pricing data, we construct a series of half-day market capitalization data via three sources and then average them to form the usable values. \href{https://developers.coinranking.com/api/documentation}{Coinranking} provides API endpoint for market capitalization, which we use as the supplement for the data from Coingecko. We also use the series of circulating supply fetched from \href{https://developers.coindesk.com}{CoinDesk API} and asset price to calculate as for the third source of market capitalization.

When constructing the four types of portfolio in Section \ref{section_chain_portfolio_construction}, we exclude liquid staking/re-staking tokens, wrapped native tokens, and stablecoins. The former two types of tokens can be regarded as "pegged" to the chain's native token, so they perfectly co-move with native tokens. \footnote{People stake their native tokens to some protocols in return for reward or interest. They would get staking/re-staking tokens representing their share of staking. Readers can check for the list of staking/re-staking tokens at: \href{https://defillama.com/lst}{https://defillama.com/lst}.} Including them would result in overestimated correlation between the chain portfolio return and the chain's native token. Stablecoins are usually pegged to fiat currency. e.g., USD or EURO, so their value remains stable all the time. Although they have accumulated substantial market capitalizations due to their widespread use, their inclusion would render portfolio returns spuriously independent of market dynamics. %Due to their widespread use, stablecoins have accumulated a large market capitalization. Including them in our portfolio would make its return spuriously independent.

\begin{figure}[H]
  \centering
  \includegraphics[width=\linewidth]{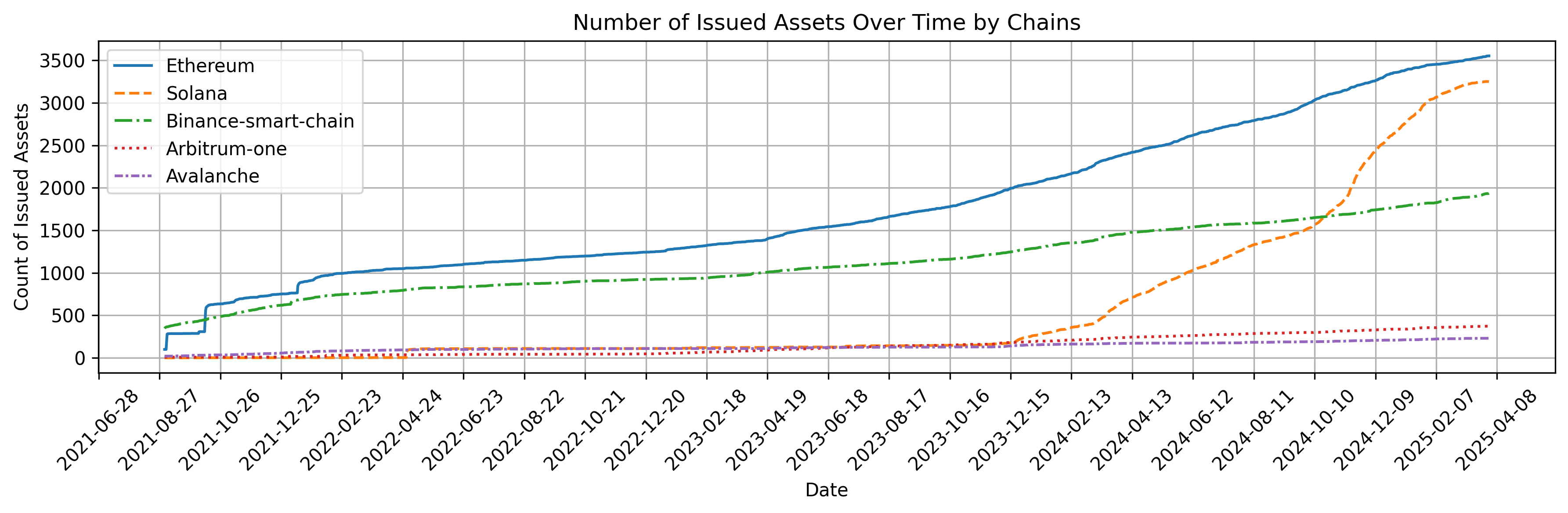}  
  \caption{Number of Issued Assets on Different Chains Over Time. The assets lists for the chains are fetched from \href{https://docs.coingecko.com/reference/introduction}{Coingecko API} and Coingecko filter assets with conditions for assets in their list. Details of the filtering conditions can be found at: \href{https://support.coingecko.com/hc/en-us/articles/4498809321369-Why-is-my-token-not-listed-on-CoinGecko}{https://support.coingecko.com/hc/en-us/articles/4498809321369-Why-is-my-token-not-listed-on-CoinGecko} }
  \label{fig:number-assets}
\end{figure}

\begin{figure}[H]
  \centering
  \includegraphics[width=\linewidth]{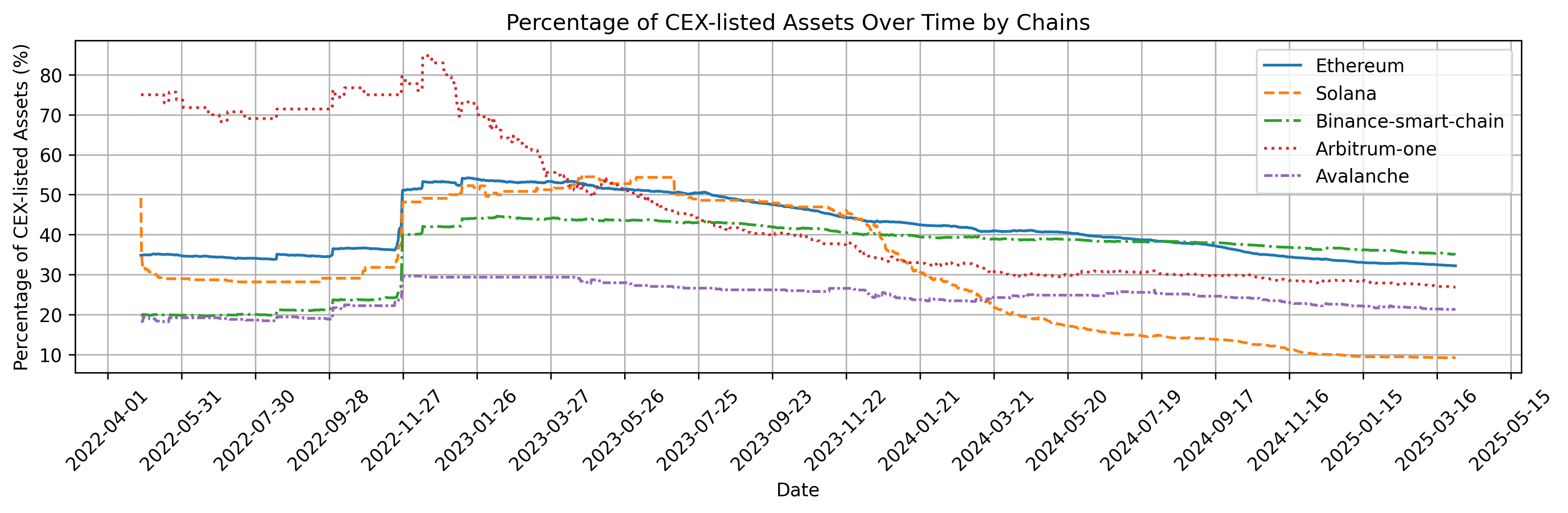}  
  \caption{Percentage of CEX-listed Assets on Different Chains Over Time. Dates of CEX-listing for the assets are fetched via \href{https://developers.coindesk.com}{CoinDesk API}, and we treat the date when the first trade on any of the largest 100 CEXs happened as the date of CEX-listing.}
  \label{fig:cex-listed-assets}
\end{figure}

\begin{figure}[H]
  \centering
  \includegraphics[width=\linewidth]{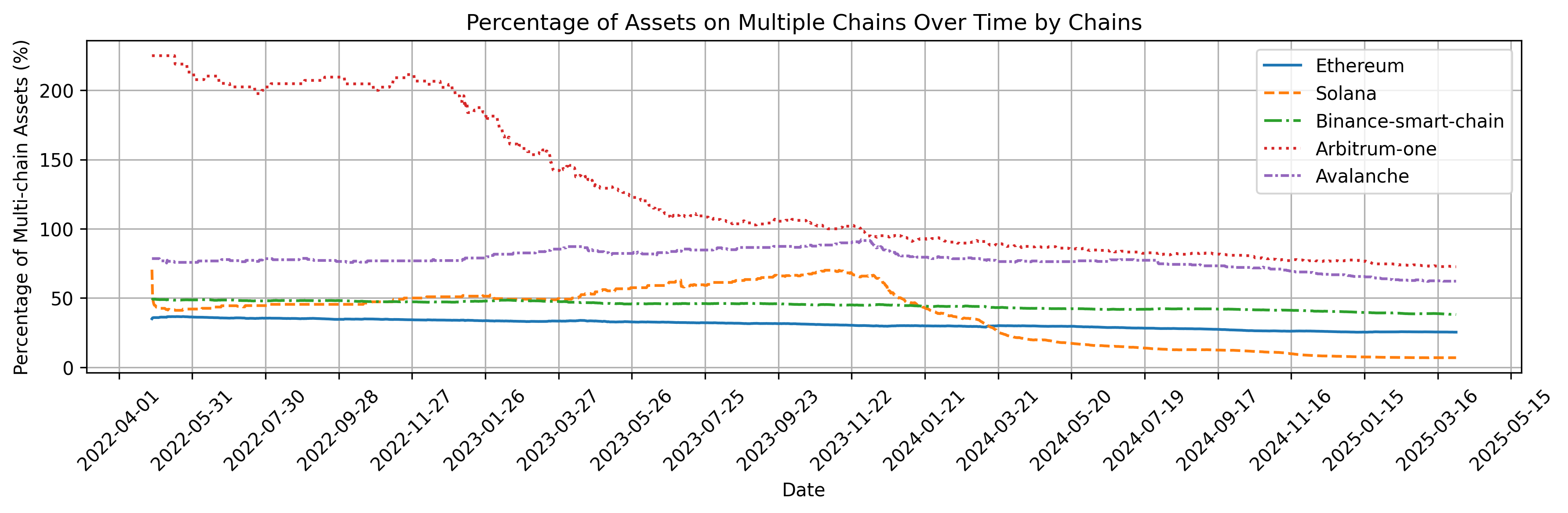}  
  \caption{Percentage Assets on Multiple Chains for Different Chains Over Time. We determine the date of issuance of an asset on a chain by the top pool creation date fetched from \href{https://docs.coingecko.com/reference/top-pools-network}{Coingecko on-chain DEX API}. Here the percentage of multi-chain assets for $chain_{i}$ is calculated as $\frac{\mathrm{Number\;of\;assets\;also\;exiting\;on\;other\;chains}}{\mathrm{Number\;of\;assets\;actively\;traded\;on\;chain_{i}}}$, so if $chain_i$ is a chain receiving assets from other chains, this percentage could be over 100\%. For $Arbitrum$, its percentage before Dec 2023 is larger than 100\%, because this chain is a layer-2 chain for Ethereum, receiving and taking over working load from Ethereum.}
  \label{fig:multichain-assets}
\end{figure}

Figure \ref{fig:number-assets} depicts the number of assets actively traded on the five chains over time in our study, while Figure \ref{fig:cex-listed-assets} plots the percentage of assets listed on CEXs for each chain. It is shown that although there are more and more assets issued on each chain, the portion of CEX-listed keeps dropping, especially after 2023, a pattern reflecting the proliferation of DEXs, as well as tradings on-chain. We also plot the percentage of assets existing on multiple chains for each chain in Figure \ref{fig:multichain-assets}. In both Figure \ref{fig:cex-listed-assets} and \ref{fig:multichain-assets}, Solana is of the lowest percentage at the end of our sampling period (31st Mar, 2025), reflecting its high level on-chain activities. %The half-day portfolio returns are calculated by weighted averaging individual assets' half-day returns in the manner of continuous compounding. The descriptive statistics for all the four types of portfolio returns are reported in Table \ref{tab:stats_portfolio}. The starting date of sampling periods for the five chains are: from 3rd Sep 2021 for Ethereum, from 28th Apr 2022 for Solana, from 1st Sep 2021 for BSC, from 31st Dec 2021 for Arbitrum, from 1st Sep 2021 for Avalanche.

Half-day portfolio returns are computed as the average of individual assets’ half-day returns weighted by market capitalization under continuous compounding. Table~\ref{tab:stats_portfolio} reports the descriptive statistics for all four types of portfolio returns. The sampling periods for the five chains begin on the following dates: September 3, 2021 for Ethereum; April 28, 2022 for Solana; September 1, 2021 for BSC; December 31, 2021 for Arbitrum; and September 1, 2021 for Avalanche.

\subsubsection{Chain-level Activity and Global Market Variables}

%We fetch the historical native token prices, i.e., for $\$\mathrm{ETH}$, $\$\mathrm{SOL}$, $\$\mathrm{BNB}$, and $\$\mathrm{AVAX}$, using \href{https://developers.coindesk.com}{CoinDesk API}. The sampling period starts from 26th Dec 2021 for $\$\mathrm{ETH}$, $\$\mathrm{SOL}$, $\$\mathrm{BNB}$, and $\$\mathrm{AVAX}$, while from 17th Mar 2023 for $\$\mathrm{ARB}$. \footnote{$\$\mathrm{ARB}$ was launched on March 2023.} Data of historical staking reward on the chains are collected from \href{https://www.stakingrewards.com/}{https://www.stakingrewards.com/}, lending rates of the native token on Arbitrum are fetched with \href{https://web3-ethereum-defi.readthedocs.io/api/aave_v3/index.html}{AAVE Protocol API}.\footnote{AAVE (\href{https://aave.com}{https://aave.com}) is the leading protocl of on-chain lending on multiple chains, such as Ethereum, BSC, and Arbitrum.} The staking reward data starts from 1st Jan 2022 for Ethereum and Solana, from 6th July 2022 for BSC, and from 16th Mar 2022 for Arbitrum and Avalanche. 

We obtain historical native token prices—$\$\mathrm{ETH}$, $\$\mathrm{SOL}$, $\$\mathrm{BNB}$, and $\$\mathrm{AVAX}$—using \href{https://developers.coindesk.com}{CoinDesk
 API}. The sampling period starts on December 26, 2021 for these four tokens, and on March 17, 2023 for $\$\mathrm{ARB}$.\footnote{$\$\mathrm{ARB}$ was launched in March 2023.} Historical staking reward data are collected from \href{https://www.stakingrewards.com/}{https://www.stakingrewards.com}
, while lending rates of native tokens on Arbitrum are obtained from \href{https://web3-ethereum-defi.readthedocs.io/api/aave_v3/index.html}{AAVE
 Protocol API}.\footnote{AAVE (\href{https://aave.com}{https://aave.com}
) is a leading on-chain lending protocol operating across multiple chains, including Ethereum, BSC, and Arbitrum.} The staking reward series begin on January 1, 2022 for Ethereum and Solana, on July 6, 2022 for BSC, and on March 16, 2022 for Arbitrum and Avalanche.

%Historical data of S\&P 500 index, FTSE 100 index, and Heng Seng Index are downloaded from \href{https://www.marketwatch.com}{https://www.marketwatch.com}. We collect series of historical Hong kong Inter-bank Offered Rate from \href{https://en.macromicro.me/collections/1626/hk-finance-relative/3752/hkd-interest-settlement-rates}{MacroMicro}, 1-month Euro Interbank Offered Rate from \href{https://www.euribor-rates.eu/en/}{https://www.euribor-rates.eu/en/}, and 1-month U.S. Treasury bill rate from \href{https://www.investing.com}{https://www.investing.com}. The sampling periods for the six global market variables all start from 4th Jan 2022 and end at 31st Mar 2025. Descriptive statistics for chain-level activity and global market variables are reported in Table \ref{tab:stats_conditions_var}.

Historical data for the S\&P 500 Index, FTSE 100 Index, and Hang Seng Index are obtained from \href{https://www.marketwatch.com}{https://www.marketwatch.com}
. We collect the Hong Kong Interbank Offered Rate (HIBOR) series from \href{https://en.macromicro.me/collections/1626/hk-finance-relative/3752/hkd-interest-settlement-rates}{MacroMicro}
, the 1-month Euro Interbank Offered Rate (EURIBOR) from \href{https://www.euribor-rates.eu/en/}{https://www.euribor-rates.eu/en/}
, and the 1-month U.S. Treasury bill rate from \href{https://www.investing.com}{https://www.investing.com}
. The sampling periods for all six global market variables span from January 4, 2022 to March 31, 2025. Descriptive statistics for the chain-level activity variables and global market variables are reported in Table~\ref{tab:stats_conditions_var}.

\subsection{Results of Regression Estimation}\label{main_section_results}

\subsubsection{Linear Models}

%Table \ref{tab:baseline} reports the estimation results for the baseline linear models illustrated by Equation (\ref{eq_baseline_all}), (\ref{eq_baseline_nonCEX}), and (\ref{eq_baseline_this_chain_only}). The results reveal the first pattern of assets across chains we find: negative spillover effects generally exist on chains. 
%If we consider results based on portfolios with all on-chain assets (Panel A), it shows negative return correlations, i.e., negative $\beta_i\,(i\in\{1,2,3,4\})$, for all the chains except Arbitrum. 
%As discussed in Section \ref{section_chain_portfolio_construction}, relationships between returns of chain portfolio returns would be better captured without bias using $R^{Local}_{\mathrm{chain_i},t}$ as the dependent variable, results of which are reported in Panel C. 
%For Ethereum (column (1) in Panel C), we obtain negative estimated $\beta_3$ (-0.140), which indicate the potential negative spillover effects from Arbitrum, i.e., assets return on Ethereum tend to drop when return of assets on Arbitrum rises. 
%Similar significantly negative $\beta_i$s are estimated for Arbitrum ($\beta_3=-0.176$) and Avalanche ($\beta_4=-0.047$) on BSC (Column (3) in Panel C), as well as Arbitrum ($\beta_4=-0.160$) on Avalanche (Column (5) in Panel C). 
%It is worth noting that although co-movements between on-chain assets also emerge, negative correlations are shown to be more prevalent than co-movements. 
%We obtain evidence of co-movements only in the case of Arbitrum, i.e., between Arbitrum and Solana ($\beta_2=0.028$) illustrated in Column (4) of Panel C.

Table~\ref{tab:baseline} reports the estimation results for the baseline linear models defined in Equations~(\ref{eq_baseline_all}), (\ref{eq_baseline_nonCEX}), and (\ref{eq_baseline_this_chain_only}). The results reveal the first key cross-chain pattern: negative spillover effects are generally present.

Panel A, which is based on portfolios including all on-chain assets, shows negative return correlations, i.e., negative $\beta_i ,(i \in \{1,2,3,4\})$, for all chains except Arbitrum. As discussed in Section~\ref{section_chain_portfolio_construction}, these relationships are better captured without bias when using $R^{Local}_{\mathrm{chain_i},t}$ as the dependent variable; the corresponding results are reported in Panel C.

For Ethereum (Column (1), Panel C), we estimate a significantly negative $\beta_3$ of –0.140, indicating potential negative spillover from Arbitrum: asset returns on Ethereum tend to decline when Arbitrum asset returns rise. Similar significantly negative $\beta_i$ estimates are found for Arbitrum ($\beta_3 = -0.176$) and Avalanche ($\beta_4 = -0.047$) with respect to BSC (Column (3), Panel C), as well as for Arbitrum ($\beta_4 = -0.160$) with respect to Avalanche (Column (5), Panel C).

Although some evidence of cross-chain co-movements also emerges, negative correlations appear more prevalent. The only instance of positive co-movement is between Arbitrum and Solana ($\beta_2 = 0.028$), as reported in Column (4) of Panel C.

%Another pattern revealed by our results is the co-movements between assets listed on CEX and those have not been CEX-listed from the same chain, as reported in Panel B of Table \ref{tab:baseline}. 

%For assets issued on Solana, those CEX-listed co-move with those have not been cex-listed, evidenced by the significantly positive estimated $\beta_0$ (0.028), while we find similar case for Avalanche with $\beta_0=0.357$. Besides, chain assets returns exhibit a reversal pattern — chain portfolio returns tend to reverse the direction half day before, evidenced by the negative $\alpha_1$ in all of the three panels.  

%To control the co-movements driven by the common systematic information from the global stock market, we add global market variables to our regression specification as illustrated by Equation (\ref{eq_linear_macro}), estimation results of which are reported in Table \ref{tab:baseline_macro}. 
%After controlling this potential source of co-movement, we find more prevalent negative correlation of returns among chains, with a larger number of chain portfolio pairs exhibiting negative spillover. 

%In Panel A, it shows that Solana ($\beta_2=-0.007$) and Arbitrum ($\beta_3=-0.088$) are with assets of negatively correlated return with BSC (Column (3) in Panel A), while not in Table \ref{tab:baseline}. 
%Also, in Panel C, $\beta_2$ starts to be estimated as negative (-0.018) for BSC in addition to $\beta_3$ and $\beta_4$, providing further evidence on the prevalence of the negative spillover effects. 

Another pattern revealed by our results is the co-movement between assets listed on CEXs and those not listed on CEXs within the same chain, as reported in Panel B of Table~\ref{tab:baseline}. For Solana-issued assets, CEX-listed tokens co-move with their non–CEX-listed counterparts, as evidenced by a significantly positive $\beta_0$ (0.028). A similar case is found for Avalanche, with $\beta_0 = 0.357$. In addition, chain asset returns exhibit a short-term reversal pattern: chain portfolio returns tend to reverse the direction from the previous half-day, as indicated by the negative $\alpha_1$ estimates across all three panels.

To account for co-movements potentially driven by common systematic information from global stock markets, we augment the specification with global market variables as in Equation~(\ref{eq_linear_macro}); the results are reported in Table~\ref{tab:baseline_macro}. After controlling for this potential source of correlation, negative return spillovers across chains become even more pronounced, with a larger number of chain portfolio pairs exhibiting significant negative correlations.

In Panel A, Solana ($\beta_2 = -0.007$) and Arbitrum ($\beta_3 = -0.088$) show negatively correlated returns with BSC (Column (3), Panel A), a pattern not observed in Table~\ref{tab:baseline}. In Panel C, $\beta_2$ for BSC is also estimated to be negative ($-0.018$), in addition to the previously negative $\beta_3$ and $\beta_4$, providing further evidence of the prevalence of negative spillover effects.

The above results are based on sampling period from 28th Apr 2022 to 31st Mar 2025, while regressions on the rest of specifications defined in Section \ref{section_models} are based on data from 17th Mar 2023 to 31st Mar 2025 because of the late launch date of \$ARB on 17th Mar 2023. 
To ensure estimations of the rest models are comparable, we redo the estimation with data from 17th Mar 2023 to 31st Mar 2025 for baseline linear models, results of which are reported by Table \ref{tab:baseline_truncated} in \nameref{appendex}. 
Although with truncated data, results reported in Table \ref{tab:baseline_truncated} are almost the same with those in Table \ref{tab:baseline}. 
The only difference is that results on truncated data exhibits more pronounced negative correlations from Ethereum. 
In Panel A, asset returns on three of the five chains tend to be negatively correlated with those on Ethereum: Solana ($\beta_1=-0.0007$ (significant at 10\% level) under Column (2)), BSC ($\beta_1=-0.017$ (significant at 5\% level) under Column (3)), and Avalanche ($\beta_1=-0.001$ (significant at 10\% level) under Column (5)). 
Similar patterns exist in Panel C, where it shows two out of the three chains with assets return negatively correlated with assets on Ethereum: Solana ($\beta_1=-0.0006$ (significant at 1\% level) under Column (2)) and Avalanche ($\beta_1=-0.007$ (significant at 5\% level) under Column (5)).

Table~\ref{tab:baseline_chain_macro} reports the estimation results for the linear models in Equation~(\ref{eq_linear_macro_crypto}), which incorporate both global market variables and chain-level activity variables. The results show that negative spillover effects persist even after controlling for chain-level activity. As expected, however, part of the negative spillover driven by investor attention to fundamental factors is absorbed by these activity variables, leading to fewer significantly negative $\beta_i$ estimates in Table~\ref{tab:baseline_chain_macro} compared with Table~\ref{tab:baseline_truncated}.

Specifically, the negative return correlations with Ethereum are largely captured by chain-level activity variables: $\beta_1$ becomes insignificant for Solana (Column (2)), BSC (Column (3)), and Avalanche (Column (5)) in Panel A of Table~\ref{tab:baseline_chain_macro}. A similar pattern holds in Panel C, where the negative return correlations with Ethereum also become insignificant for Solana (Column (2)) and Avalanche (Column (5)).

In addition, the inclusion of $R_{\$\mathrm{BTC},t}$, which serves as a leading indicator of systematic crypto market conditions, helps to absorb common market co-movements. As a result, negative spillover effects become more pronounced in certain cases. For instance, Avalanche’s portfolio return exhibits a significantly negative correlation with Ethereum’s ($\beta_4 = -0.078$, significant at the 5\% level; Column (1), Panel A).

Overall, the linear model results provide robust evidence of negative correlations among assets across chains, even after accounting for chain-level activity and global stock market performance. While some of the spillovers are explained by activity variables that proxy for investor attraction, negative spillover effects remain. In the next section, we further explore the underlying channels of these effects using non-linear models.

\subsubsection{Non-Linear Models}

Table \ref{tab:nonlinear} reports the estimation results of the baseline non-linear models illustrated by Equation (\ref{eq_nonlinear}), in which we decompose slope coefficients of chain portfolio returns into two parts: the unconditional part (with coefficient $\beta_{i0},\;i\in\{1,2,3,4\}$) and the part conditional on the change of chain-level activity variables. 

%As we discussed in Section \ref{section_chain_condition_variables}, return on the native token and staking reward rate can be use to proxy investors' preference to the underlying chain. If the chain-level activity variables can indicate investors' preference to a certain chain, then the marginal change of those variables captures whether the chain is attracting attention or losing attention, which we proxy with $\mathrm{SR}_{chain_i,t}$ and $R_{native_i,t}$, the unexpected portion of chain-level activity variables from ARIMA approach. Therefore, $\beta_{i1}\, and \,\beta_{i2},\;i\in\{1,2,3,4\}$ represent how the portfolio by assets on $chain_0$ interdependent with returns on other chains ($chain_i,\;i\in\{1,2,3,4\}$) when other chains are getting or losing attention.

As discussed in Section~\ref{section_chain_condition_variables}, the returns on the native token and the staking reward rate serve as proxies for investors’ preferences toward the underlying chain. If these chain-level activity variables capture investor attention to a given chain, then their marginal changes reflect whether the chain is gaining or losing attention. We proxy these changes with $\mathrm{SR}_{chain_i,t}$ and $R_{native_i,t}$, defined as the unexpected components of the activity variables extracted via an ARIMA approach. Accordingly, $\beta_{i1}$ and $\beta_{i2}$ ($i \in \{1,2,3,4\}$) measure how the portfolio of assets on $chain_0$ co-moves with returns on other chains ($chain_i$) when those chains are either attracting or losing investor attention.

%In both Panel A and Panel C of Table \ref{tab:nonlinear},  it is shown that most of $\beta_{i1}\, and \,\beta_{i2}\;(i\in\{1,2,3,4\})$ are estimated to be negative, although some of them are insignificant. Negative spillover effects become more prominent when other chains are attracting attention, i.e., more negative slope on portfolio returns. For investors on BSC, they are more willing to sell (buy) their assets on BSC when Arbitrum is drawing (losing) attention accompanied by price rising (dropping) of assets on Arbitrum, evidenced by the negative estimate of $\beta_{31}$ (-0.204 (significant at 5\% level)) for BSC under Column (3) in Panel A. For the more stringent examination of the correlations reported in Panel C, we find that investors on Arbitrum are among the most easily swayed. Return of the chain portfolio on Arbitrum tend to be negatively correlated with all of the other four chains. When investors are attracted by the other four chains, represented by rising $\mathrm{SR}_{chain_i,t}$ or $R_{native_i,t}$, they may sell assets on Arbitrum while buying assets on the other chains, indicated by the negative estimates such as $\beta_{32}$ (-0.670), $\beta_{22}$ (-0.158), and $\beta_{12}$ (-0.650) in Column (4) of Panel C (all of the three parameters, $\beta_{32}$ $\beta_{22}$ $\beta_{12}$, are estimated to be significantly negative in 1\% level).

In both Panel A and Panel C of Table~\ref{tab:nonlinear}, most estimates of $\beta_{i1}$ and $\beta_{i2}$ ($i \in {1,2,3,4}$) are negative, though some are statistically insignificant. Negative spillover effects become more pronounced when other chains attract investor attention, as reflected in more negative slopes for portfolio returns.

For BSC, investors tend to sell (buy) their assets when Arbitrum is drawing (losing) attention, accompanied by rising (falling) Arbitrum asset prices, evidenced by the significantly negative estimate of $\beta_{31} = -0.204$ (5\% level) in Column (3), Panel A. Under the more stringent specification in Panel C, investors on Arbitrum appear particularly susceptible to attention shifts: the return on Arbitrum’s chain portfolio is negatively correlated with all four other chains. When investors are drawn to other chains, captured by increases in $\mathrm{SR}_{chain_i,t}$ or $R_{native_i,t}$, they tend to sell assets on Arbitrum while reallocating to those chains. This is indicated by the significantly negative coefficients $\beta_{32} = -0.670$, $\beta_{22} = -0.158$, and $\beta_{12} = -0.650$ in Column (4), Panel C (all significant at the 1\% level).
 
%Besides decomposing potential negative spillover effects among chains, we also decompose the return reversal recorded in the last section. For example, as shown under Column (4) in Panel A of Table \ref{tab:baseline}, how portfolio return for Arbitrum at time $t$ can be related to its return in the last time period $t-1$ is decomposed into three parts which are captured by $\beta_{i0}$, $\beta_{i1}$ and $\beta_{i2}$ ($i\in\{1,2,3,4\}$). For BSC (Column (3) in Panel A) and Arbitrum (Column (4) in Panel A), we obtain estimated $\beta_{02}$ to be significantly positive (1.198 for BSC and 0.417 for Arbitrum), revealing that return reversal diminishes when the chain is attracting attention. 

%Table \ref{tab:nonlinear_macro} reports the estimation results of the non-linear models with global market variables as controls, as illustrated by Equation (\ref{eq_nonlinear_macro}). After controlling the common systematic information from stock markets, negative return correlations become more pronounced for BSC and Avalanche. As shown in Panel C, we detect unconditional significantly negative spillover effects to BSC (Column (3)) from both Solana ($\beta_{20}=-0.032$) and Arbitrum ($\beta_{30}=-0.239$), and greater the negative spillover for Arbitrum when Arbitrum is becoming more preferred ($\beta_{32}=-0.381$). Negative spillover effects are also shown to be accelerated for Avalanche (Column (5)) when investors are more inclined to Arbitrum ($\beta_{42}=-1.368$).
In addition to decomposing potential negative spillover effects across chains, we also examine the decomposition of the return reversal documented in the previous section. For instance, as shown in Column (4), Panel A of Table~\ref{tab:baseline}, the relationship between Arbitrum’s portfolio return at time $t$ and its return at $t-1$ can be decomposed into three components, captured by $\beta_{i0}$, $\beta_{i1}$, and $\beta_{i2}$ ($i \in \{1,2,3,4\}$). For BSC (Column (3), Panel A) and Arbitrum (Column (4), Panel A), the estimates of $\beta_{02}$ are significantly positive (1.198 for BSC and 0.417 for Arbitrum), indicating that return reversals diminish when the chain is attracting greater investor attention.

Table~\ref{tab:nonlinear_macro} presents the estimation results of the non-linear models augmented with global market variables, as specified in Equation~(\ref{eq_nonlinear_macro}). After controlling for common systematic information from global stock markets, negative return correlations become more pronounced for BSC and Avalanche. In Panel C, we detect unconditional significantly negative spillover effects on BSC (Column (3)) from both Solana ($\beta_{20} = -0.032$) and Arbitrum ($\beta_{30} = -0.239$), as well as stronger negative spillovers for Arbitrum when it becomes more preferred ($\beta_{32} = -0.381$). For Avalanche (Column (5)), negative spillover effects are further amplified when investors shift toward Arbitrum, as reflected in the strongly negative $\beta_{42} = -1.368$.

Finally, Table~\ref{tab:nonlinear_macro_extreme} reports the estimation results of the non-linear models with extreme return dummies as proxies for investor attention, as specified in Equation~(\ref{eq_nonlinear_extremeR_macro}). As before, the slope coefficients for each chain portfolio return are decomposed into unconditional and conditional parts. In addition to chain-level activity variables, we include extreme return dummies, $D^{U}_{chain_i,t}$ and $D^{L}_{chain_i,t}$, to capture the component of attention allocation beyond that reflected in activity variables. Significantly negative estimates of the coefficients on these dummies—$\beta_{i3}$ and $\beta_{i4}$ ($i \in \{1,2,3,4\}$)—indicate stronger negative spillovers when investors are drawn to other chains.

The results reveal that negative spillovers induced by attention allocation are pervasive in the crypto market. For Ethereum (Column (1), Panel A), investors tend to sell when assets on Solana and Arbitrum experience extreme upward moves, i.e., periods when investor attention shifts away from Ethereum ($\beta_{13} = -0.464$ for Solana; $\beta_{33} = -0.901$ for Arbitrum). Conversely, Ethereum investors tend to buy when Solana and Arbitrum undergo extreme downturns, i.e., when attention shifts away from those chains ($\beta_{14} = -0.473$ for Solana; $\beta_{34} = -0.345$ for Arbitrum).

A similar pattern is observed for BSC and Arbitrum. BSC investors sell (buy) when Solana experiences an extreme increase (decrease) in returns ($\beta_{23} = -0.152$; $\beta_{24} = -0.171$). Arbitrum investors respond analogously to extreme returns on Ethereum ($\beta_{13} = -0.062$; $\beta_{14} = -0.071$) and Solana ($\beta_{23} = -0.076$; $\beta_{24} = -0.077$). The more stringent specification in Panel C yields results consistent with those in Panel A.

%Although negative spillover effects due to attention allocation generally exists in the crypto market, it shows that the pattern is asymmetric. The evidence that investors on Ethereum selling assets when assets on Solana experience an extreme jump does not necessarily imply that investors on Solana would behave similarly when there comes extreme return on Ethereum. Assets on Solana are with prices adjusting in a relatively independent way, so that although extreme returns come on other chains, returns on Solana assets are not affected.

%Even though very few, we can still see positively estimated values for $\beta_{i3}\, and \,\beta_{i4}\;(i\in\{1,2,3,4\})$, such as $\beta_{43}=0.259$ and $\beta_{44}=0.263$. These cases can be ascribed to the cross-market re-balancing referred by \textcite{kodres2002rational,kyle1999contagion,fleming1998information}. To respond to shocks in one market, investors try to optimally readjust their portfolios in other markets, transmitting shocks while generating co-movement (contagion).Therefore, during the extreme days, investors need to liquidate their positions in other markets to cover the loss they encountered in the first market, therefore generating forces of co-movements for assets in different markets. This rationale may also apply to our case, in which the forces of co-movement and negative spillover effects are mixed and the signs of $\beta_{i3}$ and $\beta_{i4}$ are determined by which force influence investors larger in magnitude. 

Although negative spillover effects due to attention allocation are generally present in the crypto market, the pattern appears asymmetric. Evidence that Ethereum investors sell assets when Solana experiences extreme upward returns does not necessarily imply the reverse; Solana investors do not exhibit similar behavior when Ethereum undergoes extreme returns. Asset prices on Solana adjust in a relatively independent manner, such that extreme returns on other chains do not necessarily translate into Solana returns.

Although rare, we do observe some positive estimates of $\beta_{i3}$ and $\beta_{i4}$ ($i \in \{1,2,3,4\}$), such as $\beta_{43} = 0.259$ and $\beta_{44} = 0.263$. These cases are consistent with the cross-market rebalancing mechanism discussed by \textcite{kodres2002rational}, \textcite{kyle1999contagion}, and \textcite{fleming1998information}. When shocks occur in one market, investors optimally rebalance their portfolios in other markets, thereby transmitting shocks and generating co-movement (contagion). During extreme episodes, investors may need to liquidate positions in other markets to cover losses in the initially shocked market, creating co-movement forces alongside negative spillovers. In our context, the signs of $\beta_{i3}$ and $\beta_{i4}$ reflect the relative dominance of these two opposing forces: contagion-induced co-movement versus attention-driven negative spillover.

\section{Conclusions}\label{main_section_conclusion}

%In this paper, we explore the negative spillover effects on return between assets segmented by different blockchains. Investors tend to sell assets on one chain when assets on other chains pump. This result is more pronounced if we control changes in global stock markets and risk-free rates. Moreover, the negative spillover effects can be channeled from investors' shifting to other chains when their attention get attracted by other chains.  They sell assets on the current chain while buy assets on chains where attention-grabbing events happens, captured by our chain-level activity variables and extreme return dummies.  Our results shows opposite pattern predicted by previous studies on market contagion, which suggest that assets in different markets tend to co-move. Our study shows the possibility that in certain field, such as cryptocurrency, asset returns can move in opposite directions rather than co-move when information transmit across markets in contagion.

In this paper, we examine the negative spillover effects in returns across assets segmented by different blockchains. We find that investors tend to sell assets on one chain when assets on another chain experience price surges. This effect remains robust after controlling for global equity market movements and changes in risk-free rates.

Furthermore, we show that these negative spillovers are driven by investors reallocating capital across chains in response to attention shocks. Specifically, when salient events occur on a particular chain, investors shift their holdings by selling assets on other chains and buying those associated with the attention-grabbing chain, as captured by our chain-level activity measures and extreme return indicators.

Our findings contrast with the conventional predictions of the contagion literature, which typically document positive co-movement across markets. Instead, we demonstrate that in settings such as cryptocurrency, asset returns can diverge sharply—moving in opposite directions—when information and attention transmit across markets.
%Since our research is based on large-scale on-chain pricing data and structure of the most current crypto market, it provides the foundations of the market to which future research can refer. Despite the contributions, our study has several limitations. In particular, although we use return relationships among assets to detect channels of the negative spillover effects, we lack micro-level evidence showing how investors shift between chains and whether they actually finance their purchases by selling assets on other chains.

%Investors in the crypto market would favor our results, as assets with negatively correlated returns benefit diversification, unlike the case of co-movements. Also, it reflects the competitiveness among chains for users, a good signal for the healthy development of the crypto industry.

Our findings carry important implications for both investors and policymakers. For investors, the presence of negative return spillovers across blockchains suggests that diversification benefits can be obtained even within the cryptocurrency ecosystem, in contrast to the positive co-movements typically documented in traditional markets. Such patterns can inform portfolio construction and risk management strategies, particularly for institutions considering exposure to digital assets.

For regulators, our results highlight the need for a cross-chain perspective when evaluating systemic risk. While substitution across blockchains may indicate healthy competition and technological dynamism, it can also fragment liquidity and amplify volatility when capital rapidly migrates in response to attention shocks. Effective oversight thus requires regulatory frameworks that explicitly recognize interdependencies among seemingly segmented ecosystems, rather than treating each chain in isolation.

Moreover, our analysis is particularly relevant for the regulation of new cryptocurrency issuances. Attention shocks often arise around token launches or other high-profile events, and these can trigger capital outflows from existing ecosystems. Regulators may therefore need to scrutinize the design, disclosure, and timing of new token offerings, especially when such events have the potential to destabilize liquidity in other parts of the market. Establishing standards for transparency and investor protection in issuance processes—whether through ICOs, IEOs, or newer mechanisms—can mitigate unintended spillover risks while allowing innovation to proceed.

Finally, recognizing the competitive forces among blockchains can guide broader policy on innovation and market structure. A regulatory approach that balances investor protection with market dynamism would not only safeguard against systemic instability but also support the sustainable growth of blockchain ecosystems.

% ------------------------------------------------------ figures and tables

\newpage
\section*{Tables and Figures}

% start of the figure
\tikzset{every picture/.style={line width=0.75pt}} %set default line width to 0.75pt        
\begin{figure}[htbp]
\centering

\begin{tikzpicture}[x=0.75pt,y=0.75pt,yscale=-1,xscale=1]
%uncomment if require: \path (0,358); %set diagram left start at 0, and has height of 358

%Rounded Rect [id:dp8234025643945585] 
\draw   (231.17,161.52) .. controls (231.17,157.92) and (234.08,155) .. (237.68,155) -- (301.5,155) .. controls (305.1,155) and (308.02,157.92) .. (308.02,161.52) -- (308.02,181.07) .. controls (308.02,184.67) and (305.1,187.58) .. (301.5,187.58) -- (237.68,187.58) .. controls (234.08,187.58) and (231.17,184.67) .. (231.17,181.07) -- cycle ;

%Rounded Rect [id:dp8711317746716261] 
\draw   (231,117.68) .. controls (231,114.08) and (233.92,111.17) .. (237.52,111.17) -- (301.33,111.17) .. controls (304.93,111.17) and (307.85,114.08) .. (307.85,117.68) -- (307.85,137.23) .. controls (307.85,140.83) and (304.93,143.75) .. (301.33,143.75) -- (237.52,143.75) .. controls (233.92,143.75) and (231,140.83) .. (231,137.23) -- cycle ;

%Shape: Rectangle [id:dp10354600789430113] 
\draw   (112.17,163.83) -- (185.8,163.83) -- (185.8,193.68) -- (112.17,193.68) -- cycle ;

%Straight Lines [id:da5726905531316524] 
\draw    (186.14,179) -- (229.78,128.31) ;
\draw [shift={(231.08,126.79)}, rotate = 130.72] [color={rgb, 255:red, 0; green, 0; blue, 0 }  ][line width=0.75]    (10.93,-3.29) .. controls (6.95,-1.4) and (3.31,-0.3) .. (0,0) .. controls (3.31,0.3) and (6.95,1.4) .. (10.93,3.29)   ;
%Straight Lines [id:da738874591651617] 
\draw    (186.14,179) -- (229.23,171.75) ;
\draw [shift={(231.2,171.42)}, rotate = 170.45] [color={rgb, 255:red, 0; green, 0; blue, 0 }  ][line width=0.75]    (10.93,-3.29) .. controls (6.95,-1.4) and (3.31,-0.3) .. (0,0) .. controls (3.31,0.3) and (6.95,1.4) .. (10.93,3.29)   ;
%Rounded Rect [id:dp404028321590745] 
\draw   (231.17,228.52) .. controls (231.17,224.92) and (234.08,222) .. (237.68,222) -- (301.5,222) .. controls (305.1,222) and (308.02,224.92) .. (308.02,228.52) -- (308.02,248.07) .. controls (308.02,251.67) and (305.1,254.58) .. (301.5,254.58) -- (237.68,254.58) .. controls (234.08,254.58) and (231.17,251.67) .. (231.17,248.07) -- cycle ;

%Straight Lines [id:da5879201527845034] 
\draw    (186.14,179) -- (230.08,237.01) ;
\draw [shift={(231.29,238.61)}, rotate = 232.86] [color={rgb, 255:red, 0; green, 0; blue, 0 }  ][line width=0.75]    (10.93,-3.29) .. controls (6.95,-1.4) and (3.31,-0.3) .. (0,0) .. controls (3.31,0.3) and (6.95,1.4) .. (10.93,3.29)   ;
%Rounded Rect [id:dp6456661030417404] 
\draw  [fill={rgb, 255:red, 74; green, 144; blue, 226 }  ,fill opacity=1 ] (385.35,106.37) .. controls (385.35,103.31) and (387.82,100.83) .. (390.88,100.83) -- (481.78,100.83) .. controls (484.84,100.83) and (487.32,103.31) .. (487.32,106.37) -- (487.32,122.97) .. controls (487.32,126.02) and (484.84,128.5) .. (481.78,128.5) -- (390.88,128.5) .. controls (387.82,128.5) and (385.35,126.02) .. (385.35,122.97) -- cycle ;

%Rounded Rect [id:dp3884655848330105] 
\draw  [fill={rgb, 255:red, 74; green, 144; blue, 226 }  ,fill opacity=1 ] (385.35,143.7) .. controls (385.35,140.64) and (387.82,138.17) .. (390.88,138.17) -- (481.78,138.17) .. controls (484.84,138.17) and (487.32,140.64) .. (487.32,143.7) -- (487.32,160.3) .. controls (487.32,163.36) and (484.84,165.83) .. (481.78,165.83) -- (390.88,165.83) .. controls (387.82,165.83) and (385.35,163.36) .. (385.35,160.3) -- cycle ;

%Rounded Rect [id:dp7039240711085313] 
\draw  [fill={rgb, 255:red, 74; green, 144; blue, 226 }  ,fill opacity=1 ] (385.09,182.37) .. controls (385.09,179.31) and (387.57,176.83) .. (390.62,176.83) -- (481.52,176.83) .. controls (484.58,176.83) and (487.06,179.31) .. (487.06,182.37) -- (487.06,198.97) .. controls (487.06,202.02) and (484.58,204.5) .. (481.52,204.5) -- (390.62,204.5) .. controls (387.57,204.5) and (385.09,202.02) .. (385.09,198.97) -- cycle ;

%Rounded Rect [id:dp9510563545583117] 
\draw  [fill={rgb, 255:red, 74; green, 144; blue, 226 }  ,fill opacity=1 ] (385.35,242.37) .. controls (385.35,239.31) and (387.82,236.83) .. (390.88,236.83) -- (481.78,236.83) .. controls (484.84,236.83) and (487.32,239.31) .. (487.32,242.37) -- (487.32,258.97) .. controls (487.32,262.02) and (484.84,264.5) .. (481.78,264.5) -- (390.88,264.5) .. controls (387.82,264.5) and (385.35,262.02) .. (385.35,258.97) -- cycle ;

%Straight Lines [id:da4221980696993505] 
\draw    (308.73,127.8) -- (383.17,114.26) ;
\draw [shift={(385.14,113.9)}, rotate = 169.69] [color={rgb, 255:red, 0; green, 0; blue, 0 }  ][line width=0.75]    (10.93,-3.29) .. controls (6.95,-1.4) and (3.31,-0.3) .. (0,0) .. controls (3.31,0.3) and (6.95,1.4) .. (10.93,3.29)   ;
%Straight Lines [id:da8148462398902359] 
\draw    (308.07,171.17) -- (383.54,115.09) ;
\draw [shift={(385.14,113.9)}, rotate = 143.38] [color={rgb, 255:red, 0; green, 0; blue, 0 }  ][line width=0.75]    (10.93,-3.29) .. controls (6.95,-1.4) and (3.31,-0.3) .. (0,0) .. controls (3.31,0.3) and (6.95,1.4) .. (10.93,3.29)   ;
%Straight Lines [id:da9855215742983985] 
\draw    (308.7,238.22) -- (384.09,115.6) ;
\draw [shift={(385.14,113.9)}, rotate = 121.59] [color={rgb, 255:red, 0; green, 0; blue, 0 }  ][line width=0.75]    (10.93,-3.29) .. controls (6.95,-1.4) and (3.31,-0.3) .. (0,0) .. controls (3.31,0.3) and (6.95,1.4) .. (10.93,3.29)   ;
%Straight Lines [id:da6355454018720286] 
\draw    (308.49,127.8) -- (383.23,150.81) ;
\draw [shift={(385.14,151.4)}, rotate = 197.11] [color={rgb, 255:red, 0; green, 0; blue, 0 }  ][line width=0.75]    (10.93,-3.29) .. controls (6.95,-1.4) and (3.31,-0.3) .. (0,0) .. controls (3.31,0.3) and (6.95,1.4) .. (10.93,3.29)   ;
%Straight Lines [id:da3487498723556811] 
\draw    (308.49,127.8) -- (383.1,189.13) ;
\draw [shift={(384.64,190.4)}, rotate = 219.42] [color={rgb, 255:red, 0; green, 0; blue, 0 }  ][line width=0.75]    (10.93,-3.29) .. controls (6.95,-1.4) and (3.31,-0.3) .. (0,0) .. controls (3.31,0.3) and (6.95,1.4) .. (10.93,3.29)   ;
%Straight Lines [id:da931600800101507] 
\draw    (308.07,171.17) -- (383.2,151.9) ;
\draw [shift={(385.14,151.4)}, rotate = 165.61] [color={rgb, 255:red, 0; green, 0; blue, 0 }  ][line width=0.75]    (10.93,-3.29) .. controls (6.95,-1.4) and (3.31,-0.3) .. (0,0) .. controls (3.31,0.3) and (6.95,1.4) .. (10.93,3.29)   ;
%Straight Lines [id:da02740372778191802] 
\draw    (308.07,171.17) -- (382.7,189.91) ;
\draw [shift={(384.64,190.4)}, rotate = 194.1] [color={rgb, 255:red, 0; green, 0; blue, 0 }  ][line width=0.75]    (10.93,-3.29) .. controls (6.95,-1.4) and (3.31,-0.3) .. (0,0) .. controls (3.31,0.3) and (6.95,1.4) .. (10.93,3.29)   ;
%Straight Lines [id:da8718523610734256] 
\draw    (308.73,127.8) -- (383.59,248.7) ;
\draw [shift={(384.64,250.4)}, rotate = 238.24] [color={rgb, 255:red, 0; green, 0; blue, 0 }  ][line width=0.75]    (10.93,-3.29) .. controls (6.95,-1.4) and (3.31,-0.3) .. (0,0) .. controls (3.31,0.3) and (6.95,1.4) .. (10.93,3.29)   ;
%Straight Lines [id:da9772555817812114] 
\draw    (308.07,171.17) -- (383.25,248.96) ;
\draw [shift={(384.64,250.4)}, rotate = 225.98] [color={rgb, 255:red, 0; green, 0; blue, 0 }  ][line width=0.75]    (10.93,-3.29) .. controls (6.95,-1.4) and (3.31,-0.3) .. (0,0) .. controls (3.31,0.3) and (6.95,1.4) .. (10.93,3.29)   ;
%Straight Lines [id:da8515920754581] 
\draw    (308.47,238.22) -- (383.82,152.9) ;
\draw [shift={(385.14,151.4)}, rotate = 131.45] [color={rgb, 255:red, 0; green, 0; blue, 0 }  ][line width=0.75]    (10.93,-3.29) .. controls (6.95,-1.4) and (3.31,-0.3) .. (0,0) .. controls (3.31,0.3) and (6.95,1.4) .. (10.93,3.29)   ;
%Straight Lines [id:da2430772807419994] 
\draw    (308.47,238.22) -- (382.95,191.46) ;
\draw [shift={(384.64,190.4)}, rotate = 147.88] [color={rgb, 255:red, 0; green, 0; blue, 0 }  ][line width=0.75]    (10.93,-3.29) .. controls (6.95,-1.4) and (3.31,-0.3) .. (0,0) .. controls (3.31,0.3) and (6.95,1.4) .. (10.93,3.29)   ;
%Straight Lines [id:da2824193584790492] 
\draw    (308.7,238.22) -- (382.67,250.08) ;
\draw [shift={(384.64,250.4)}, rotate = 189.11] [color={rgb, 255:red, 0; green, 0; blue, 0 }  ][line width=0.75]    (10.93,-3.29) .. controls (6.95,-1.4) and (3.31,-0.3) .. (0,0) .. controls (3.31,0.3) and (6.95,1.4) .. (10.93,3.29)   ;
%Rounded Rect [id:dp1027170486716289] 
\draw  [fill={rgb, 255:red, 74; green, 144; blue, 226 }  ,fill opacity=1 ] (578.82,105.87) .. controls (578.82,102.81) and (581.3,100.33) .. (584.35,100.33) -- (643.78,100.33) .. controls (646.84,100.33) and (649.32,102.81) .. (649.32,105.87) -- (649.32,122.47) .. controls (649.32,125.52) and (646.84,128) .. (643.78,128) -- (584.35,128) .. controls (581.3,128) and (578.82,125.52) .. (578.82,122.47) -- cycle ;

%Straight Lines [id:da5829203131137534] 
\draw    (487.27,115.3) -- (575.67,115.08) ;
\draw [shift={(577.67,115.07)}, rotate = 179.86] [color={rgb, 255:red, 0; green, 0; blue, 0 }  ][line width=0.75]    (10.93,-3.29) .. controls (6.95,-1.4) and (3.31,-0.3) .. (0,0) .. controls (3.31,0.3) and (6.95,1.4) .. (10.93,3.29)   ;
%Rounded Rect [id:dp928991779251715] 
\draw  [fill={rgb, 255:red, 74; green, 144; blue, 226 }  ,fill opacity=1 ] (578.82,143.37) .. controls (578.82,140.31) and (581.3,137.83) .. (584.35,137.83) -- (643.78,137.83) .. controls (646.84,137.83) and (649.32,140.31) .. (649.32,143.37) -- (649.32,159.97) .. controls (649.32,163.02) and (646.84,165.5) .. (643.78,165.5) -- (584.35,165.5) .. controls (581.3,165.5) and (578.82,163.02) .. (578.82,159.97) -- cycle ;

%Rounded Rect [id:dp6567241183949538] 
\draw  [fill={rgb, 255:red, 74; green, 144; blue, 226 }  ,fill opacity=1 ] (578.42,181.37) .. controls (578.42,178.31) and (580.9,175.83) .. (583.96,175.83) -- (643.39,175.83) .. controls (646.44,175.83) and (648.92,178.31) .. (648.92,181.37) -- (648.92,197.97) .. controls (648.92,201.02) and (646.44,203.5) .. (643.39,203.5) -- (583.96,203.5) .. controls (580.9,203.5) and (578.42,201.02) .. (578.42,197.97) -- cycle ;

%Rounded Rect [id:dp6194167230232639] 
\draw  [fill={rgb, 255:red, 74; green, 144; blue, 226 }  ,fill opacity=1 ] (578.42,240.87) .. controls (578.42,237.81) and (580.9,235.33) .. (583.96,235.33) -- (643.39,235.33) .. controls (646.44,235.33) and (648.92,237.81) .. (648.92,240.87) -- (648.92,257.47) .. controls (648.92,260.52) and (646.44,263) .. (643.39,263) -- (583.96,263) .. controls (580.9,263) and (578.42,260.52) .. (578.42,257.47) -- cycle ;

%Straight Lines [id:da5458731152158239] 
\draw    (487.27,152.8) -- (575.67,152.58) ;
\draw [shift={(577.67,152.57)}, rotate = 179.86] [color={rgb, 255:red, 0; green, 0; blue, 0 }  ][line width=0.75]    (10.93,-3.29) .. controls (6.95,-1.4) and (3.31,-0.3) .. (0,0) .. controls (3.31,0.3) and (6.95,1.4) .. (10.93,3.29)   ;
%Straight Lines [id:da486501621357018] 
\draw    (486.77,190.3) -- (575.17,190.08) ;
\draw [shift={(577.17,190.07)}, rotate = 179.86] [color={rgb, 255:red, 0; green, 0; blue, 0 }  ][line width=0.75]    (10.93,-3.29) .. controls (6.95,-1.4) and (3.31,-0.3) .. (0,0) .. controls (3.31,0.3) and (6.95,1.4) .. (10.93,3.29)   ;
%Straight Lines [id:da4441732594038611] 
\draw    (487.27,250.8) -- (575.67,250.58) ;
\draw [shift={(577.67,250.57)}, rotate = 179.86] [color={rgb, 255:red, 0; green, 0; blue, 0 }  ][line width=0.75]    (10.93,-3.29) .. controls (6.95,-1.4) and (3.31,-0.3) .. (0,0) .. controls (3.31,0.3) and (6.95,1.4) .. (10.93,3.29)   ;

% Text Node
\draw (248.83,164.9) node [anchor=north west][inner sep=0.75pt]  [font=\scriptsize] [align=left] {CEX(2)};
% Text Node
\draw (248.67,121.07) node [anchor=north west][inner sep=0.75pt]  [font=\scriptsize] [align=left] {CEX(1)};
% Text Node
\draw (282.06,191.85) node [anchor=north west][inner sep=0.75pt]  [rotate=-89.85] [align=left] {......};
% Text Node
\draw (130.17,170.83) node [anchor=north west][inner sep=0.75pt]  [font=\small] [align=left] {Users};
% Text Node
\draw (158.63,137.83) node [anchor=north west][inner sep=0.75pt]  [font=\tiny] [align=left] {Fiat money};
% Text Node
\draw (248.83,231.9) node [anchor=north west][inner sep=0.75pt]  [font=\scriptsize] [align=left] {CEX(n)};
% Text Node
\draw (402.09,108.33) node [anchor=north west][inner sep=0.75pt]  [font=\scriptsize] [align=left] {Blockchain(1)};
% Text Node
\draw (402.09,145.67) node [anchor=north west][inner sep=0.75pt]  [font=\scriptsize] [align=left] {Blockchain(2)};
% Text Node
\draw (401.83,184.33) node [anchor=north west][inner sep=0.75pt]  [font=\scriptsize] [align=left] {Blockchain(3)};
% Text Node
\draw (448.46,207.51) node [anchor=north west][inner sep=0.75pt]  [rotate=-89.85] [align=left] {......};
% Text Node
\draw (401.75,244.33) node [anchor=north west][inner sep=0.75pt]  [font=\scriptsize] [align=left] {Blockchain(m)};
% Text Node
\draw (290.14,97) node [anchor=north west][inner sep=0.75pt]  [font=\tiny] [align=left] {Chain native token (i)};
% Text Node
\draw (592.55,108.33) node [anchor=north west][inner sep=0.75pt]  [font=\scriptsize] [align=left] {DEXs(1)};
% Text Node
\draw (497.81,102.5) node [anchor=north west][inner sep=0.75pt]  [font=\tiny] [align=left] {Native token (1)};
% Text Node
\draw (592.55,145.83) node [anchor=north west][inner sep=0.75pt]  [font=\scriptsize] [align=left] {DEXs(2)};
% Text Node
\draw (592.16,183.83) node [anchor=north west][inner sep=0.75pt]  [font=\scriptsize] [align=left] {DEXs(3)};
% Text Node
\draw (625.65,207.01) node [anchor=north west][inner sep=0.75pt]  [rotate=-89.85] [align=left] {......};
% Text Node
\draw (591.74,243.33) node [anchor=north west][inner sep=0.75pt]  [font=\scriptsize] [align=left] {DEXs(m)};
% Text Node
\draw (497.81,140) node [anchor=north west][inner sep=0.75pt]  [font=\tiny] [align=left] {Native token (2)};
% Text Node
\draw (497.31,177.5) node [anchor=north west][inner sep=0.75pt]  [font=\tiny] [align=left] {Native token (3)};
% Text Node
\draw (495.81,238) node [anchor=north west][inner sep=0.75pt]  [font=\tiny] [align=left] {Native token (m)};

\end{tikzpicture}
\caption{The conponents and liquidity flow of the crypto market.}
\label{fig:crypto_market}
\end{figure}
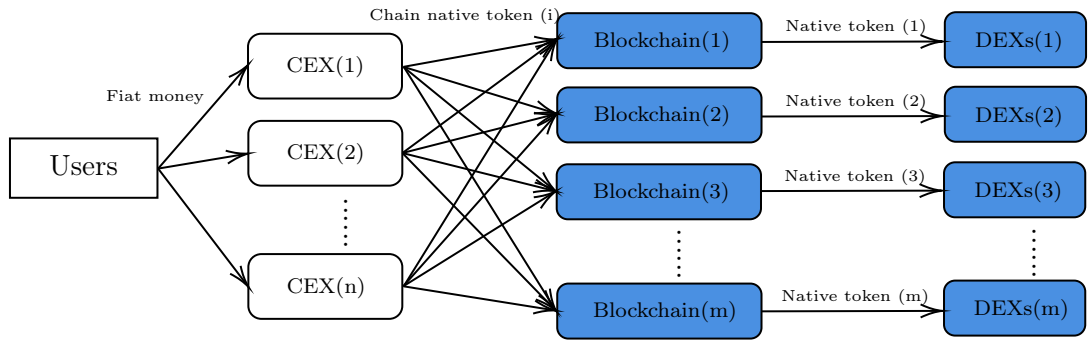
% end of the figure

% ---------------------- figure & table of stats

\newpage

\newpage

\begin{landscape}
\begin{singlespace}

% ---------------------------- table of stats for chain portfolio returns

\begin{ThreePartTable}
  %──────────────────────────────────────────────────────────────────────────────
  % Notes, flush‐left and without indent
  \begin{TableNotes}[para,flushleft]
    \footnotesize
    Notes: 
    This table shows descriptive statistics for the four types of chain portfolio returns discussed in Section \ref{section_chain_portfolio_construction}: $R^{All}_{\mathrm{chain}_{i},t}$, $R^{CEX}_{\mathrm{chain}_{i},t}$, $R^{non-CEX}_{\mathrm{chain}_{i},t}$, and $R^{Local}_{\mathrm{chain}_{i},t}$ while $chain_i\in\{\mathrm{Ethereum},\,\mathrm{Solana},\,\mathrm{BSC},\,\mathrm{Arbitrum},\,\mathrm{Avalanche}\}$.\\ 
    We report results of Augmented Dickey-Fuller (ADF) test to determine if the return series are stationary, with the null hypothesis: the time series has a unit root (i.e., it is non-stationary).\\
      ***/**/* denote significance at 1\%/5\%/10\%.
      
  \end{TableNotes}
  %──────────────────────────────────────────────────────────────────────────────

  % Force the table to span the full text width
  \setlength\LTleft{0pt}
  \setlength\LTright{0pt}
 {\small
  \begin{tabularx}{\linewidth}{@{}l*{6}{Y}@{}}
    \caption{Descriptive statistics of chain portfolio returns\label{tab:stats_portfolio}}\\
    % insert notes at top
    \insertTableNotes\\
    \addlinespace
    \toprule
      Variables 
        & Mean & Std & Skewness & Kurtosis & Jarque-Bera &  ADF \\
    \midrule
    \endfirsthead

    \caption[]{Descriptive statistics of chain portfolio returns (continued)}\\
    \toprule
      Variables 
        & Mean & Std & Skewness & Kurtosis & Jarque-Bera &  ADF \\
    \midrule
    \endhead

    \midrule
    \multicolumn{7}{r}{\emph{Continued on next page}}
    \endfoot

    \bottomrule
    \endlastfoot

    %--------------- your data rows here ----------------
    \multicolumn{7}{l}{Panel A: chain portfolios for $Ethereum$}\\
    \addlinespace
    $R^{All}_{\mathrm{Ethereum},t}$ 
      &-0.0035  & 0.0749  & 3.5004  & 175.4621  & 3241141*** & -18.1133*** \\
    $R^{\mathrm{CEX}}_{\mathrm{Ethereum},t}$
      & -0.0011 & 0.0183 & -0.1418  & 34.4507  & 107620*** & -42.3431*** \\
    $R^{non-CEX}_{\mathrm{Ethereum},t}$
      & -0.0036 & 0.0846 & 3.0142 & 131.7192 & 1806481*** & -18.2003*** \\
    $R^{Local}_{\mathrm{Ethereum},t}$
      & -0.0036 & 0.0771 & 3.4143 & 165.5336 & 2879041*** & -18.1030***\\
    \midrule

    \multicolumn{7}{l}{Panel B: chain portfolios for $Solana$}\\
    \addlinespace
    $R^{All}_{\mathrm{Solana},t}$ 
      & 0.0003 & 0.0466 & 6.2327 & 111.1673 & 1055637*** & -5.8254***\\
    $R^{\mathrm{CEX}}_{\mathrm{Solana},t}$
      & 0.0004 & 0.0482 & 5.4046 & 99.4193 & 838195*** & -5.7933*** \\
    $R^{non-CEX}_{\mathrm{Solana},t}$
      & -0.0005 & 0.0368 & 1.8871 &  32.1328  & 76840*** &  -15.0705*** \\
    $R^{Local}_{\mathrm{Solana},t}$
      & 0.0002 & 0.0474 & 5.8846 & 104.8182  & 935424*** &  -5.9388*** \\
    \midrule
    
    \multicolumn{7}{l}{Panel C: chain portfolios for $BSC$}\\
    \addlinespace
    $R^{All}_{\mathrm{BSC},t}$ 
      & 0.0031 & 0.2401 & 49.6794 & 2514.9908 & 688613374*** & -51.1032*** \\
    $R^{\mathrm{CEX}}_{\mathrm{BSC},t}$
      & 0.0028 & 0.2457 & 49.3559 & 2494.3277 & 677334856*** & -51.1082***\\
    $R^{non-CEX}_{\mathrm{BSC},t}$
      & -0.0010 & 0.0420 & 21.3755  & 801.8520  & 69732472*** & -17.8865*** \\
    $R^{Local}_{\mathrm{BSC},t}$
      & 0.0042 & 0.2584 & 49.801646 & 2523.2892 & 693168788*** & -51.1390*** \\
    \midrule
    
    \multicolumn{7}{l}{Panel D: chain portfolios for $Arbitrum$}\\
    \addlinespace
    $R^{All}_{\mathrm{Arbitrum},t}$ 
      &  -0.0005 & 0.0143 & -1.6981  & 34.2486  & 97731*** & -7.4762*** \\
    $R^{\mathrm{CEX}}_{\mathrm{Arbitrum},t}$
      & -0.0014 & 0.0240 & -0.4021 & 22.6392  & 38216*** & -27.9908*** \\
    $R^{non-CEX}_{\mathrm{Arbitrum},t}$
      & 0.0001 & 0.0153 &  0.4371 &  38.3280  & 123530*** & -8.0880***\\
    $R^{Local}_{\mathrm{Arbitrum},t}$
      & -0.0001 & 0.0184 & -14.1864 & 378.6432 & 14037543*** & -10.1153*** \\
      \midrule
      
    \multicolumn{7}{l}{Panel E: chain portfolios for $Avalanche$}\\
    \addlinespace
    $R^{All}_{\mathrm{Avalanche},t}$ 
      & -0.0013 & 0.0252 & -1.5589 & 37.8110 &  133095*** & -11.6112*** \\
    $R^{\mathrm{CEX}}_{\mathrm{Avalanche},t}$
      & -0.0017 & 0.0296 & -4.7800 & 124.9607 & 1598223*** & -12.2269*** \\
    $R^{non_CEX}_{\mathrm{Avalanche},t}$
      & -0.0006 & 0.0283 & 0.1605 & 10.994610 & 6975*** & -11.2165*** \\
    $R^{Local}_{\mathrm{Avalanche},t}$
      & -0.0015 & 0.0341 & -0.5572 &  19.2809 &  29017*** & -18.0678*** \\

  \end{tabularx}
  }

\end{ThreePartTable}
\end{singlespace}

% \vspace{1cm}
% ---------------------------- table of stats for chain & macro conditions
\newpage
\renewcommand{\tabularxcolumn}[1]{>{\hspace{0pt}\arraybackslash}m{#1}}
\setlength{\tabcolsep}{-2pt}
\begin{ThreePartTable}
  %──────────────────────────────────────────────────────────────────────────────
  % Notes, flush‐left and without indent
  \begin{TableNotes}[para,flushleft]
    \footnotesize
    Notes: 
    This table shows descriptive statistics for the chain-level activity and global market variables discussed in Section \ref{section_chain_condition_variables} and \ref{section_global_market_var}.\\
    Among the variables, we use ARIMA approach to decompose while regard the residual as the unexpected portion for staking reward rates and risk-free rates. In this table, we report the statistics of the original series for these two types of variables and the ARIMA model parameters (p,d,q), as well as the Akaike Information Criterion (AIC).\\
    We report results of Augmented Dickey-Fuller (ADF) test to determine if the return series are stationary, with the null hypothesis: the time series has a unit root (i.e., it is non-stationary).\\
      ***/**/* denote significance at 1\%/5\%/10\%.
      
  \end{TableNotes}
  %──────────────────────────────────────────────────────────────────────────────

  % Force the table to span the full text width
  \setlength\LTleft{0pt}
  \setlength\LTright{0pt}
 {\small
  \begin{tabularx}{\linewidth}{@{}l*{8}{Y}@{}}
    \caption{Descriptive statistics of chain portfolio returns\label{tab:stats_conditions_var}}\\
    % insert notes at top
    \insertTableNotes\\
    \addlinespace
    \toprule
      Variables 
        & Mean & Std & Skewness & Kurtosis & Jarque-Bera &  ADF & ARIMA & AIC \\
    \midrule
    \endfirsthead

    \caption[]{Descriptive statistics of chain portfolio returns (continued)}\\
    \toprule
      Variables 
        & Mean & Std & Skewness & Kurtosis & Jarque-Bera &  ADF & ARIMA & AIC \\
    \midrule
    \endhead

    \midrule
    \multicolumn{9}{r}{\emph{Continued on next page}}
    \endfoot

    \bottomrule
    \endlastfoot

    %--------------- your data rows here ----------------
    \multicolumn{7}{l}{Panel A: chain-level activity variables}\\
    \addlinespace
    $R_{\$\mathrm{BTC}, t}$ 
      &0.0002  & 0.0200  & -0.0777  & 8.1540  & 2659*** & -16.1415*** & - & - \\
    $R_{\$\mathrm{ETH}, t}$
      & -0.0003 & 0.0257 & -0.2639  & 8.5921  & 3155*** & -35.8198*** & - & -\\
    $R_{\$\mathrm{SOL}, t}$
      & -0.0002 &  0.0379 & -0.1757 & 9.36044 &  4058*** & -35.1694*** & - & -\\
    $R_{\$\mathrm{BNB}, t}$
      & 0.0000 & 0.0223 & -0.3450 & 9.2069 & 3900*** & -21.1245*** & - & -\\
    $R_{\$\mathrm{ARB}, t}$
      & -0.0021 & 0.0614 & -21.1660 & 672.2105 & 28177105*** & -19.5454*** & - & -\\
    $R_{\$\mathrm{AVAX}, t}$
      & -0.0007 & 0.0374 & 0.0699 & 6.7547 & 1412*** & -20.6619*** & - & -\\
    $\mathrm{SR}_{\mathrm{Ethereum}, t}$
      & 4.0000 & 0.8266 & 2.5648 & 12.8588 &  6000*** & -2.9379** & (3, 1, 3) & 696.9752\\
    $\mathrm{SR}_{\mathrm{Solana}, t}$
      & 6.6661 & 0.8754 & 0.5048 & 6.2423 & 522*** & -2.6240*  & (2, 1, 3) & 616.3317\\
    $\mathrm{SR}_{\mathrm{BSC}, t}$
      & 4.8409 & 0.1166 & 3.2758 & 14.1802 & 7221*** & -6.3175*** & (2, 0, 2) & -2377.8480\\ 
    $\mathrm{SR}_{\mathrm{Arbitrum}, t}$
      & 2.5041 & 0.0091 & 3.0802 & 41.6286 & 145807*** & -4.8149*** & (1, 0, 3) &  -17806.7355\\
    $\mathrm{SR}_{\mathrm{Avalanche}, t}$
      & 4.8168 & 0.0132 & 2.2801 & 16.6033 & 19615*** & -3.6049*** & (1, 0, 3) & -16376.1333\\
    \midrule

    \multicolumn{7}{l}{Panel B: global market variables}\\
    \addlinespace
    $\mathrm{SPR}_t$ 
      & 0.0002 & 0.0069 & -0.3435 & 6.5414 & 1151*** & -19.8052*** & - & -\\
    $\mathrm{HSR}_t$
      & -0.0001 & 0.0110 & 0.1769 &  7.2296 & 1561*** & -32.2747*** & - & -\\
    $\mathrm{FTSER}_t$
      & 0.0001 & 0.0056 & -0.5926 &  12.9491  &  6873*** &  -40.6051*** & - & -\\
    $\mathrm{HIBOR}_t$ 
      & 1.1094 & 1.4404 & 1.5625 & 4.2084 & 1971*** & -1.3886 & (3, 1, 3) & -10069.4847\\
    $\mathrm{EURIBOR}_t$
      & 1.6372 & 1.8723 &  -0.1199 & 1.2328 & 144*** & -1.1865 & (1, 1, 3) & -5387.5456\\
    $\mathrm{TREA}_t$
      & 3.9643 & 1.7839 & -1.1703 &  2.9465  & 193*** &  1.3064 & (3, 1, 2) & 3137.4210\\

  \end{tabularx}
  }

\end{ThreePartTable}

\end{landscape}

% -------------------------------- table for baseline models
\newpage

\begin{singlespace}
\begin{ThreePartTable}
  %──────────────────────────────────────────────────────────────────────────────
  % Notes, flush‐left and without indent
  \begin{TableNotes}[para,flushleft]
    \scriptsize
    Notes: This table shows estimates of the baseline linear models, i.e., as illustrated by Equation (\ref{eq_baseline_all}), (\ref{eq_baseline_nonCEX}), and (\ref{eq_baseline_this_chain_only}), for the five chain portfolios in our study.  Panel A reports the estimated parameters of the models for all the five chains based on Equation (\ref{eq_baseline_all}), which are as following:
      \begin{equation*}
      R^{All}_{\mathrm{Ethereum},t}
        = \alpha_{0}
          + \alpha_{1}R^{All}_{\mathrm{Ethereum},t-1}
          + \beta_{0}R^{CEX}_{\mathrm{Ethereum},t}
          + \beta_{1}R^{All}_{\mathrm{Solana},t}
          + \beta_{2}R^{All}_{\mathrm{BSC},t}
          + \beta_{3}R^{All}_{\mathrm{Arbitrum},t}
          + \beta_{4}R^{All}_{\mathrm{Avalanche},t}
          + e_{t},
      \end{equation*}
      \begin{equation*}
      R^{All}_{\mathrm{Solana},t}
        = \alpha_{0}
          + \alpha_{1}R^{All}_{\mathrm{Solana},t-1}
          + \beta_{0}R^{CEX}_{\mathrm{Solana},t}
          + \beta_{1}R^{All}_{\mathrm{Ethereum},t}
          + \beta_{2}R^{All}_{\mathrm{BSC},t}
          + \beta_{3}R^{All}_{\mathrm{Arbitrum},t}
          + \beta_{4}R^{All}_{\mathrm{Avalanche},t}
          + e_{t},
      \end{equation*}
      \begin{equation*}
      R^{All}_{\mathrm{BSC},t}
        = \alpha_{0}
          + \alpha_{1}R^{All}_{\mathrm{BSC},t-1}
          + \beta_{0}R^{CEX}_{\mathrm{BSC},t}
          + \beta_{1}R^{All}_{\mathrm{Ethereum},t}
          + \beta_{2}R^{All}_{\mathrm{Solana},t}
          + \beta_{3}R^{All}_{\mathrm{Arbitrum},t}
          + \beta_{4}R^{All}_{\mathrm{Avalanche},t}
          + e_{t},
      \end{equation*}
      \begin{equation*}
      R^{All}_{\mathrm{Arbitrum},t}
        = \alpha_{0}
          + \alpha_{1}R^{All}_{\mathrm{Arbitrum},t-1}
          + \beta_{0}R^{CEX}_{\mathrm{Arbitrum},t}
          + \beta_{1}R^{All}_{\mathrm{Ethereum},t}
          + \beta_{2}R^{All}_{\mathrm{Solana},t}
          + \beta_{3}R^{All}_{\mathrm{BSC},t}
          + \beta_{4}R^{All}_{\mathrm{Avalanche},t}
          + e_{t},
      \end{equation*}
      \begin{equation*}
      R^{All}_{\mathrm{Avalanche},t}
        = \alpha_{0}
          + \alpha_{1}R^{All}_{\mathrm{Avalanche},t-1}
          + \beta_{0}R^{CEX}_{\mathrm{Avalanche},t}
          + \beta_{1}R^{All}_{\mathrm{Ethereum},t}
          + \beta_{2}R^{All}_{\mathrm{Solana},t}
          + \beta_{3}R^{All}_{\mathrm{BSC},t}
          + \beta_{4}R^{All}_{\mathrm{Arbitrum},t}
          + e_{t}.
      \end{equation*}
    Panel B reports the estimated parameters of the models for all the five chains based on Equation (\ref{eq_baseline_nonCEX}), which are as following:
      \begin{equation*}
      R^{non-CEX}_{\mathrm{Ethereum},t}
        = \alpha_{0}
          + \alpha_{1}R^{non-CEX}_{\mathrm{Ethereum},t-1}
          + \beta_{0}R^{CEX}_{\mathrm{Ethereum},t}
          + \beta_{1}R^{All}_{\mathrm{Solana},t}
          + \beta_{2}R^{All}_{\mathrm{BSC},t}
          + \beta_{3}R^{All}_{\mathrm{Arbitrum},t}
          + \beta_{4}R^{All}_{\mathrm{Avalanche},t}
          + e_{t},
      \end{equation*}
      \begin{equation*}
      R^{non-CEX}_{\mathrm{Solana},t}
        = \alpha_{0}
          + \alpha_{1}R^{non-CEX}_{\mathrm{Solana},t-1}
          + \beta_{0}R^{CEX}_{\mathrm{Solana},t}
          + \beta_{1}R^{All}_{\mathrm{Ethereum},t}
          + \beta_{2}R^{All}_{\mathrm{BSC},t}
          + \beta_{3}R^{All}_{\mathrm{Arbitrum},t}
          + \beta_{4}R^{All}_{\mathrm{Avalanche},t}
          + e_{t},
      \end{equation*}
      \begin{equation*}
      R^{non-CEX}_{\mathrm{BSC},t}
        = \alpha_{0}
          + \alpha_{1}R^{non-CEX}_{\mathrm{BSC},t-1}
          + \beta_{0}R^{CEX}_{\mathrm{BSC},t}
          + \beta_{1}R^{All}_{\mathrm{Ethereum},t}
          + \beta_{2}R^{All}_{\mathrm{Solana},t}
          + \beta_{3}R^{All}_{\mathrm{Arbitrum},t}
          + \beta_{4}R^{All}_{\mathrm{Avalanche},t}
          + e_{t},
      \end{equation*}
      \begin{equation*}
      R^{non-CEX}_{\mathrm{Arbitrum},t}
        = \alpha_{0}
          + \alpha_{1}R^{non-CEX}_{\mathrm{Arbitrum},t-1}
          + \beta_{0}R^{CEX}_{\mathrm{Arbitrum},t}
          + \beta_{1}R^{All}_{\mathrm{Ethereum},t}
          + \beta_{2}R^{All}_{\mathrm{Solana},t}
          + \beta_{3}R^{All}_{\mathrm{BSC},t}
          + \beta_{4}R^{All}_{\mathrm{Avalanche},t}
          + e_{t},
      \end{equation*}
      \begin{equation*}
      R^{non-CEX}_{\mathrm{Avalanche},t}
        = \alpha_{0}
          + \alpha_{1}R^{non-CEX}_{\mathrm{Avalanche},t-1}
          + \beta_{0}R^{CEX}_{\mathrm{Avalanche},t}
          + \beta_{1}R^{All}_{\mathrm{Ethereum},t}
          + \beta_{2}R^{All}_{\mathrm{Solana},t}
          + \beta_{3}R^{All}_{\mathrm{BSC},t}
          + \beta_{4}R^{All}_{\mathrm{Arbitrum},t}
          + e_{t}.
      \end{equation*}
    Panel C reports the estimated parameters of the models for all the five chains based on Equation (\ref{eq_baseline_this_chain_only}), which are as following:
      \begin{equation*}
      R^{Local}_{\mathrm{Ethereum},t}
        = \alpha_{0}
          + \alpha_{1}R^{Local}_{\mathrm{Ethereum},t-1}
          + \beta_{0}R^{CEX}_{\mathrm{Ethereum},t}
          + \beta_{1}R^{All}_{\mathrm{Solana},t}
          + \beta_{2}R^{All}_{\mathrm{BSC},t}
          + \beta_{3}R^{All}_{\mathrm{Arbitrum},t}
          + \beta_{4}R^{All}_{\mathrm{Avalanche},t}
          + e_{t},
      \end{equation*}
      \begin{equation*}
      R^{Local}_{\mathrm{Solana},t}
        = \alpha_{0}
          + \alpha_{1}R^{Local}_{\mathrm{Solana},t-1}
          + \beta_{0}R^{CEX}_{\mathrm{Solana},t}
          + \beta_{1}R^{All}_{\mathrm{Ethereum},t}
          + \beta_{2}R^{All}_{\mathrm{BSC},t}
          + \beta_{3}R^{All}_{\mathrm{Arbitrum},t}
          + \beta_{4}R^{All}_{\mathrm{Avalanche},t}
          + e_{t},
      \end{equation*}
      \begin{equation*}
      R^{Local}_{\mathrm{BSC},t}
        = \alpha_{0}
          + \alpha_{1}R^{Local}_{\mathrm{BSC},t-1}
          + \beta_{0}R^{CEX}_{\mathrm{BSC},t}
          + \beta_{1}R^{All}_{\mathrm{Ethereum},t}
          + \beta_{2}R^{All}_{\mathrm{Solana},t}
          + \beta_{3}R^{All}_{\mathrm{Arbitrum},t}
          + \beta_{4}R^{All}_{\mathrm{Avalanche},t}
          + e_{t},
      \end{equation*}
      \begin{equation*}
      R^{Local}_{\mathrm{Arbitrum},t}
        = \alpha_{0}
          + \alpha_{1}R^{Local}_{\mathrm{Arbitrum},t-1}
          + \beta_{0}R^{CEX}_{\mathrm{Arbitrum},t}
          + \beta_{1}R^{All}_{\mathrm{Ethereum},t}
          + \beta_{2}R^{All}_{\mathrm{Solana},t}
          + \beta_{3}R^{All}_{\mathrm{BSC},t}
          + \beta_{4}R^{All}_{\mathrm{Avalanche},t}
          + e_{t},
      \end{equation*}
      \begin{equation*}
      R^{Local}_{\mathrm{Avalanche},t}
        = \alpha_{0}
          + \alpha_{1}R^{Local}_{\mathrm{Avalanche},t-1}
          + \beta_{0}R^{CEX}_{\mathrm{Avalanche},t}
          + \beta_{1}R^{All}_{\mathrm{Ethereum},t}
          + \beta_{2}R^{All}_{\mathrm{Solana},t}
          + \beta_{3}R^{All}_{\mathrm{BSC},t}
          + \beta_{4}R^{All}_{\mathrm{Arbitrum},t}
          + e_{t},
      \end{equation*}
      while the series of residuals for all the modeled in the three panels are modeled by Glosten-Jagannathan-Runkle (GJR) asymmetric GARCH approach in Equation (\ref{garch_model})
    and the optimal combination of $p$, $o$, and $q$ for each model is reported in this table. \\
    Results in this table are based on unified sampling period from 28th Apr 2022 to 31st Mar 2025.\\
    t-statistics are in parentheses. \\
      ***/**/* denote significance at 1\%/5\%/10\%.
  \end{TableNotes}
  %──────────────────────────────────────────────────────────────────────────────

  % Force the table to span the full text width
  \setlength\LTleft{0pt}
  \setlength\LTright{0pt}

  \begin{tabularx}{\textwidth}{@{}l*{5}{Y}@{}}
    \caption{Estimates of the baseline linear models.\label{tab:baseline}}\\
    % insert notes at top
    \insertTableNotes\\
    \addlinespace
    \toprule
      Model: 
        & (1) & (2) & (3) & (4) & (5) \\
      Chain portfolios: 
        & $\mathrm{Ethereum}$ & $\mathrm{Solana}$ & $\mathrm{BSC}$ & $\mathrm{Arbitrum}$ & $\mathrm{Avalanche}$ \\
    \midrule
    \endfirsthead

    \caption[]{Estimates of the baseline linear models. (continued)}\\
    \toprule
      Model: 
        & (1) & (2) & (3) & (4) & (5) \\
      Chain portfolios: 
        & $\mathrm{Ethereum}$ & $\mathrm{Solana}$ & $\mathrm{BSC}$ & $\mathrm{Arbitrum}$ & $\mathrm{Avalanche}$ \\
    \midrule
    \endhead

    \midrule
    \multicolumn{6}{r}{\emph{Continued on next page}}
    \endfoot

    \bottomrule
    \endlastfoot

    %--------------- Panel A ----------------
    \multicolumn{6}{l}{Panel A: $R^{All}_{\mathrm{chain}_{i},t}$ as the dependent variable}\\
    \addlinespace

    $\alpha_0$ 
      & -0.0007**  & -0.0002** & 0.0003 & 0.0003  & 0.00004 \\
      & (-1.648)    & (-2.559) & (0.251)  & (0.403)  & (0.512)  \\

    $\alpha_1$
      & -0.016      & -0.003 & -0.0002 & -0.0007  & -0.006 \\
      & (-0.658)     & (-0.270) & (-1.224)  & (-0.018)  & (-0.902)  \\

    $\beta_0$
      & 0.655***    & 0.868*** & 0.978***  & 0.397***  & 0.846*** \\
      & (16.583)    & (84.495) & (3363.669)  & (12.247)  & (80.691)  \\

    $\beta_1$
      & 0.040       & 0.002* & -0.015*   & 0.0002  & -0.003*** \\
      & (1.359)     & (1.767) & (-1.769)  & (0.043)  & (-5.029)  \\

    $\beta_2$
      & -0.001***   & 0.0006*** & -0.002   & 0.017 & -0.005 \\
      & (-2.633) & (0.006) & (-0.613) & (1.156) & (-0.900)  \\

    $\beta_3$
      & -0.047    & -0.034* & -0.08   & -0.0003  & 0.00009** \\
      & (-1.144) & (-1.661) & (-0.581) & (-0.292)  & (2.138)  \\

    $\beta_4$
      & -0.108    & -0.019* & -0.012   & -0.017  & -0.022** \\
      & (-1.230) & (-1.796) & (-0.416) & (-0.257)  & (-2.537)  \\

    \addlinespace

    p   & 2 & 1 & 2 & 1 & 1 \\
    o   & 2 & 2 & 2 & 1 & 0 \\
    q   & 3 & 3 & 3 & 3 & 1  \\
    \addlinespace

    \(R^2\) & 0.008 & 0.804 & 0.994 & 0.564 & 0.909 \\
  
  \midrule
  %--------------- Panel B ----------------
    \multicolumn{6}{l}{Panel B: $R^{\mathrm{non-CEX}}_{\mathrm{chain}_{i},t}$ as the dependent variable}\\
    \addlinespace
    $\alpha_0$ 
      & -0.0002  & -0.001** & 0.0005 & 0.0002  & 0.0004 \\
      & (-0.156)  & (-2.419) & (0.297)  & (1.179)  & (1.036)  \\

    $\alpha_1$
      & -0.153*  & -0.103*** & -0.006 & 0.050*  & -0.048* \\
      & (-1.645)   & (-3.024) & (0.071)  & (1.792)  & (-1.868)  \\

    $\beta_0$
      & 0.173    & 0.028* & 0.0003  & 0.016  & 0.357*** \\
      & (1.576)  & (1.646) & (0.907)  & (1.461)  & (13.718)  \\

    $\beta_1$
      & -0.042   & -0.0003 & -0.038   & -0.0003  & -0.013*** \\
      & (-0.970) & (-0.023) & (-0.856)  & (-0.221)  & (-3.191)  \\

    $\beta_2$
      & -0.002** & 0.002*** & -0.006   & 0.021*** & -0.013 \\
      & (-2.350) & (5.435) & (-0.847) & (3.373) & (-0.356)  \\

    $\beta_3$
      & -0.0815  & 0.219*** & -0.098   & 0.0001  & 0.001 \\
      & (-1.043) & (3.566) & (-1.526) & (-0.836)  & (0.449)  \\

    $\beta_4$
      & -0.133    & 0.073 & 0.034   & -0.011  & -0.106** \\
      & (-0.784) & (1.412) & (1.500) & (-0.828)  & (-2.203)  \\

    \addlinespace

    p   & 2 & 1 & 1 & 1 & 1 \\
    o   & 2 & 0 & 1 & 0 & 0 \\
    q   & 3 & 2 & 3 & 1 & 1  \\
    \addlinespace

    \(R^2\) & -0.020 & 0.018 & 0.004 & 0.013 & 0.117 \\

  \midrule
  %--------------- Panel C ----------------
    \multicolumn{6}{l}{Panel C: $R^{\mathrm{Local}}_{\mathrm{chain}_{i},t}$ as the dependent variable}\\
    \addlinespace
    $\alpha_0$ 
      & -0.0007  & 0.00003 & 0.001** & 0.0005**  & -0.00004 \\
      & (-0.985)  & (0.280) & (2.081)  & (2.022)  & (-0.125)  \\

    $\alpha_1$
      & -0.022  & -0.005 & -0.0002* & -0.017  & -0.090*** \\
      & (-0.686) & (-0.454) & (-1.723)  & (-0.270)  & (-3.255)  \\

    $\beta_0$
      & 0.639*** & 0.880*** & 1.048***  & 0.101  & 0.602*** \\
      & (8.162)  & (55.996) & (117.205)  & (1.090)  & (9.698)  \\

    $\beta_1$
      & -0.034   & 0.002   & -0.042   & 0.014  & -0.009*** \\
      & (-0.695) & (0.788) & (-0.969)  & (1.367)  & (-3.388)  \\

    $\beta_2$
      & -0.001   & 0.0001  & -0.007   & 0.028** & -0.00008 \\
      & (-1.628) & (1.423) & (-0.890) & (2.433) & (-0.004)  \\

    $\beta_3$
      & -0.140**  & -0.009 & -0.176**   & -0.003  & 0.0002 \\
      & (-2.491) & (-0.293) & (-2.452) & (-1.173)  & (0.160)  \\

    $\beta_4$
      & -0.002    & -0.017 & -0.047*   & -0.024  & -0.160*** \\
      & (-0.001) & (-0.937) & (-1.652) & (-1.511)  & (-2.848)  \\

    \addlinespace

    p   & 2 & 1 & 2 & 1 & 2 \\
    o   & 2 & 2 & 2 & 1 & 1 \\
    q   & 3 & 3 & 3 & 3 & 3  \\
    \addlinespace

    \(R^2\) & 0.020 & 0.800 & 0.987 & 0.049 & 0.260 \\

  \end{tabularx}

\end{ThreePartTable}
\end{singlespace}
% --------------------------------------------------

% -------------------------------- table for baseline models with global var
\newpage

\begin{singlespace}
\begin{ThreePartTable}
  %──────────────────────────────────────────────────────────────────────────────
  % Notes, flush‐left and without indent
  \begin{TableNotes}[para,flushleft]
    \scriptsize
    Notes: This table shows estimates of the baseline linear models with global market variables, i.e., as illustrated by Equation (\ref{eq_linear_macro}), for the five chain portfolios in our study.  Panel A reports the estimated parameters of the models for all the five chains based on Equation Equation (\ref{eq_linear_macro}), which are as following:
      \begin{equation*}
      \begin{split}
      R^{All}_{\mathrm{Ethereum},t}
        &= \alpha_{0}
          + \alpha_{1}R^{All}_{\mathrm{Ethereum},t-1}
          + \beta_{0}R^{CEX}_{\mathrm{Ethereum},t}
          + \beta_{1}R^{All}_{\mathrm{Solana},t}
          + \beta_{2}R^{All}_{\mathrm{BSC},t}
          + \beta_{3}R^{All}_{\mathrm{Arbitrum},t}
          + \beta_{4}R^{All}_{\mathrm{Avalanche},t}\\
        & + \theta_{0} \, \mathrm{FTSER}_{t}
          + \theta_{1} \, \mathrm{HSR}_{t}
          + \theta_{2} \, \mathrm{SPR}_{t}
          + \theta_{3} \, \mathrm{EURIBOR}_{t}
          + \theta_{4} \, \mathrm{HIBOR}_{t}
          + \theta_{5} \, \mathrm{TREA}_{t}
          + e_{t},
          \end{split}
      \end{equation*}
      \begin{equation*}
      \begin{split}
      R^{All}_{\mathrm{Solana},t}
        & = \alpha_{0}
          + \alpha_{1}R^{All}_{\mathrm{Solana},t-1}
          + \beta_{0}R^{CEX}_{\mathrm{Solana},t}
          + \beta_{1}R^{All}_{\mathrm{Ethereum},t}
          + \beta_{2}R^{All}_{\mathrm{BSC},t}
          + \beta_{3}R^{All}_{\mathrm{Arbitrum},t}
          + \beta_{4}R^{All}_{\mathrm{Avalanche},t}\\
          & + \theta_{0} \, \mathrm{FTSER}_{t}
          + \theta_{1} \, \mathrm{HSR}_{t}
          + \theta_{2} \, \mathrm{SPR}_{t}
          + \theta_{3} \, \mathrm{EURIBOR}_{t}
          + \theta_{4} \, \mathrm{HIBOR}_{t}
          + \theta_{5} \, \mathrm{TREA}_{t}
          + e_{t},
      \end{split}
      \end{equation*}
      \begin{equation*}
      \begin{split}
      R^{All}_{\mathrm{BSC},t}
        & = \alpha_{0}
          + \alpha_{1}R^{All}_{\mathrm{BSC},t-1}
          + \beta_{0}R^{CEX}_{\mathrm{BSC},t}
          + \beta_{1}R^{All}_{\mathrm{Ethereum},t}
          + \beta_{2}R^{All}_{\mathrm{Solana},t}
          + \beta_{3}R^{All}_{\mathrm{Arbitrum},t}
          + \beta_{4}R^{All}_{\mathrm{Avalanche},t}\\
          & + \theta_{0} \, \mathrm{FTSER}_{t}
          + \theta_{1} \, \mathrm{HSR}_{t}
          + \theta_{2} \, \mathrm{SPR}_{t}
          + \theta_{3} \, \mathrm{EURIBOR}_{t}
          + \theta_{4} \, \mathrm{HIBOR}_{t}
          + \theta_{5} \, \mathrm{TREA}_{t}
          + e_{t},
      \end{split}
      \end{equation*}
      \begin{equation*}
      \begin{split}
      R^{All}_{\mathrm{Arbitrum},t}
        & = \alpha_{0}
          + \alpha_{1}R^{All}_{\mathrm{Arbitrum},t-1}
          + \beta_{0}R^{CEX}_{\mathrm{Arbitrum},t}
          + \beta_{1}R^{All}_{\mathrm{Ethereum},t}
          + \beta_{2}R^{All}_{\mathrm{Solana},t}
          + \beta_{3}R^{All}_{\mathrm{BSC},t}
          + \beta_{4}R^{All}_{\mathrm{Avalanche},t} \\
          & + \theta_{0} \, \mathrm{FTSER}_{t}
          + \theta_{1} \, \mathrm{HSR}_{t}
          + \theta_{2} \, \mathrm{SPR}_{t}
          + \theta_{3} \, \mathrm{EURIBOR}_{t}
          + \theta_{4} \, \mathrm{HIBOR}_{t}
          + \theta_{5} \, \mathrm{TREA}_{t}
          + e_{t},
      \end{split}
      \end{equation*}
      \begin{equation*}
      \begin{split}
      R^{All}_{\mathrm{Avalanche},t}
        & = \alpha_{0}
          + \alpha_{1}R^{All}_{\mathrm{Avalanche},t-1}
          + \beta_{0}R^{CEX}_{\mathrm{Avalanche},t}
          + \beta_{1}R^{All}_{\mathrm{Ethereum},t}
          + \beta_{2}R^{All}_{\mathrm{Solana},t}
          + \beta_{3}R^{All}_{\mathrm{BSC},t}
          + \beta_{4}R^{All}_{\mathrm{Arbitrum},t} \\
          & + \theta_{0} \, \mathrm{FTSER}_{t}
          + \theta_{1} \, \mathrm{HSR}_{t}
          + \theta_{2} \, \mathrm{SPR}_{t}
          + \theta_{3} \, \mathrm{EURIBOR}_{t}
          + \theta_{4} \, \mathrm{HIBOR}_{t}
          + \theta_{5} \, \mathrm{TREA}_{t}
          + e_{t}.
      \end{split}
      \end{equation*}
      while the series of residuals for all the modeled in the three panels are modeled by Glosten-Jagannathan-Runkle (GJR) asymmetric GARCH approach:
    \begin{equation}
    \begin{split}
        e_{t} \sim \mathcal{N}(0, \sigma_t^2), \;
        \sigma_t^2 = \omega + \sum_{i=1}^p \alpha_i e_{t-i}^2 + \sum_{j=1}^o \gamma_j e_{t-j}^2 \cdot \mathbb{I}_{\{e_{t-j} < 0\}} + \sum_{k=1}^q \beta_k \sigma_{t-k}^2,
    \end{split}
    \end{equation}
    and the optimal combination of $p$, $o$, and $q$ for each model is reported in this table. \\
    Panel B reports the estimated parameters of the models for all the five chains based on the specification with $R^{\mathrm{non-CEX}}_{\mathrm{chain}_{i},t}$ as the dependent variable and $R^{\mathrm{non-CEX}}_{\mathrm{chain}_{i},t-1}$ as the lagged return. So $\alpha_1$ stands for the coefficient of $R^{\mathrm{non-CEX}}_{\mathrm{chain}_{i},t-1}$. \\
    Panel C reports the estimated parameters of the models for all the five chains based on the specification with $R^{\mathrm{Local}}_{\mathrm{chain}_{i},t}$ as the dependent variable and $R^{\mathrm{Local}}_{\mathrm{chain}_{i},t-1}$ as the lagged return. So $\alpha_1$ stands for the coefficient of $R^{\mathrm{Local}}_{\mathrm{chain}_{i},t-1}$. \\
    Results in this table are based on unified sampling period from 28th Apr 2022 to 31st Mar 2025.\\
    t-statistics are in parentheses. \\
      ***/**/* denote significance at 1\%/5\%/10\%.
  \end{TableNotes}
  %──────────────────────────────────────────────────────────────────────────────

  % Force the table to span the full text width
  \setlength\LTleft{0pt}
  \setlength\LTright{0pt}

  \begin{tabularx}{\textwidth}{@{}l*{5}{Y}@{}}
    \caption{Estimates of the linear models with global market variables.\label{tab:baseline_macro}}\\
    % insert notes at top
    \insertTableNotes\\
    \addlinespace
    \toprule
      Model: 
        & (1) & (2) & (3) & (4) & (5) \\
      Chain portfolios: 
        & $\mathrm{Ethereum}$ & $\mathrm{Solana}$ & $\mathrm{BSC}$ & $\mathrm{Arbitrum}$ & $\mathrm{Avalanche}$ \\
    \midrule
    \endfirsthead

    \caption[]{Estimates of the linear models with global market variables. (continued)}\\
    \toprule
      Model: 
        & (1) & (2) & (3) & (4) & (5) \\
      Chain portfolios: 
        & $\mathrm{Ethereum}$ & $\mathrm{Solana}$ & $\mathrm{BSC}$ & $\mathrm{Arbitrum}$ & $\mathrm{Avalanche}$ \\
    \midrule
    \endhead

    \midrule
    \multicolumn{6}{r}{\emph{Continued on next page}}
    \endfoot

    \bottomrule
    \endlastfoot

    %--------------- Panel A ----------------
    \multicolumn{6}{l}{Panel A: $R^{All}_{\mathrm{chain}_{i},t}$ as the dependent variable}\\
    \addlinespace

    $\alpha_0$ 
      & -0.001***  & -0.001*** & 0.0001 & 0.00009  & 0.00008 \\
      & (-2.512)   & (-4.617) & (0.224) & (0.589)  & (0.816)  \\

    $\alpha_1$
      & -0.032    & 0.012  & 0.0000002  & -0.006  & -0.008 \\
      & (-1.327)  & (0.766) & (0.002)  & (-0.299)  & (-0.965)  \\

    $\beta_0$
      & 0.445***  & 0.893*** & 0.978***  & 0.398***  & 0.867*** \\
      & (6.737)   & (15.850) & (771.591)  & (28.361)  & (72.215)  \\

    $\beta_1$
      & -0.008    & 0.0003 & 0.016   & -0.0007  & -0.003* \\
      & (-0.127)  & (0.169) & (1.239)  & (-0.239)  & (-1.852)  \\

    $\beta_2$
      & -0.0008*** & 0.0004 & -0.007*** & 0.016*** & -0.002 \\
      & (-2.005)   & (0.885) & (-2.387) & (2.955) & (-0.304)  \\

    $\beta_3$
      &-0.025    & 0.012   & -0.088***   & -0.0003  & 0.0002 \\
      & (-0.436) & (0.274) & (-2.416) & (-0.852)  & (0.240)  \\

    $\beta_4$
      & -0.017    & -0.020  & -0.011   & -0.020  & -0.018* \\
      & (-0.653) & (-1.118) & (-0.754) & (-1.563)  & (-1.856)  \\

    $\theta_0$
      & 0.012    & -0.097*** & 0.015   & 0.012  & -0.079*** \\
      & (0.161) & (-4.393) & (0.206) & (0.236)  & (-4.500)  \\

    $\theta_1$
      & 0.188*** & -0.026 & 0.090**   & -0.050***  & 0.020** \\
      & (3.029) & (-0.377) & (1.860) & (-3.055)  & (2.221)  \\

    $\theta_2$
      & 0.023    & -0.026 & 0.047   & 0.001  & -0.067*** \\
      & (-0.653) & (-0.211) & (0.544) & (0.026)  & (-4.160)  \\

    $\theta_3$
      & 0.043***   & -0.055*** & 0.026   & 0.004  & -0.005 \\
      & (1.960) & (-6.205) & (1.052) & (0.649)  & (-1.070)  \\

    $\theta_4$
      & -0.028*** & -0.002  & 0.003   & -0.001  & -0.001 \\
      & (-2.914) & (-0.709) & (0.863) & (-0.827)  & (0.957)  \\

    $\theta_5$
      & -0.024*** & 0.003*** & -0.010**   & -0.002  & -0.012* \\
      & (-2.922) & (2.455) & (-2.388) & (-0.786)  & (-1.666)  \\

    \addlinespace

    p   & 2 & 1 & 1 & 1 & 1 \\
    o   & 2 & 1 & 1 & 0 & 0 \\
    q   & 2 & 2 & 1 & 2 & 1  \\
    \addlinespace

    Adjusted \(R^2\) & 0.007 & 0.847 & 0.996 & 0.538 & 0.909 \\
  
  \midrule
  %--------------- Panel B ----------------
    \multicolumn{6}{l}{Panel B: $R^{\mathrm{non-CEX}}_{\mathrm{chain}_{i},t}$ as the dependent variable}\\
    \addlinespace
    $\alpha_0$ 
      & -0.002  & -0.002** & 0.00007 & 0.0003  & 0.0007 \\
      & (-1.584)  & (-2.173) & (0.120)  & (1.406)  & (1.419)  \\

    $\alpha_1$
      & -0.053*  & -0.063   & 0.039  & 0.043  & -0.063** \\
      & (-1.853) & (-1.313) & (0.543) & (1.300)  & (-2.083)  \\

    $\beta_0$
      & 0.088    & 0.001 & 0.00004  & 0.016  & 0.383*** \\
      & (1.154)  & (0.106) & (0.200)  & (1.195)  & (8.405)  \\

    $\beta_1$
      &  -0.024   & -0.004 & 0.018   & 0.002  & -0.007 \\
      & (-0.464) & (-0.427) & (-0.856)  & (0.547)  & (-1.135)  \\

    $\beta_2$
      & -0.002*** & 0.002*** & -0.009   & 0.023*** & 0.004 \\
      & (-4.115) & (5.496) & (-0.696) & (2.832) & (0.176)  \\

    $\beta_3$
      & -0.047  & 0.206** & -0.120**   & -0.0002  & 0.004 \\
      & (-0.478) & (1.960) & (-1.983) & (-0.550)  & (1.321)  \\

    $\beta_4$
      & 0.041    & -0.047 & 0.031   & -0.016  & -0.162*** \\
      & (0.717) & (-0.593) & (1.201) & (-0.807)  & (-2.897)  \\

    $\theta_0$
      & 0.165    & -0.170 & -0.117   & -0.011  & -0.178* \\
      & (1.397) & (-1.201) & (-0.651) & (-0.156)  & (-1.731)  \\

    $\theta_1$
      & 0.089   & 0.004   & -0.047   & -0.049**  & -0.056 \\
      & (0.907) & (0.04) & (-0.884) & (-2.299)  & (-1.288)  \\

    $\theta_2$
      & -0.083    & 0.0004 & 0.253   & 0.040  & -0.395*** \\
      & (-0.636) & (0.003) & (2.076) & (0.994)  & (-4.927)  \\

    $\theta_3$
      & 0.071   & -0.032 & 0.0010   & -0.006  & -0.017 \\
      & (1.642) & (-0.978) & (0.04) & (-0.644)  & (-0.733)  \\

    $\theta_4$
      &  -0.011 & 0.006  & 0.002   & -0.0009  & -0.0007 \\
      & (-0.728) & (0.497) & (0.299) & (-0.359)  & (-0.112)  \\

    $\theta_5$
      & -0.011* & -0.033 & -0.005   & -0.008  & -0.005 \\
      & (-1.712) & (-1.000) & (-0.531) & (-1.634)  & (-1.126)  \\

    \addlinespace

    p   & 2 & 3 & 1 & 1 & 2 \\
    o   & 0 & 2 & 1 & 0 & 0 \\
    q   & 3 & 3 & 3 & 2 & 2  \\
    \addlinespace

    \(R^2\)       & 0.002 & 0.008 & 0.008 & 0.016 & 0.161 \\

  \midrule
  %--------------- Panel C ----------------
    \multicolumn{6}{l}{Panel C: $R^{\mathrm{Local}}_{\mathrm{chain}_{i},t}$ as the dependent variable}\\
    \addlinespace
    $\alpha_0$ 
      & -0.002    & -0.0002  & 0.002    & -0.0001  & -0.0004 \\
      & (-0.672)  & (-0.590) & (0.939)  & (-0.458)  & (-0.904)  \\

    $\alpha_1$
      & -0.051  & 0.009    & 0.00003 & 0.014  & -0.0004 \\
      & (-1.521) & (0.557) & (0.10)  & (0.223)  & (-0.013)  \\

    $\beta_0$
      & 0.528*** & 0.904*** & 1.054***  & -0.040  & 0.680*** \\
      & (4.610)  & (15.400) & (1890.655)  & (-0.601)  & (12.256)  \\

    $\beta_1$
      & -0.023   & 0.003*   & 0.019   & 0.017***  & -0.0003 \\
      & (-0.407) & (1.789) & (0.656)  & (3.377)  & (-0.053)  \\

    $\beta_2$
      & -0.002**   & -0.00002  & -0.018*** & 0.033*** & 0.008 \\
      & (-2.094) & (-0.094) & (-3.865) & (3.246) & (0.428)  \\

    $\beta_3$
      & -0.102  & 0.040    & -0.133**   & -0.0008  & -0.00004 \\
      & (-1.082) & (1.440) & (-2.036) & (-0.868)  & (-0.018)  \\

    $\beta_4$
      & 0.012    & -0.012  & -0.080*** & -0.011  & -0.224*** \\
      & (0.313) & (-0.757) & (-3.067) & (-0.541)  & (-3.676)  \\

    $\theta_0$
      & 0.062    & -0.127** & 0.067   & -0.002  &-0.390*** \\
      & (0.668) & (-2.462) & (0.358) & (-0.03)  & (-4.016)  \\

    $\theta_1$
      & 0.176**   & -0.011 & 0.135*   & -0.056**  & 0.043 \\
      & (2.460) & (-0.293) & (1.704) & (-2.500)  & (0.851)  \\

    $\theta_2$
      & -0.026    & -0.032  & -0.055   & -0.081  & -0.168* \\
      & (-0.289) & (-0.642) & (-0.309) & (-1.053)  & (-1.662)  \\

    $\theta_3$
      & 0.061**   & -0.056*** & 0.002   & 0.063***  & 0.004 \\
      & (2.169) & (-4.090) & (0.016) & (2.605)  & (0.226)  \\

    $\theta_4$
      &  -0.025* & -0.006***  & 0.032   & 0.011***  & -0.0008 \\
      & (-1.816) & (-2.779) & (0.597) & (3.651)  & (-0.131)  \\

    $\theta_5$
      & -0.015   & 0.005*** & -0.023  & -0.013***  & -0.010* \\
      & (-1.506) & (2.888) & (-1.449) & (-2.828)  & (-1.959)  \\

    \addlinespace

    p   & 2 & 3 & 2 & 2 & 1 \\
    o   & 2 & 2 & 0 & 0 & 0 \\
    q   & 3 & 1 & 2 & 3 & 1  \\
    \addlinespace

    Adjusted \(R^2\) & 0.026 & 0.844 & 0.993 & -0.029 & 0.273 \\

  \end{tabularx}

\end{ThreePartTable}
\end{singlespace}

% --------------------------------------------------------------

% -------------------------------- table for baseline models with global var and market condition var
\newpage

\begin{singlespace}
\begin{ThreePartTable}
  %──────────────────────────────────────────────────────────────────────────────
  % Notes, flush‐left and without indent
  \begin{TableNotes}[para,flushleft]
    \scriptsize
    Notes: This table shows estimates of the baseline linear models with both global market and chain-level activity variables, i.e., as illustrated by Equation (\ref{eq_linear_macro_crypto}), for the five chain portfolios in our study.  Panel A reports the estimated parameters of the models for all the five chains based on Equation Equation (\ref{eq_linear_macro_crypto}), which are as following:
      \begin{equation*}
      \begin{split}
      R^{All}_{\mathrm{Ethereum},t}
        &= \alpha_{0}
          + \alpha_{1}R^{All}_{\mathrm{Ethereum},t-1}
          + \beta_{0}R^{CEX}_{\mathrm{Ethereum},t}
          + \beta_{1}R^{All}_{\mathrm{Solana},t}
          + \beta_{2}R^{All}_{\mathrm{BSC},t}
          + \beta_{3}R^{All}_{\mathrm{Arbitrum},t}
          + \beta_{4}R^{All}_{\mathrm{Avalanche},t}\\
        & + \theta_{0} \, \mathrm{FTSER}_{t}
          + \theta_{1} \, \mathrm{HSR}_{t}
          + \theta_{2} \, \mathrm{SPR}_{t}
          + \theta_{3} \, \mathrm{EURIBOR}_{t}
          + \theta_{4} \, \mathrm{HIBOR}_{t}
          + \theta_{5} \, \mathrm{TREA}_{t}
          + \gamma_{0} \, R_{\$\mathrm{BTC},t} \\
        & + \gamma_{1} \, R_{\$\mathrm{ETH},t}
          + \gamma_{2} \, R_{\$\mathrm{SOL},t}
          + \gamma_{3} \, R_{\$\mathrm{BNB},t}
          + \gamma_{4} \, R_{\$\mathrm{ARB},t}
          + \gamma_{5} \, R_{\$\mathrm{AVAX},t}
          + \gamma_{6} \, \mathrm{SR}_{\mathrm{Ethereum},t} \\
        & + \gamma_{7} \, \mathrm{SR}_{\mathrm{Solana},t}
          + \gamma_{8} \, \mathrm{SR}_{\mathrm{BSC},t}
          + \gamma_{9} \, \mathrm{SR}_{\mathrm{Arbitrum},t}
          + \gamma_{10} \, \mathrm{SR}_{\mathrm{Avalanche},t}
          + e_{t},
          \end{split}
      \end{equation*}
      \begin{equation*}
      \begin{split}
      R^{All}_{\mathrm{Solana},t}
        & = \alpha_{0}
          + \alpha_{1}R^{All}_{\mathrm{Solana},t-1}
          + \beta_{0}R^{CEX}_{\mathrm{Solana},t}
          + \beta_{1}R^{All}_{\mathrm{Ethereum},t}
          + \beta_{2}R^{All}_{\mathrm{BSC},t}
          + \beta_{3}R^{All}_{\mathrm{Arbitrum},t}
          + \beta_{4}R^{All}_{\mathrm{Avalanche},t}\\
          & + \theta_{0} \, \mathrm{FTSER}_{t}
          + \theta_{1} \, \mathrm{HSR}_{t}
          + \theta_{2} \, \mathrm{SPR}_{t}
          + \theta_{3} \, \mathrm{EURIBOR}_{t}
          + \theta_{4} \, \mathrm{HIBOR}_{t}
          + \theta_{5} \, \mathrm{TREA}_{t}
          + \gamma_{0} \, R_{\$\mathrm{BTC},t} \\
        & + \gamma_{1} \, R_{\$\mathrm{ETH},t}
          + \gamma_{2} \, R_{\$\mathrm{SOL},t}
          + \gamma_{3} \, R_{\$\mathrm{BNB},t}
          + \gamma_{4} \, R_{\$\mathrm{ARB},t}
          + \gamma_{5} \, R_{\$\mathrm{AVAX},t}
          + \gamma_{6} \, \mathrm{SR}_{\mathrm{Ethereum},t} \\
        & + \gamma_{7} \, \mathrm{SR}_{\mathrm{Solana},t}
          + \gamma_{8} \, \mathrm{SR}_{\mathrm{BSC},t}
          + \gamma_{9} \, \mathrm{SR}_{\mathrm{Arbitrum},t}
          + \gamma_{10} \, \mathrm{SR}_{\mathrm{Avalanche},t}
          + e_{t},
      \end{split}
      \end{equation*}
      \begin{equation*}
      \begin{split}
      R^{All}_{\mathrm{BSC},t}
        & = \alpha_{0}
          + \alpha_{1}R^{All}_{\mathrm{BSC},t-1}
          + \beta_{0}R^{CEX}_{\mathrm{BSC},t}
          + \beta_{1}R^{All}_{\mathrm{Ethereum},t}
          + \beta_{2}R^{All}_{\mathrm{Solana},t}
          + \beta_{3}R^{All}_{\mathrm{Arbitrum},t}
          + \beta_{4}R^{All}_{\mathrm{Avalanche},t}\\
          & + \theta_{0} \, \mathrm{FTSER}_{t}
          + \theta_{1} \, \mathrm{HSR}_{t}
          + \theta_{2} \, \mathrm{SPR}_{t}
          + \theta_{3} \, \mathrm{EURIBOR}_{t}
          + \theta_{4} \, \mathrm{HIBOR}_{t}
          + \theta_{5} \, \mathrm{TREA}_{t}
          + \gamma_{0} \, R_{\$\mathrm{BTC},t} \\
        & + \gamma_{1} \, R_{\$\mathrm{ETH},t}
          + \gamma_{2} \, R_{\$\mathrm{SOL},t}
          + \gamma_{3} \, R_{\$\mathrm{BNB},t}
          + \gamma_{4} \, R_{\$\mathrm{ARB},t}
          + \gamma_{5} \, R_{\$\mathrm{AVAX},t}
          + \gamma_{6} \, \mathrm{SR}_{\mathrm{Ethereum},t} \\
        & + \gamma_{7} \, \mathrm{SR}_{\mathrm{Solana},t}
          + \gamma_{8} \, \mathrm{SR}_{\mathrm{BSC},t}
          + \gamma_{9} \, \mathrm{SR}_{\mathrm{Arbitrum},t}
          + \gamma_{10} \, \mathrm{SR}_{\mathrm{Avalanche},t}
          + e_{t},
      \end{split}
      \end{equation*}
      \begin{equation*}
      \begin{split}
      R^{All}_{\mathrm{Arbitrum},t}
        & = \alpha_{0}
          + \alpha_{1}R^{All}_{\mathrm{Arbitrum},t-1}
          + \beta_{0}R^{CEX}_{\mathrm{Arbitrum},t}
          + \beta_{1}R^{All}_{\mathrm{Ethereum},t}
          + \beta_{2}R^{All}_{\mathrm{Solana},t}
          + \beta_{3}R^{All}_{\mathrm{BSC},t}
          + \beta_{4}R^{All}_{\mathrm{Avalanche},t} \\
          & + \theta_{0} \, \mathrm{FTSER}_{t}
          + \theta_{1} \, \mathrm{HSR}_{t}
          + \theta_{2} \, \mathrm{SPR}_{t}
          + \theta_{3} \, \mathrm{EURIBOR}_{t}
          + \theta_{4} \, \mathrm{HIBOR}_{t}
          + \theta_{5} \, \mathrm{TREA}_{t}
          + \gamma_{0} \, R_{\$\mathrm{BTC},t} \\
        & + \gamma_{1} \, R_{\$\mathrm{ETH},t}
          + \gamma_{2} \, R_{\$\mathrm{SOL},t}
          + \gamma_{3} \, R_{\$\mathrm{BNB},t}
          + \gamma_{4} \, R_{\$\mathrm{ARB},t}
          + \gamma_{5} \, R_{\$\mathrm{AVAX},t}
          + \gamma_{6} \, \mathrm{SR}_{\mathrm{Ethereum},t} \\
        & + \gamma_{7} \, \mathrm{SR}_{\mathrm{Solana},t}
          + \gamma_{8} \, \mathrm{SR}_{\mathrm{BSC},t}
          + \gamma_{9} \, \mathrm{SR}_{\mathrm{Arbitrum},t}
          + \gamma_{10} \, \mathrm{SR}_{\mathrm{Avalanche},t}
          + e_{t},
      \end{split}
      \end{equation*}
      \begin{equation*}
      \begin{split}
      R^{All}_{\mathrm{Avalanche},t}
        & = \alpha_{0}
          + \alpha_{1}R^{All}_{\mathrm{Avalanche},t-1}
          + \beta_{0}R^{CEX}_{\mathrm{Avalanche},t}
          + \beta_{1}R^{All}_{\mathrm{Ethereum},t}
          + \beta_{2}R^{All}_{\mathrm{Solana},t}
          + \beta_{3}R^{All}_{\mathrm{BSC},t}
          + \beta_{4}R^{All}_{\mathrm{Arbitrum},t} \\
          & + \theta_{0} \, \mathrm{FTSER}_{t}
          + \theta_{1} \, \mathrm{HSR}_{t}
          + \theta_{2} \, \mathrm{SPR}_{t}
          + \theta_{3} \, \mathrm{EURIBOR}_{t}
          + \theta_{4} \, \mathrm{HIBOR}_{t}
          + \theta_{5} \, \mathrm{TREA}_{t}
          + \gamma_{0} \, R_{\$\mathrm{BTC},t} \\
        & + \gamma_{1} \, R_{\$\mathrm{ETH},t}
          + \gamma_{2} \, R_{\$\mathrm{SOL},t}
          + \gamma_{3} \, R_{\$\mathrm{BNB},t}
          + \gamma_{4} \, R_{\$\mathrm{ARB},t}
          + \gamma_{5} \, R_{\$\mathrm{AVAX},t}
          + \gamma_{6} \, \mathrm{SR}_{\mathrm{Ethereum},t} \\
        & + \gamma_{7} \, \mathrm{SR}_{\mathrm{Solana},t}
          + \gamma_{8} \, \mathrm{SR}_{\mathrm{BSC},t}
          + \gamma_{9} \, \mathrm{SR}_{\mathrm{Arbitrum},t}
          + \gamma_{10} \, \mathrm{SR}_{\mathrm{Avalanche},t}
          + e_{t}.
      \end{split}
      \end{equation*}
      while the series of residuals for all the modeled in the three panels are modeled by Glosten-Jagannathan-Runkle (GJR) asymmetric GARCH approach:
    \begin{equation}
    \begin{split}
        e_{t} \sim \mathcal{N}(0, \sigma_t^2), \;
        \sigma_t^2 = \omega + \sum_{i=1}^p \alpha_i e_{t-i}^2 + \sum_{j=1}^o \gamma_j e_{t-j}^2 \cdot \mathbb{I}_{\{e_{t-j} < 0\}} + \sum_{k=1}^q \beta_k \sigma_{t-k}^2,
    \end{split}
    \end{equation}
    and the optimal combination of $p$, $o$, and $q$ for each model is reported in this table. \\
    Panel B reports the estimated parameters of the models for all the five chains based on the specification with $R^{\mathrm{non-CEX}}_{\mathrm{chain}_{i},t}$ as the dependent variable and $R^{\mathrm{non-CEX}}_{\mathrm{chain}_{i},t-1}$ as the lagged return. So $\alpha_1$ stands for the coefficient of $R^{\mathrm{non-CEX}}_{\mathrm{chain}_{i},t-1}$. \\
    Panel C reports the estimated parameters of the models for all the five chains based on the specification with $R^{\mathrm{Local}}_{\mathrm{chain}_{i},t}$ as the dependent variable and $R^{\mathrm{Local}}_{\mathrm{chain}_{i},t-1}$ as the lagged return. So $\alpha_1$ stands for the coefficient of $R^{\mathrm{Local}}_{\mathrm{chain}_{i},t-1}$. \\
    Results in this table are based on unified sampling period from 17th Mar 2023 to 31st Mar 2025.\\
    t-statistics are in parentheses. \\
      ***/**/* denote significance at 1\%/5\%/10\%.
  \end{TableNotes}
  %──────────────────────────────────────────────────────────────────────────────

  % Force the table to span the full text width
  \setlength\LTleft{0pt}
  \setlength\LTright{0pt}

  \begin{tabularx}{\textwidth}{@{}l*{5}{Y}@{}}
    \caption{Estimates of the linear models with global market and chain-level variables.\label{tab:baseline_chain_macro}}\\
    % insert notes at top
    \insertTableNotes\\
    \addlinespace
    \toprule
      Model: 
        & (1) & (2) & (3) & (4) & (5) \\
      Chain portfolios: 
        & $\mathrm{Ethereum}$ & $\mathrm{Solana}$ & $\mathrm{BSC}$ & $\mathrm{Arbitrum}$ & $\mathrm{Avalanche}$ \\
    \midrule
    \endfirsthead

    \caption[]{Estimates of the linear models with global market and chain-level activity variables. (continued)}\\
    \toprule
      Model: 
        & (1) & (2) & (3) & (4) & (5) \\
      Chain portfolios: 
        & $\mathrm{Ethereum}$ & $\mathrm{Solana}$ & $\mathrm{BSC}$ & $\mathrm{Arbitrum}$ & $\mathrm{Avalanche}$ \\
    \midrule
    \endhead

    \midrule
    \multicolumn{6}{r}{\emph{Continued on next page}}
    \endfoot

    \bottomrule
    \endlastfoot

    %--------------- Panel A ----------------
    \multicolumn{6}{l}{Panel A: $R^{All}_{\mathrm{chain}_{i},t}$ as the dependent variable}\\
    \addlinespace

    $\alpha_0$ 
      & -0.001*  & -0.0002 & -0.0001 & 0.00003  & -0.00003 \\
      & ( -1.764)   & (-0.301) & (-0.230) & (0.070)  & (-0.270)  \\

    $\alpha_1$
      & -0.026    & 0.002  & -0.0001  & 0.017  & 0.005 \\
      & (-1.290)  & (0.601) & (-1.444)  & (1.506)  & (0.458)  \\

    $\beta_0$
      & 0.608***  & 0.990*** & 0.978***  & 0.399***  & 0.785*** \\
      & (17.542)   & (14.410) & (1478.514)  & (10.630)  & (33.187)  \\

    $\beta_1$
      & -0.030    & -0.0002 & 0.005   & -0.002  & -0.001 \\
      & (-0.703)  & (-0.272) & (0.291)  & (-1.287)  & (-1.050)  \\

    $\beta_2$
      & -0.001*** & 0.0003 & 0.002 & 0.005*** & 0.003 \\
      & (-4.409)   & (0.492) & (0.636) & (2.955) & (0.534)  \\

    $\beta_3$
      &0.121***    & 0.006   & -0.102*   & 0.00003  & 0.0007 \\
      & (3.141)   & (0.248) & (-1.929) & (0.691)  & (0.809)  \\

    $\beta_4$
      & -0.078**    & 0.013  & 0.044**   & -0.004  & -0.029* \\
      & (-2.024) & (0.237) & (2.000) & (-0.227)  & (-1.917)  \\

    $\theta_0$
      & 0.141    & 0.027 & -0.107**   & 0.027  & -0.041 \\
      & (1.640) & (0.338) & (-2.047) & (1.247)  & (-1.493)  \\

    $\theta_1$
      & -0.019 & -0.020 & 0.046   & -0.021  & 0.027** \\
      & (-0.354) & (-0.336) & (0.982) & (-1.248)  & (1.976)  \\

    $\theta_2$
      & -0.245*** & 0.031 & 0.274**   & -0.010  & 0.025 \\
      & (-3.175) & (0.297) & (2.064) & (-0.369)  & (1.250)  \\

    $\theta_3$
      & 0.036*   &  -0.013 & 0.026   & 0.002  & -0.011* \\
      & (1.902) & (-0.359) & (1.018) & (0.260)  & (-1.700)  \\

    $\theta_4$
      & -0.029*** & -0.002  & 0.003   & -0.001  & 0.0009 \\
      & (-3.203) & (-0.279) & (1.119) & (-0.820)  & (0.825)  \\

    $\theta_5$
      & -0.007 & 0.001 & -0.003   & -0.001  & -0.0005 \\
      & (-1.106) & (-0.261) & (-0.757) & (-0.544)  & (-0.564)  \\

    $\gamma_0$
      & 0.043 &  -0.011 & 0.048   & 0.295***  & 0.099*** \\
      & (0.811) & (-0.224) & (1.160) & (18.845)  & (3.199)  \\

    $\gamma_1$
      & 0.124** & -0.007 & -0.026   & -0.045**  & 0.026 \\
      & (2.298) & (-0.367) & (-0.651) & (-2.486)  & (0.913)  \\

    $\gamma_2$
      & 0.011 & -0.015 & -0.010   & -0.004  & -0.0003 \\
      & (0.448) & (-0.357) & (-0.679) & (-0.463)  & (-0.02)  \\

    $\gamma_3$
      & -0.051* & 0.014 & -0.068**   & 0.014*  & 0.006 \\
      & (-1.725) & (1.170) & (-2.443) & (1.874)  & (0.479)  \\

    $\gamma_4$
      & 0.004 & 0.0005 & 0.001   & 0.005**  & 0.003*** \\
      & (0.595) & (0.269) & (0.621) & (2.480)  & (5.372)  \\

    $\gamma_5$
      & -0.084** & -0.003 & 0.010   & -0.007  & -0.121*** \\
      & (-2.388) & (-0.177) & (0.558) & (-0.741)  & (-6.161)  \\

    $\gamma_6$
      & 0.001 & -0.00008 & -0.0005   & 0.0001  & -0.0001 \\
      & (1.630) & (-0.307) & (-0.539) & (0.346)  & (-0.338)  \\

    $\gamma_7$
      & -0.001 & -0.001 & -0.001   & -0.0004*  & -0.0002 \\
      & (-1.034) & (-0.307) & (-1.170) & (-1.745)  & (-0.563)  \\

    $\gamma_8$
      & 0.006 & 0.005 & -0.003   & 0.0003  & 0.002 \\
      & (0.552) & (0.318) & (-1.272) & (0.300)  & (0.994)  \\

    $\gamma_9$
      & 0.035 & -0.015 & -0.104   & 0.0006  & 0.026** \\
      & (0.948) & (-0.193) & (-1.062) & (0.04)  & (2.462)  \\

    $\gamma_{10}$
      & 0.458** & 0.052 & 0.088   & 0.008  & 0.041* \\
      & (2.192) & (0.342) & (1.237) & (0.181)  & (1.909)  \\

    \addlinespace

    p   & 2 & 2 & 1 & 2 & 1 \\
    o   & 2 & 1 & 0 & 0 & 0 \\
    q   & 3 & 3 & 1 & 1 & 1  \\
    \addlinespace

    Adjusted \(R^2\) & 0.027 & 0.986 & 0.997 & 0.913 & 0.920 \\
  
  \midrule
  %--------------- Panel B ----------------
    \multicolumn{6}{l}{Panel B: $R^{\mathrm{non-CEX}}_{\mathrm{chain}_{i},t}$ as the dependent variable}\\
    \addlinespace
    $\alpha_0$ 
      & 0.0003  & -0.0003 & -0.0002 & -0.00007  & 0.0004 \\
      & (0.255)  & (-0.337) & (-0.282)  & (-0.704)  & (0.914)  \\

    $\alpha_1$
      & -0.003*  & -0.047   & 0.031  & 0.051***  & 0.014 \\
      & (1.818) & (-1.032) & (0.252) & (5.014)  & (0.385)  \\

    $\beta_0$
      & -0.015    & 0.005 & -0.0002  & -0.007*  & 0.136** \\
      & (-0.456)  & (0.345) & (-1.043)  & (-1.789)  & (2.252)  \\

    $\beta_1$
      &  -0.004   & -0.002 & 0.020   & -0.0008  & 0.00004 \\
      & (-0.714) & (-0.295) & (0.968)  & (-0.647)  & (0.007)  \\

    $\beta_2$
      & -0.002** & 0.002 & -0.001   & 0.008*** & 0.015 \\
      & (-1.986) & (0.893) & (-0.128) & (3.206) & (0.413)  \\

    $\beta_3$
      & 0.088  & 0.084 & 0.023   & -0.0004  & 0.004* \\
      & (1.054) & (0.314) & (0.363) & (-1.019)  & (1.802)  \\

    $\beta_4$
      & 0.060    & 0.281 & 0.035   & -0.003  & -0.081 \\
      & (0.866) & (2.449) & (1.136) & (-0.301)  & (-0.830)  \\

    $\theta_0$
      & 0.025    & 0.140 & -0.275    & 0.017  & -0.052 \\
      & (0.199) & (0.831) & (-1.336) & (0.957)  & (-0.335)  \\

    $\theta_1$
      & -0.030   & -0.048   & -0.106   & -0.022**  & 0.033 \\
      & (-0.543) & (-0.249) & (-1.587) & (-2.168)  & (0.713)  \\

    $\theta_2$
      & 0.034   & -0.006   & 0.592   & 0.013  & 0.042 \\
      & (0.810) & (-0.042) & (1.759) & (0.940)  & (0.394)  \\

    $\theta_3$
      & -0.008   & -0.044   & -0.005   & -0.005  & -0.018 \\
      & (-0.295) & (-1.150) & (-0.218) & (-1.515)  & (-0.615)  \\

    $\theta_4$
      &  -0.003  & 0.004   & 0.0003   & 0.0005  & -0.0003 \\
      & (-0.689) & (0.389) & (0.08) & (0.553)  & (-0.056)  \\

    $\theta_5$
      & -0.006** & -0.040   & -0.0005   & -0.003  & -0.003 \\
      & (-2.216) & (-0.745) & (-0.057) & (-1.421)  & (-0.579)  \\

    $\gamma_0$
      & -0.053 &  -0.173   & -0.093   & 0.482***  & 0.237** \\
      & (-0.732) & (-1.426) & (-0.966) & (43.368)  & (2.371)  \\

    $\gamma_1$
      & 0.013   & 0.119   & 0.090   & -0.094***  & 0.142 \\
      & (0.240) & (1.034) & (1.552) & (-7.743)  & (1.474)  \\

    $\gamma_2$
      & -0.030** & -0.084* & -0.024   & -0.0007  & 0.016 \\
      & (-2.448) & (-1.841) & (-0.264) & (-0.128)  & (0.429)  \\

    $\gamma_3$
      & -0.015   & 0.104   & -0.328*** & 0.030***  & 0.024 \\
      & (-0.580) & (1.555) & (-2.784) & (5.079)  & (0.593)  \\

    $\gamma_4$
      & -0.004   & 0.043   &  -0.005** & 0.015***  & 0.008 \\
      & (-1.376) & (0.699) & (-2.103)  & (3.583)  & (0.836)  \\

    $\gamma_5$
      & 0.060** & 0.157** & 0.044   & -0.0004  & -0.477*** \\
      & (2.445) & (2.336) & (0.702) & (-0.066)  & (-6.632)  \\

    $\gamma_6$
      & 0.0004  & 0.003   & -0.001   & 0.001***  & 0.0003 \\
      & (0.599) & (1.313) & (-0.957) & (3.202)  & (0.175)  \\

    $\gamma_7$
      & 0.0003  & -0.003   & -0.0002   & -0.00005  & -0.0008 \\
      & (0.175) & (-0.129) & (-0.189) & (-0.159)  & (-0.429)  \\

    $\gamma_8$
      & 0.010   & 0.030   &  0.006   & 0.001  & 0.011 \\
      & (1.565) & (1.007) & (0.855) & (0.897)  & (1.156)  \\

    $\gamma_9$
      & -0.007  & -0.314*  & -0.312   & -0.005  & 0.089 \\
      & (-0.03) & (-2.034) & (-0.344) & (-0.660)  & (0.425)  \\

    $\gamma_{10}$
      & 0.035   &  0.164  & -0.069   & 0.003  & 0.240*** \\
      & (0.397) & (0.719) & (-0.286) & (0.178)  & (2.619)  \\

    \addlinespace

    p   & 2 & 2 & 1 & 1 & 1 \\
    o   & 0 & 0 & 0 & 2 & 1 \\
    q   & 3 & 3 & 2 & 1 & 1  \\
    \addlinespace

    \(R^2\)       & -0.009 & 0.048 & 0.065 & 0.775 & 0.361 \\

  \midrule
  %--------------- Panel C ----------------
    \multicolumn{6}{l}{Panel C: $R^{\mathrm{Local}}_{\mathrm{chain}_{i},t}$ as the dependent variable}\\
    \addlinespace
    $\alpha_0$ 
      & -0.002*   & 0.0005***  & 0.002***  & 0.0002**  & -0.0006 \\
      & (-1.822)  & (2.751) & (3.451)  & (1.995)  & (-1.055)  \\

    $\alpha_1$
      & -0.073***  & -0.002   & 0.00010 & 0.035***  & -0.028 \\
      & (-2.742)   & (-0.586) & (0.347)  & (3.536)  & (-0.626)  \\

    $\beta_0$
      & 0.825*** & 0.999      & 1.053***  & 0.006  & 0.7128*** \\
      & (15.100)  & (472.378) & (595.983)  & (1.269)  & (4.902)  \\

    $\beta_1$
      & -0.039   & 0.0004   & 0.047   & -0.0007  & -0.003 \\
      & (-0.711) & (0.139) & (0.602)  & (-0.563)  & (-0.442)  \\

    $\beta_2$
      & -0.002***   & -0.0009  & -0.013 & 0.005** & 0.012 \\
      & (-5.629) & (-0.602) & (-1.581) & (2.382) & (0.422)  \\

    $\beta_3$
      & 0.039  & 0.021    & -0.103***   & -0.00004  & -0.0004 \\
      & (0.515) & (0.460) & (-3.748) & (-0.793)  & (-0.145)  \\

    $\beta_4$
      & -0.064    & 0.033  & 0.013 & -0.003  & -0.210*** \\
      & (-1.057) & (0.456) & (0.421) & (-0.343)  & (-2.780)  \\

    $\theta_0$
      & 0.200    & -0.025 & -0.095   & 0.039*  & -0.117 \\
      & (1.284) & (-0.194) & (-0.452) & (1.849)  & (-0.899)  \\

    $\theta_1$
      & -0.055   & -0.005 & 0.040   & -0.024**  & 0.016 \\
      & (-0.732) & (-0.036) & (0.881) & (-2.025)  & (0.235)  \\

    $\theta_2$
      & -0.320**    & -0.029  & 0.297   & 0.005  & 0.058 \\
      & (-2.516) & (-0.566) & (1.167) & (0.330)  & (0.530)  \\

    $\theta_3$
      & 0.043   & -0.010 & -0.002   & -0.005  & -0.043* \\
      & (1.588) & (-0.399) & (-0.003) & (-1.166)  & (-1.835)  \\

    $\theta_4$
      &  -0.040*** & -0.002  & 0.026   & 0.0008  & 0.007 \\
      & (-3.404) & (-0.469) & (1.587) & (0.657)  & (0.964)  \\

    $\theta_5$
      & -0.008   & -0.002 & -0.027  & -0.002  & -0.012 \\
      & (-0.977) & (-1.297) & (-0.543) & ( -0.734)  & (-1.554)  \\

    $\gamma_0$
      & 0.052 &  -0.003   & -0.046   & 0.471***  & 0.249*** \\
      & (0.563) & (-0.018) & (-0.532) & (35.349)  & (3.186)  \\

    $\gamma_1$
      & 0.194*** & -0.024   & -0.021   & -0.103***  & -0.194** \\
      & (2.881)  & (-0.516) & (-0.513) & (-8.410)  & (-2.159)  \\

    $\gamma_2$
      & -0.003  & 0.031 & 0.003   & -0.0006  & 0.015 \\
      & (-0.077) & (1.503) & (0.138) & (-0.112)  & (0.445)  \\

    $\gamma_3$
      & -0.088*   & 0.008   & -0.059 & 0.032***  & -0.068 \\
      & (-1.880) & (0.687) & (-1.360) & (4.264)  & (-1.346)  \\

    $\gamma_4$
      & -0.013** & -0.00005 &  -0.010 & 0.015***  & 0.008 \\
      & (-2.481) & (-0.075) & (-1.283)  & (5.602)  & (0.883)  \\

    $\gamma_5$
      & -0.121** & -0.009 & -0.021   & -0.004  & -0.072 \\
      & (-2.374) & (-0.631) & (-1.029) & (-0.657)  & (-0.914)  \\

    $\gamma_6$
      & 0.0005  & -0.0002   & 0.001   & 0.0004  & -0.0008 \\
      & (0.555) & (-0.094) & (0.864) & (1.012)  & (-0.542)  \\

    $\gamma_7$
      & -0.0008  & -0.0007   & 0.0001   & -0.0003  & 0.0008 \\
      & (-0.529) & (-0.132) & (0.229) & (-1.144)  & (0.363)  \\

    $\gamma_8$
      & 0.020   & 0.004   &  -0.0004   & 0.0003  & -0.0010 \\
      & (1.234) & (0.712) & (-0.097) & (0.168)  & (-0.083)  \\

    $\gamma_9$
      & 0.081  & -0.044  & -0.083**   & -0.006  & -0.150 \\
      & (1.417) & (-1.040) & (-2.054) & (-0.602)  & (-1.417)  \\

    $\gamma_{10}$
      & 0.491*   &  0.071***  & 0.265**   & 0.016  & 0.229** \\
      & (1.930) & (3.076) & (2.494) & (0.829)  & (2.035)  \\

    \addlinespace

    p   & 2 & 2 & 3 & 2 & 1 \\
    o   & 2 & 2 & 0 & 1 & 0 \\
    q   & 3 & 3 & 3 & 1 & 1  \\
    \addlinespace

    Adjusted \(R^2\) & 0.041 & 0.982 & 0.995 & 0.798 & 0.408 \\

  \end{tabularx}

\end{ThreePartTable}
\end{singlespace}

% -------------------------------- table for non-limear models
\newpage

\begin{singlespace}
\begin{ThreePartTable}
  %──────────────────────────────────────────────────────────────────────────────
  % Notes, flush‐left and without indent
  \begin{TableNotes}[para,flushleft]
    \scriptsize
    Notes: This table shows estimates of the non-linear models, i.e., as illustrated by Equation (\ref{eq_nonlinear}), for the five chain portfolios in our study.  Panel A reports the estimated parameters of the models for all the five chains based on Equation Equation (\ref{eq_nonlinear}), which are as following:
      \begin{equation*}
      \begin{split}
      R^{All}_{\mathrm{Ethereum},t}
        &= \alpha_{0}
          + \alpha_{1}R^{CEX}_{\mathrm{Ethereum},t}
          + (\beta_{00}
             + \beta_{01}\,\mathrm{SR}_{\mathrm{Ethereum},t-1}
             + \beta_{02}\,R_{\$\mathrm{ETH},t-1})\,
             R^{All}_{\mathrm{Ethereum},t-1}
          + (\beta_{10}\\
           & + \beta_{11}\,\mathrm{SR}_{\mathrm{Solana},t}
             + \beta_{12}\,R_{\$\mathrm{SOL},t})\,
             R^{All}_{\mathrm{Solana},t} 
          + (\beta_{20}
             + \beta_{21}\,\mathrm{SR}_{\mathrm{BSC},t}
             + \beta_{22}\,R_{\$\mathrm{BNB},t})\,
             R^{All}_{\mathrm{BSC},t} 
          + (\beta_{30}\\
           & + \beta_{31}\,\mathrm{SR}_{\mathrm{Arbitrum},t}
             + \beta_{32}\,R_{\$\mathrm{ARB},t})\,
             R^{All}_{\mathrm{Arbitrum},t}
          + (\beta_{40}
             + \beta_{41}\,\mathrm{SR}_{\mathrm{Avalanche},t}
             + \beta_{42}\,R_{\$\mathrm{AVAX},t})\,
             R^{All}_{\mathrm{Avalanche},t}\\
          & + e_{t},
          \end{split}
      \end{equation*}
      \begin{equation*}
      \begin{split}
      R^{All}_{\mathrm{Solana},t}
        &= \alpha_{0}
          + \alpha_{1}R^{CEX}_{\mathrm{Solana},t}
          + (\beta_{00}
             + \beta_{01}\,\mathrm{SR}_{\mathrm{SoLana},t-1}
             + \beta_{02}\,R_{\$\mathrm{SOL},t-1})\,
             R^{All}_{\mathrm{Solana},t-1}
          + (\beta_{10}
             + \beta_{11}\,\mathrm{SR}_{\mathrm{Ethereum},t}\\
           & + \beta_{12}\,R_{\$\mathrm{ETH},t})\,
             R^{All}_{\mathrm{Ethereum},t} 
          + (\beta_{20}
             + \beta_{21}\,\mathrm{SR}_{\mathrm{BSC},t}
             + \beta_{22}\,R_{\$\mathrm{BNB},t})\,
             R^{All}_{\mathrm{BSC},t} 
          + (\beta_{30}
             + \beta_{31}\,\mathrm{SR}_{\mathrm{Arbitrum},t} \\
           & + \beta_{32}\,R_{\$\mathrm{ARB},t})\,
             R^{All}_{\mathrm{Arbitrum},t}
          + (\beta_{40}
             + \beta_{41}\,\mathrm{SR}_{\mathrm{Avalanche},t}
             + \beta_{42}\,R_{\$\mathrm{AVAX},t})\,
             R^{All}_{\mathrm{Avalanche},t}+ e_{t},
      \end{split}
      \end{equation*}
      \begin{equation*}
      \begin{split}
      R^{All}_{\mathrm{BSC},t}
        &= \alpha_{0}
          + \alpha_{1}R^{CEX}_{\mathrm{BSC},t}
          + (\beta_{00}
             + \beta_{01}\,\mathrm{SR}_{\mathrm{BSC},t-1}
             + \beta_{02}\,R_{\$\mathrm{BNB},t-1})\,
             R^{All}_{\mathrm{BSC},t-1}
          + (\beta_{10}
             + \beta_{11}\,\mathrm{SR}_{\mathrm{Ethereum},t}\\
          &  + \beta_{12}\,R_{\$\mathrm{ETH},t})\,
             R^{All}_{\mathrm{Ethereum},t} 
          + (\beta_{20}
             + \beta_{21}\,\mathrm{SR}_{\mathrm{Solana},t}
             + \beta_{22}\,R_{\$\mathrm{SOL},t})\,
             R^{All}_{\mathrm{Solana},t} 
          + (\beta_{30}
             + \beta_{31}\,\mathrm{SR}_{\mathrm{Arbitrum},t}\\
           & + \beta_{32}\,R_{\$\mathrm{ARB},t})\,
             R^{All}_{\mathrm{Arbitrum},t}
          + (\beta_{40}
             + \beta_{41}\,\mathrm{SR}_{\mathrm{Avalanche},t}
             + \beta_{42}\,R_{\$\mathrm{AVAX},t})\,
             R^{All}_{\mathrm{Avalanche},t}+ e_{t},
      \end{split}
      \end{equation*}
      \begin{equation*}
      \begin{split}
      R^{All}_{\mathrm{Arbitrum},t}
        &= \alpha_{0}
          + \alpha_{1}R^{CEX}_{\mathrm{Arbitrum},t}
          + (\beta_{00}
             + \beta_{01}\,\mathrm{SR}_{\mathrm{Arbitrum},t-1}
             + \beta_{02}\,R_{\$\mathrm{ARB},t-1})\,
             R^{All}_{\mathrm{Arbitrum},t-1}
          + (\beta_{10} \\
          &  + \beta_{11}\,\mathrm{SR}_{\mathrm{Ethereum},t}
             + \beta_{12}\,R_{\$\mathrm{ETH},t})\,
             R^{All}_{\mathrm{Ethereum},t} 
          + (\beta_{20}
             + \beta_{21}\,\mathrm{SR}_{\mathrm{Solana},t}
             + \beta_{22}\,R_{\$\mathrm{SOL},t})\,
             R^{All}_{\mathrm{Solana},t} +\\ 
             & (\beta_{30}
             + \beta_{31}\,\mathrm{SR}_{\mathrm{BSC},t}
             + \beta_{32}\,R_{\$\mathrm{BSC},t})\,
             R^{All}_{\mathrm{BSC},t}
          + (\beta_{40}
             + \beta_{41}\,\mathrm{SR}_{\mathrm{Avalanche},t}
             + \beta_{42}\,R_{\$\mathrm{AVAX},t})\,
             R^{All}_{\mathrm{Avalanche},t}+ e_{t},
      \end{split}
      \end{equation*}
      \begin{equation*}
      \begin{split}
      R^{All}_{\mathrm{Avalanche},t}
        &= \alpha_{0}
          + \alpha_{1}R^{CEX}_{\mathrm{Avalanche},t}
          + (\beta_{00}
             + \beta_{01}\,\mathrm{SR}_{\mathrm{Avalanche},t-1}
             + \beta_{02}\,R_{\$\mathrm{AVAX},t-1})\,
             R^{All}_{\mathrm{Avalanche},t-1}
          + (\beta_{10} \\
          &  + \beta_{11}\,\mathrm{SR}_{\mathrm{Ethereum},t}
             + \beta_{12}\,R_{\$\mathrm{ETH},t})\,
             R^{All}_{\mathrm{Ethereum},t} 
          + (\beta_{20}
             + \beta_{21}\,\mathrm{SR}_{\mathrm{Solana},t}
             + \beta_{22}\,R_{\$\mathrm{SOL},t})\,
             R^{All}_{\mathrm{Solana},t} +\\ 
             & (\beta_{30}
             + \beta_{31}\,\mathrm{SR}_{\mathrm{BSC},t}
             + \beta_{32}\,R_{\$\mathrm{BSC},t})\,
             R^{All}_{\mathrm{BSC},t}
          + (\beta_{40}
             + \beta_{41}\,\mathrm{SR}_{\mathrm{Arbitrum},t}
             + \beta_{42}\,R_{\$\mathrm{ARB},t})\,
             R^{All}_{\mathrm{Arbitrum},t}+ e_{t},
      \end{split}
      \end{equation*}
      while the series of residuals for all the modeled in the three panels are modeled by Glosten-Jagannathan-Runkle (GJR) asymmetric GARCH approach:
    \begin{equation}
    \begin{split}
        e_{t} \sim \mathcal{N}(0, \sigma_t^2), \;
        \sigma_t^2 = \omega + \sum_{i=1}^p \alpha_i e_{t-i}^2 + \sum_{j=1}^o \gamma_j e_{t-j}^2 \cdot \mathbb{I}_{\{e_{t-j} < 0\}} + \sum_{k=1}^q \beta_k \sigma_{t-k}^2,
    \end{split}
    \end{equation}
    and the optimal combination of $p$, $o$, and $q$ for each model is reported in this table. \\
    Panel B reports the estimated parameters of the models for all the five chains based on the specification with $R^{\mathrm{non-CEX}}_{\mathrm{chain}_{i},t}$ as the dependent variable and $R^{\mathrm{non-CEX}}_{\mathrm{chain}_{i},t-1}$ as the lagged return.\\
    Panel C reports the estimated parameters of the models for all the five chains based on the specification with $R^{\mathrm{Local}}_{\mathrm{chain}_{i},t}$ as the dependent variable and $R^{\mathrm{Local}}_{\mathrm{chain}_{i},t-1}$ as the lagged return. \\
    Results in this table are based on unified sampling period from 17th Mar 2023 to 31st Mar 2025.\\
    t-statistics are in parentheses. \\
      ***/**/* denote significance at 1\%/5\%/10\%.
  \end{TableNotes}
  %──────────────────────────────────────────────────────────────────────────────

  % Force the table to span the full text width
  \setlength\LTleft{0pt}
  \setlength\LTright{0pt}

  \begin{tabularx}{\textwidth}{@{}l*{5}{Y}@{}}
    \caption{Estimates of the baseline non-linear models.\label{tab:nonlinear}}\\
    % insert notes at top
    \insertTableNotes\\
    \addlinespace
    \toprule
      Model: 
        & (1) & (2) & (3) & (4) & (5) \\
      Chain portfolios: 
        & $\mathrm{Ethereum}$ & $\mathrm{Solana}$ & $\mathrm{BSC}$ & $\mathrm{Arbitrum}$ & $\mathrm{Avalanche}$ \\
    \midrule
    \endfirsthead

    \caption[]{Estimates of the baseline non-linear models. (continued)}\\
    \toprule
      Model: 
        & (1) & (2) & (3) & (4) & (5) \\
      Chain portfolios: 
        & $\mathrm{Ethereum}$ & $\mathrm{Solana}$ & $\mathrm{BSC}$ & $\mathrm{Arbitrum}$ & $\mathrm{Avalanche}$ \\
    \midrule
    \endhead

    \midrule
    \multicolumn{6}{r}{\emph{Continued on next page}}
    \endfoot

    \bottomrule
    \endlastfoot

    %--------------- Panel A ----------------
    \multicolumn{6}{l}{Panel A: $R^{All}_{\mathrm{chain}_{i},t}$ as the dependent variable}\\
    \addlinespace

    $\alpha_0$ 
      & -0.0009  & 0.0004 & 0.0005  & 0.0003**  & 0.00005 \\
      & (-1.351) & (1.332) & (0.601) & (2.092)  & (0.593)  \\

    $\alpha_1$
      & 0.645***  & 0.986*** & 0.977***  & 0.430***  & 0.835*** \\
      & (16.454)  & (54.193) & (400.929) & (39.265)  & (61.010)  \\

    $\beta_{00}$
      & -0.009     & 0.003    & 0.001    & -0.007  & -0.001 \\
      & (-0.214)   & (0.383) & (0.600)  & (-0.445)  & (-0.161)  \\

    $\beta_{01}$
      & -0.012    & -0.00002  & -0.016   & -0.014   & 0.026 \\
      & (-0.292)  & (-0.011) & (-0.983) & (-0.459) & (0.662)  \\

    $\beta_{02}$
      & 0.150     & -0.031   & 1.198** & 0.417** & 0.134 \\
      & (0.238)   & (-0.538) & (2.222)  & (2.533) & (0.961)  \\

    $\beta_{10}$
      & -0.029    & -0.0006   & -0.023** & 0.001  & -0.002*** \\
      & (-0.866)  & (-0.944) & (-3.045)   & (0.853)  & (-2.767)  \\

    $\beta_{11}$
      & 0.175**   & 0.005   & -0.037   & -0.002   & -0.002 \\
      & (2.042)   & (0.664) & (-1.497) & (-0.248)  & (-0.386)  \\

    $\beta_{12}$
      & -0.723    & 0.041  & -0.069   & -0.650***  & 0.116 \\
      & (-1.140) & (0.333) & (-0.149) & (-3.101)  & (0.971)  \\

    $\beta_{20}$
      & -0.017*  & -0.00005   & 0.004   & 0.022***    & -0.004 \\
      & (-1.707)   & (-0.026) & (0.847) & (4.884)  & (-0.726)  \\

    $\beta_{21}$
      & -0.027    & -0.016   & -0.021   & -0.002   & 0.026 \\
      & (-0.430)  & (-0.532) & ( -1.269)  & (-0.158) & (1.088)  \\

    $\beta_{22}$
      & -4.087*** & 0.623    & -0.028   & -0.158***   & 0.020 \\
      & (-4.015)   & (0.452) & (-0.422) & (-2.818) & (0.284)  \\

    $\beta_{30}$
      & 0.098**     & 0.027*   & -0.111**    & -0.002   & 0.002 \\
      & (1.961)   & (1.817) & ( -2.280) & (-1.469) & (0.873)  \\

    $\beta_{31}$
      & 0.099    & -0.042*   & -0.204**   & 0.002  & 0.016\\
      & (0.757)  & (-1.727) & (-1.984)  & (0.419)  & (1.375)  \\

    $\beta_{32}$
      & 1.506   & 0.063   & -0.543   & -0.670***   & -0.051 \\
      & (1.584)  & (1.270) & (-1.137) & (-2.640) & (-0.230)  \\

    $\beta_{40}$
      & -0.044   & 0.029   & 0.024  & -0.015*  & -0.075** \\
      & (-0.683) & (1.397) & (1.217)  & (-1.705)  & (-5.944)  \\

    $\beta_{41}$
      & 0.070    & 0.011 & -0.051   & 0.020      & 0.100*** \\
      & (0.963)  & (0.017) & (-1.274)  & (1.107) & (2.750)  \\

    $\beta_{42}$
      & -0.113    & 0.003   & -0.149  & 0.182   & 0.047 \\
      & (-0.034)  & (0.005) & (-0.470) & (0.892) & (0.378)  \\

    \addlinespace

    p   & 2 & 2 & 1 & 1 & 2 \\
    o   & 2 & 2 & 1 & 0 & 0 \\
    q   & 3 & 3 & 1 & 1 & 2  \\
    \addlinespace

    Adjusted \(R^2\) & 0.020 & 0.981 & 0.995 & 0.818 & 0.890 \\
  
  \midrule
  %--------------- Panel B ----------------
    \multicolumn{6}{l}{Panel B: $R^{\mathrm{non-CEX}}_{\mathrm{chain}_{i},t}$ as the dependent variable}\\
    \addlinespace
    $\alpha_0$ 
      & -0.002  & -0.001   & 0.0009  & 0.0006***  & 0.0001 \\
      & (-1.021) & (-0.307) & (0.683) & (2.918)  & (0.254)  \\

    $\alpha_1$
      & 0.090    & 0.002  &  0.0005  & 0.046***  & 0.306*** \\
      & (1.406)  & (0.133) & (1.080) & (3.943)  & (9.818)  \\

    $\beta_{00}$
      & 0.002     & 0.022    & 0.00008  & -0.016  & 0.008 \\
      & (0.696)   & (1.526) & (0.223)  & (-0.662)  & (0.224)  \\

    $\beta_{01}$
      & 0.039    & 0.035   & -0.025   & -0.097   & 0.038 \\
      & (0.861)  & (0.175) & (-0.057) & (-1.126) & (0.427)  \\

    $\beta_{02}$
      & 0.073     & 1.092   & 0.185   & 0.971***   & 0.139 \\
      & (0.134)   & (0.833) & (0.139)  & (3.015) & (0.255)  \\

    $\beta_{10}$
      & -0.007    & -0.007   & -0.052   & 0.003  & -0.014*** \\
      & (-0.339)  & (-0.286) & (-1.305) & (1.406)  & (-5.457)  \\

    $\beta_{11}$
      & -0.049   & -0.004     & 0.007   & -0.007   & -0.019 \\
      & (-0.240)   & (-0.026) & (0.201) & (-0.532)  & (-0.624)  \\

    $\beta_{12}$
      & -1.046    & 0.037  & 0.431   & -1.02***    & 0.814 \\
      & (-0.978) & (0.061) & (0.237) & (-3.522)  & (1.356)  \\

    $\beta_{20}$
      & 0.007    & 0.009   & -0.014   & 0.035*** & -0.026 \\
      & (0.210)  & (0.268) & (-0.984) & (4.839)  & (-0.958)  \\

    $\beta_{21}$
      & 0.163    & 0.036   & -0.042   & -0.010   & 0.166 \\
      & (1.038)  & (0.606) & (-1.291)  & (-0.465) & (1.378)  \\

    $\beta_{22}$
      & -4.372     & 0.290    & 0.129   & -0.256*** & 0.152 \\
      & (-0.682)   & (0.038) & (0.816) & (-3.069) & (0.388)  \\

    $\beta_{30}$
      & -0.144     & 0.475   & -0.641*  & -0.002   & 0.006 \\
      & (-0.516)   & (0.451) & (-1.671)  & (-1.375) & (1.048)  \\

    $\beta_{31}$
      & 0.733    & -0.817    & 0.070   & 0.011*   & 0.006 \\
      & (1.453)  & (-1.273)  & (0.389)  & (1.853)  & (0.220)  \\

    $\beta_{32}$
      & -0.814   & -0.974  & -0.952   & -1.259***   & 0.477 \\
      & (-0.527) & (-0.036) & (-0.315) & (-3.096) & (0.548)  \\

    $\beta_{40}$
      & -0.095   & 0.080   & 0.011  & -0.027**  & -0.328*** \\
      & (-0.886) & (0.438)  & (0.312)  & (-2.100)  & (-5.597)  \\

    $\beta_{41}$
      & -0.142    & 0.139   & -0.058   &  0.047*     & 0.260 \\
      & ( -0.532)  & (0.509) & (-0.530)  & (1.800) & (1.179)  \\

    $\beta_{42}$
      & -9.127    & 0.612   & -0.497   & 0.259   & -0.214 \\
      & (-1.157)  & (0.196) & (-0.732) & (0.747) & (-0.279)  \\

    \addlinespace

    p   & 3 & 1 & 1 & 1 & 1 \\
    o   & 2 & 0 & 1 & 0 & 1 \\
    q   & 2 & 1 & 2 & 1 & 1  \\
    \addlinespace

    Adjusted \(R^2\) & -0.015 & 0.024 & -0.010 & 0.097 & 0.156 \\

  \midrule
  %--------------- Panel C ----------------
    \multicolumn{6}{l}{Panel C: $R^{\mathrm{Local}}_{\mathrm{chain}_{i},t}$ as the dependent variable}\\
    \addlinespace
    $\alpha_0$ 
      & -0.001   & 0.0003  & 0.001*  & 0.0005**  & 0.0004 \\
      & (-1.474) & (0.575) & (1.676) & (2.555)  & (0.562)  \\

    $\alpha_1$
      & 0.906***  & 1.000*** & 1.050***  & 0.065*    & 0.800*** \\
      & (17.564)  & (738.327) & (121.116) & (5.196)  & (8.314)  \\

    $\beta_{00}$
      & -0.012     & -0.0003    & 0.002***  & -0.033  & -0.092*** \\
      & (-0.223)   & (-0.339) & (6.034)  & (-1.343)  & (-2.657)  \\

    $\beta_{01}$
      & -0.044    & 0.008  & 0.003     & -0.106   & 0.077 \\
      & (-0.895)  & (0.908) & (0.182)  & (-1.224) & (0.833)  \\

    $\beta_{02}$
      & -0.362     & -0.023   & 0.512   & 0.308 & 0.049 \\
      & (-0.429)   & (-0.875) & (0.756)  & (1.346) & (0.044)  \\

    $\beta_{10}$
      & -0.087*   & -0.0001   & -0.045   & 0.003*  & -0.009*** \\
      & (-1.802)  & (-0.190) & (-1.160)  & (1.747)  & (-3.247)  \\

    $\beta_{11}$
      & 0.122   & 0.005     & -0.027   & -0.008   & -0.001 \\
      & (1.099)   & (0.153) & (-0.521) & (-0.509)  & (-0.034)  \\

    $\beta_{12}$
      & -0.690    & -0.017  & -0.564   & -0.985***  & 0.434 \\
      & (-0.939) & (-0.133) & (-0.068) & (-3.244)  & (0.906)  \\

    $\beta_{20}$
      & -0.009  & -0.0009   &  -0.014  & 0.035***  & -0.003 \\
      & (-1.387)   & (-0.092) & (-1.571) & (4.768)  & (-0.189)  \\

    $\beta_{21}$
      & 0.035    & -0.012  &  -0.013   & -0.001   & 0.037 \\
      & (0.995)  & (-0.377) & (-0.731)  & (-0.042) & (0.428)  \\

    $\beta_{22}$
      & -4.143***  & 0.133    & 0.068   & -0.264***   & 0.174 \\
      & (-2.674)   & (0.202) & ( 0.431) & (-3.193) & (0.795)  \\

    $\beta_{30}$
      & -0.014    & 0.075   & -0.262    & -0.003*   & 0.007** \\
      & (-0.193)   & (1.151) & (-1.348) & (-1.656) & (2.043)  \\

    $\beta_{31}$
      & 0.105    & -0.094   & 0.052   & 0.010   & 0.070*** \\
      & (0.555)  & (-0.527) & (0.257)  & (1.596)  & (3.334)  \\

    $\beta_{32}$
      & 1.761    & -0.019    & -0.248   & -1.357*** & -0.776 \\
      & (1.574)  & (-0.143) & (-0.300) & (-3.141) & (-1.168)  \\

    $\beta_{40}$
      & -0.128*   & 0.030   & 0.032   & -0.031**  & -0.256*** \\
      & (-1.820) & (0.425) & (0.336)  & (-2.196)  & (-4.260)  \\

    $\beta_{41}$
      & 0.059    & -0.008   & -0.081   & 0.049*      & 0.021 \\
      & (0.574)  & (-0.081) & (-1.193)  & (1.928) & (0.120)  \\

    $\beta_{42}$
      & -0.200    & 0.141   & -0.074   & 0.298 & -1.276 \\
      & (-0.051)  & (0.879) & (-0.059) & (0.871) & (-1.602)  \\

    \addlinespace

    p   & 2 & 2 & 2 & 1 & 1 \\
    o   & 2 & 2 & 2 & 0 & 0 \\
    q   & 2 & 3 & 3 & 1 & 1  \\
    \addlinespace

    Adjusted \(R^2\) & 0.026 & 0.975 & 0.990 & 0.105 & 0.400 \\

  \end{tabularx}

\end{ThreePartTable}
\end{singlespace}

% -------------------------------- table for non-limear models with macro
\newpage

\begin{singlespace}
\begin{ThreePartTable}
  %──────────────────────────────────────────────────────────────────────────────
  % Notes, flush‐left and without indent
  \begin{TableNotes}[para,flushleft]
    \scriptsize
    Notes: This table shows estimates of the non-linear models, i.e., as illustrated by Equation (\ref{eq_nonlinear_macro}), for the five chain portfolios in our study.  Panel A reports the estimated parameters of the models for all the five chains based on Equation Equation (\ref{eq_nonlinear_macro}), which are as following:
      \begin{equation*}
      \begin{split}
      R^{All}_{\mathrm{Ethereum},t}
        &= \alpha_{0}
          + \alpha_{1}R^{CEX}_{\mathrm{Ethereum},t}
          + (\beta_{00}
             + \beta_{01}\,\mathrm{SR}_{\mathrm{Ethereum},t-1}
             + \beta_{02}\,R_{\$\mathrm{ETH},t-1})\,
             R^{All}_{\mathrm{Ethereum},t-1}
          + (\beta_{10}\\
           & + \beta_{11}\,\mathrm{SR}_{\mathrm{Solana},t}
             + \beta_{12}\,R_{\$\mathrm{SOL},t})\,
             R^{All}_{\mathrm{Solana},t} 
          + (\beta_{20}
             + \beta_{21}\,\mathrm{SR}_{\mathrm{BSC},t}
             + \beta_{22}\,R_{\$\mathrm{BNB},t})\,
             R^{All}_{\mathrm{BSC},t} 
          + (\beta_{30}\\
           & + \beta_{31}\,\mathrm{SR}_{\mathrm{Arbitrum},t}
             + \beta_{32}\,R_{\$\mathrm{ARB},t})\,
             R^{All}_{\mathrm{Arbitrum},t}
          + (\beta_{40}
             + \beta_{41}\,\mathrm{SR}_{\mathrm{Avalanche},t}
             + \beta_{42}\,R_{\$\mathrm{AVAX},t})\,
             R^{All}_{\mathrm{Avalanche},t}\\
          &  + \theta_{0} \, \mathrm{FTSER}_{t}
          + \theta_{1} \, \mathrm{HSR}_{t}
          + \theta_{2} \, \mathrm{SPR}_{t}
          + \theta_{3} \, \mathrm{EURIBOR}_{t}
          + \theta_{4} \, \mathrm{HIBOR}_{t}
          + \theta_{5} \, \mathrm{TREA}_{t}
          + e_{t},
          \end{split}
      \end{equation*}
      \begin{equation*}
      \begin{split}
      R^{All}_{\mathrm{Solana},t}
        &= \alpha_{0}
          + \alpha_{1}R^{CEX}_{\mathrm{Solana},t}
          + (\beta_{00}
             + \beta_{01}\,\mathrm{SR}_{\mathrm{SoLana},t-1}
             + \beta_{02}\,R_{\$\mathrm{SOL},t-1})\,
             R^{All}_{\mathrm{Solana},t-1}
          + (\beta_{10}
             + \beta_{11}\,\mathrm{SR}_{\mathrm{Ethereum},t}\\
           & + \beta_{12}\,R_{\$\mathrm{ETH},t})\,
             R^{All}_{\mathrm{Ethereum},t} 
          + (\beta_{20}
             + \beta_{21}\,\mathrm{SR}_{\mathrm{BSC},t}
             + \beta_{22}\,R_{\$\mathrm{BNB},t})\,
             R^{All}_{\mathrm{BSC},t} 
          + (\beta_{30}
             + \beta_{31}\,\mathrm{SR}_{\mathrm{Arbitrum},t} \\
           & + \beta_{32}\,R_{\$\mathrm{ARB},t})\,
             R^{All}_{\mathrm{Arbitrum},t}
          + (\beta_{40}
             + \beta_{41}\,\mathrm{SR}_{\mathrm{Avalanche},t}
             + \beta_{42}\,R_{\$\mathrm{AVAX},t})\,
             R^{All}_{\mathrm{Avalanche},t} \\
          & + \theta_{0} \, \mathrm{FTSER}_{t}
          + \theta_{1} \, \mathrm{HSR}_{t}
          + \theta_{2} \, \mathrm{SPR}_{t}
          + \theta_{3} \, \mathrm{EURIBOR}_{t}
          + \theta_{4} \, \mathrm{HIBOR}_{t}
          + \theta_{5} \, \mathrm{TREA}_{t}
          + e_{t},
      \end{split}
      \end{equation*}
      \begin{equation*}
      \begin{split}
      R^{All}_{\mathrm{BSC},t}
        &= \alpha_{0}
          + \alpha_{1}R^{CEX}_{\mathrm{BSC},t}
          + (\beta_{00}
             + \beta_{01}\,\mathrm{SR}_{\mathrm{BSC},t-1}
             + \beta_{02}\,R_{\$\mathrm{BNB},t-1})\,
             R^{All}_{\mathrm{BSC},t-1}
          + (\beta_{10}
             + \beta_{11}\,\mathrm{SR}_{\mathrm{Ethereum},t}\\
          &  + \beta_{12}\,R_{\$\mathrm{ETH},t})\,
             R^{All}_{\mathrm{Ethereum},t} 
          + (\beta_{20}
             + \beta_{21}\,\mathrm{SR}_{\mathrm{Solana},t}
             + \beta_{22}\,R_{\$\mathrm{SOL},t})\,
             R^{All}_{\mathrm{Solana},t} 
          + (\beta_{30}
             + \beta_{31}\,\mathrm{SR}_{\mathrm{Arbitrum},t}\\
           & + \beta_{32}\,R_{\$\mathrm{ARB},t})\,
             R^{All}_{\mathrm{Arbitrum},t}
          + (\beta_{40}
             + \beta_{41}\,\mathrm{SR}_{\mathrm{Avalanche},t}
             + \beta_{42}\,R_{\$\mathrm{AVAX},t})\,
             R^{All}_{\mathrm{Avalanche},t} \\
          & + \theta_{0} \, \mathrm{FTSER}_{t}
          + \theta_{1} \, \mathrm{HSR}_{t}
          + \theta_{2} \, \mathrm{SPR}_{t}
          + \theta_{3} \, \mathrm{EURIBOR}_{t}
          + \theta_{4} \, \mathrm{HIBOR}_{t}
          + \theta_{5} \, \mathrm{TREA}_{t}
          + e_{t},
      \end{split}
      \end{equation*}
      \begin{equation*}
      \begin{split}
      R^{All}_{\mathrm{Arbitrum},t}
        &= \alpha_{0}
          + \alpha_{1}R^{CEX}_{\mathrm{Arbitrum},t}
          + (\beta_{00}
             + \beta_{01}\,\mathrm{SR}_{\mathrm{Arbitrum},t-1}
             + \beta_{02}\,R_{\$\mathrm{ARB},t-1})\,
             R^{All}_{\mathrm{Arbitrum},t-1}
          + (\beta_{10} \\
          &  + \beta_{11}\,\mathrm{SR}_{\mathrm{Ethereum},t}
             + \beta_{12}\,R_{\$\mathrm{ETH},t})\,
             R^{All}_{\mathrm{Ethereum},t} 
          + (\beta_{20}
             + \beta_{21}\,\mathrm{SR}_{\mathrm{Solana},t}
             + \beta_{22}\,R_{\$\mathrm{SOL},t})\,
             R^{All}_{\mathrm{Solana},t} +\\ 
             & (\beta_{30}
             + \beta_{31}\,\mathrm{SR}_{\mathrm{BSC},t}
             + \beta_{32}\,R_{\$\mathrm{BSC},t})\,
             R^{All}_{\mathrm{BSC},t}
          + (\beta_{40}
             + \beta_{41}\,\mathrm{SR}_{\mathrm{Avalanche},t}
             + \beta_{42}\,R_{\$\mathrm{AVAX},t})\,
             R^{All}_{\mathrm{Avalanche},t}\\
        & + \theta_{0} \, \mathrm{FTSER}_{t}
          + \theta_{1} \, \mathrm{HSR}_{t}
          + \theta_{2} \, \mathrm{SPR}_{t}
          + \theta_{3} \, \mathrm{EURIBOR}_{t}
          + \theta_{4} \, \mathrm{HIBOR}_{t}
          + \theta_{5} \, \mathrm{TREA}_{t}
          + e_{t},
      \end{split}
      \end{equation*}
      \begin{equation*}
      \begin{split}
      R^{All}_{\mathrm{Avalanche},t}
        &= \alpha_{0}
          + \alpha_{1}R^{CEX}_{\mathrm{Avalanche},t}
          + (\beta_{00}
             + \beta_{01}\,\mathrm{SR}_{\mathrm{Avalanche},t-1}
             + \beta_{02}\,R_{\$\mathrm{AVAX},t-1})\,
             R^{All}_{\mathrm{Avalanche},t-1}
          + (\beta_{10} \\
          &  + \beta_{11}\,\mathrm{SR}_{\mathrm{Ethereum},t}
             + \beta_{12}\,R_{\$\mathrm{ETH},t})\,
             R^{All}_{\mathrm{Ethereum},t} 
          + (\beta_{20}
             + \beta_{21}\,\mathrm{SR}_{\mathrm{Solana},t}
             + \beta_{22}\,R_{\$\mathrm{SOL},t})\,
             R^{All}_{\mathrm{Solana},t} +\\ 
             & (\beta_{30}
             + \beta_{31}\,\mathrm{SR}_{\mathrm{BSC},t}
             + \beta_{32}\,R_{\$\mathrm{BSC},t})\,
             R^{All}_{\mathrm{BSC},t}
          + (\beta_{40}
             + \beta_{41}\,\mathrm{SR}_{\mathrm{Arbitrum},t}
             + \beta_{42}\,R_{\$\mathrm{ARB},t})\,
             R^{All}_{\mathrm{Arbitrum},t}\\ 
        & + \theta_{0} \, \mathrm{FTSER}_{t}
          + \theta_{1} \, \mathrm{HSR}_{t}
          + \theta_{2} \, \mathrm{SPR}_{t}
          + \theta_{3} \, \mathrm{EURIBOR}_{t}
          + \theta_{4} \, \mathrm{HIBOR}_{t}
          + \theta_{5} \, \mathrm{TREA}_{t}
          + e_{t},
      \end{split}
      \end{equation*}
      while the series of residuals for all the modeled in the three panels are modeled by Glosten-Jagannathan-Runkle (GJR) asymmetric GARCH approach:
    \begin{equation}
    \begin{split}
        e_{t} \sim \mathcal{N}(0, \sigma_t^2), \;
        \sigma_t^2 = \omega + \sum_{i=1}^p \alpha_i e_{t-i}^2 + \sum_{j=1}^o \gamma_j e_{t-j}^2 \cdot \mathbb{I}_{\{e_{t-j} < 0\}} + \sum_{k=1}^q \beta_k \sigma_{t-k}^2,
    \end{split}
    \end{equation}
    and the optimal combination of $p$, $o$, and $q$ for each model is reported in this table. \\
    Panel B reports the estimated parameters of the models for all the five chains based on the specification with $R^{\mathrm{non-CEX}}_{\mathrm{chain}_{i},t}$ as the dependent variable and $R^{\mathrm{non-CEX}}_{\mathrm{chain}_{i},t-1}$ as the lagged return.\\
    Panel C reports the estimated parameters of the models for all the five chains based on the specification with $R^{\mathrm{Local}}_{\mathrm{chain}_{i},t}$ as the dependent variable and $R^{\mathrm{Local}}_{\mathrm{chain}_{i},t-1}$ as the lagged return. \\
    Results in this table are based on unified sampling period from 17th Mar 2023 to 31st Mar 2025.\\
    t-statistics are in parentheses. \\
      ***/**/* denote significance at 1\%/5\%/10\%.
  \end{TableNotes}
  %──────────────────────────────────────────────────────────────────────────────

  % Force the table to span the full text width
  \setlength\LTleft{0pt}
  \setlength\LTright{0pt}

  \begin{tabularx}{\textwidth}{@{}l*{5}{Y}@{}}
    \caption{Estimates of the non-linear models with global market variables.\label{tab:nonlinear_macro}}\\
    % insert notes at top
    \insertTableNotes\\
    \addlinespace
    \toprule
      Model: 
        & (1) & (2) & (3) & (4) & (5) \\
      Chain portfolios: 
        & $\mathrm{Ethereum}$ & $\mathrm{Solana}$ & $\mathrm{BSC}$ & $\mathrm{Arbitrum}$ & $\mathrm{Avalanche}$ \\
    \midrule
    \endfirsthead

    \caption[]{Estimates of the non-linear models with global market variables. (continued)}\\
    \toprule
      Model: 
        & (1) & (2) & (3) & (4) & (5) \\
      Chain portfolios: 
        & $\mathrm{Ethereum}$ & $\mathrm{Solana}$ & $\mathrm{BSC}$ & $\mathrm{Arbitrum}$ & $\mathrm{Avalanche}$ \\
    \midrule
    \endhead

    \midrule
    \multicolumn{6}{r}{\emph{Continued on next page}}
    \endfoot

    \bottomrule
    \endlastfoot

    %--------------- Panel A ----------------
    \multicolumn{6}{l}{Panel A: $R^{All}_{\mathrm{chain}_{i},t}$ as the dependent variable}\\
    \addlinespace

    $\alpha_0$ 
      & -0.001  & 0.0006* & 0.0002  & 0.0002  & 0.0001 \\
      & (-1.496) & (1.953) & (0.413) & (1.237)  & (1.094)  \\

    $\alpha_1$
      & 0.531***  &  0.987*** & 0.978***  & 0.416***  & 0.867*** \\
      & (12.530)  & (70.659) & (1102.918) & (24.589)  & (64.574)  \\

    $\beta_{00}$
      & -0.026   & 0.002  & 0.0006    & 0.0005  & 0.0005 \\
      & (-1.386) & (0.648) & (0.249)  & (0.024)  & (0.044)  \\

    $\beta_{01}$
      &  -0.026    & -0.0004  & -0.018   & -0.003   & 0.050 \\
      & (-0.434)  & (-0.037) & (-0.933) & (-0.074) & (0.574)  \\

    $\beta_{02}$
      & 0.160     & 0.038   & 1.100** & 0.350**    & -0.191 \\
      & (0.226)   & (0.334) & (2.056)  & (2.163) & (-0.781)  \\

    $\beta_{10}$
      & -0.012    & -0.0009  & 0.005    & 0.003  & -0.003* \\
      & (-0.320)  & (-0.819) & (0.323)  & (1.138)  & (-1.791)  \\

    $\beta_{11}$
      & -0.008     & -0.001   & -0.063**   & -0.007   & -0.006 \\
      & (-0.141)  & (-0.325)  & (-2.440) & (-0.512)  & (-0.862)  \\

    $\beta_{12}$
      & -0.492    & -0.031  & -0.928   & -0.658***    & 0.052 \\
      & (-1.309) & (-0.301) & (-1.593) & (-2.848)  & (0.419)  \\

    $\beta_{20}$
      & 0.010***  & -0.0004   & -0.0004  & 0.023***    & -0.006 \\
      & (3.219)   & (-0.245) & (-0.067) & (4.040)  & (-0.672)  \\

    $\beta_{21}$
      & -0.012    & -0.014   & -0.032   & -0.009   & 0.029 \\
      & (-0.659)  & (-1.585) & (-1.350)  & (-0.377) & (0.690)  \\

    $\beta_{22}$
      &  3.960*** & 0.471    & -0.007   & -0.183*** & 0.015 \\
      & (4.060)   & (0.999) & (-0.116) & (-2.596) & (0.158)  \\

    $\beta_{30}$
      & 0.123**   & 0.021   & -0.126*** & -0.001   & 0.003 \\
      & (2.321)   & (1.173) & (-3.154) & (-1.011) & (0.898)  \\

    $\beta_{31}$
      & 0.032    & -0.067   &  -0.277**   & 0.0003    & 0.015 \\
      & (0.286)  & (-1.564) & (-2.012)  & (0.679)  & (1.202)  \\

    $\beta_{32}$
      & 0.230    & 0.005   & -0.352   & -0.595*   & 0.033 \\
      & (0.597)  & (0.121) & (-1.441) & (-1.872) & (0.129)  \\

    $\beta_{40}$
      & -0.010   & 0.025*** & 0.056***  & -0.017  & -0.070*** \\
      & (-0.354) & (2.859)  & (2.579)  & (-1.382)  & (-4.843)  \\

    $\beta_{41}$
      & 0.074    & 0.017   & -0.074   & 0.022    & 0.104* \\
      & (1.214)  & (1.071) & (-1.074)  & (0.792) & (1.768)  \\

    $\beta_{42}$
      & 1.022*    & 0.089   & -0.414   & 0.128 & -0.055 \\
      & (1.693)  & (0.483) & (-1.436) & (0.461) & (-0.619)  \\

    $\theta_0$
      & 0.135*  & 0.044** &  -0.099*  & -0.052  & -0.058** \\
      & (1.652) & (2.043) & (-1.900) & (-1.269)  & (-2.370)  \\

    $\theta_1$
      & 0.018   & -0.025   & 0.053   & -0.009  & 0.027** \\
      & (0.359) & (-0.942) & (1.189) & (-0.620)  & (2.137)  \\

    $\theta_2$
      & -0.130*    & -0.004  & 0.219*   & 0.247***  & -0.023 \\
      & (-1.843) & (-0.246) & (1.682) & (1.839)  & (-1.042)  \\

    $\theta_3$
      & 0.053***   & -0.012* & 0.027   & 0.006    & -0.008 \\
      & (2.678) & (-1.655) & (0.926) & (0.898)  & (-1.521)  \\

    $\theta_4$
      &  -0.030*** & -0.0007  & 0.0008   & -0.002  & 0.001 \\
      & (-3.229) & (-1.271) & (0.258) & (-1.195)  & (1.189)  \\

    $\theta_5$
      & -0.009   & -0.0003 & -0.004  & 0.0009  & -0.0007 \\
      & (-1.389) & (-0.665) & (-1.247) & (0.433)  & (-0.718)  \\

    \addlinespace

    p   & 2 & 2 & 1 & 1 & 2 \\
    o   & 2 & 0 & 0 & 0 & 0 \\
    q   & 3 & 3 & 1 & 1 & 1  \\
    \addlinespace

    Adjusted \(R^2\) & 0.025 & 0.983 & 0.997 & 0.728 & 0.893 \\
  
  \midrule
  %--------------- Panel B ----------------
    \multicolumn{6}{l}{Panel B: $R^{\mathrm{non-CEX}}_{\mathrm{chain}_{i},t}$ as the dependent variable}\\
    \addlinespace
    $\alpha_0$ 
      & 0.0009   & -0.002***   & -0.0002  & 0.0004  & 0.0009* \\
      & (0.659) & (-2.803) & (-0.233) & (2.772)    & (1.707)  \\

    $\alpha_1$
      & 0.121    & 0.006  & -0.00003  & -0.014   & 0.402*** \\
      & (1.099)  & (0.436) & (-0.144) & (-1.177)  & (7.111)  \\

    $\beta_{00}$
      & -0.001     & 0.001   & 0.0003***   & 0.010    & 0.002\\
      & (-0.180)   & (0.059) & (2.856)  & (0.358)  & (0.027)  \\

    $\beta_{01}$
      & 0.075    & -0.049   & 0.177   & -0.105   & 0.075 \\
      & (1.308)  & (-0.264) & (1.078) & (-1.090) & (0.053)  \\

    $\beta_{02}$
      & 0.215    & 0.986   & 0.180    & 0.897***  & -0.490 \\
      & (0.577)   & (1.036) & (0.107) & (2.650)   & (-0.466)  \\

    $\beta_{10}$
      & 0.077    & -0.003   & 0.011    & 0.006    & -0.008 \\
      & (0.562)  & (-0.135) & (0.643)  & (1.533)  & (-0.590)  \\

    $\beta_{11}$
      & -0.436   & -0.147     & -0.079**  & -0.019  & -0.056 \\
      & (-0.630)   & (-0.535) & (-2.094) & (-0.977)  & (-0.633)  \\

    $\beta_{12}$
      & -1.430    & -0.442   & -0.073   & -0.995*** & 0.498 \\
      & (-0.443)  & (-0.504) & (-0.040) & (-3.330)  & (0.724)  \\

    $\beta_{20}$
      & -0.0007    & -0.0003   & -0.014   & 0.036*** & -0.016\\
      & (-0.050)  & (-0.039) & (-0.850) & (4.147)  & (-0.408)  \\

    $\beta_{21}$
      & 0.011    & 0.055 & -0.032    & -0.016      & 0.188 \\
      & (0.118)  & (1.043)  & (-0.795)  & (-0.530) & (0.393)  \\

    $\beta_{22}$
      & -0.084     & -3.325  & 0.145   & -0.292***   & 0.127 \\
      & (-0.036)   & (-1.612) & (0.781) & (-2.764) & (0.287)  \\

    $\beta_{30}$
      & -0.047     & 0.266   & -0.254***   & -0.002   & 0.008 \\
      & (-0.472)   & (1.180) & (-2.655)  & (-1.293) & (0.656)  \\

    $\beta_{31}$
      & 1.374    & -0.381    & 0.082     & 0.014*    & 0.007 \\
      & (1.341)  & (-0.442)  & (0.686)  & (1.932)  & (0.129)  \\

    $\beta_{32}$
      & -0.029   & -1.129   & -0.783   & -1.252**  & 0.797 \\
      & (-0.072) & (-0.540) & (-1.118) & (-2.531) & (0.551)  \\

    $\beta_{40}$
      & 0.055   & 0.069    & 0.025    & -0.036*  & -0.311*** \\
      & (0.519) & (0.597)  & (0.777)  & (-1.900)  & (-3.646)  \\

    $\beta_{41}$
      & -0.624    & -0.131   & -0.098    & 0.046  &  0.305 \\
      & (-1.209)  & (-0.773) & (-1.348)  & (1.236) & (0.303)  \\

    $\beta_{42}$
      & -2.640    & -0.133   & -0.489   & 0.182   & -0.594 \\
      & (-1.217)  & (-0.101) & (-0.721) & (0.384) & (-0.405)  \\

    $\theta_0$
      & -0.116   & -0.060   & -0.260   & -0.098    & -0.101 \\
      & (-0.655) & (-0.274) & (-1.318) & (-1.522)  & (-0.786)  \\

    $\theta_1$
      & 0.075   & -0.014   &  -0.133*   & -0.002  & -0.015 \\
      & (0.562) & (-0.100) & (-1.887) & (-0.089)  & (-0.148)  \\

    $\theta_2$
      & -0.790***& 0.231   & 0.456***   & 0.415*** &  -0.174 \\
      & (-3.617) & (1.163) & (2.797) & (8.909)     & (-1.523)  \\

    $\theta_3$
      & -0.014   & -0.078*   & -0.014   & 0.002   & -0.024 \\
      & (-0.281) & (-1.669) & (-0.481) & (0.196)  & (-0.352)  \\

    $\theta_4$
      &  -0.013 & 0.007    & 0.001   & -0.003    & 0.002 \\
      & (-2.165) & (0.611) & (0.227)  & (-0.973)  & (0.107)  \\

    $\theta_5$
      & -0.005   & -0.138**   & 0.00008  & 0.00001    & -0.007 \\
      & (-0.886) & (-2.053) & (0.011) & (0.004)  & (-0.408)  \\

    \addlinespace

    p   & 2 & 2 & 1 & 1 & 2 \\
    o   & 0 & 0 & 1 & 0 & 0 \\
    q   & 3 & 3 & 1 & 1 & 2  \\
    \addlinespace

    Adjusted \(R^2\) & -0.016 & -0.032 & 0.023 & 0.173 & 0.204 \\

  \midrule
  %--------------- Panel C ----------------
    \multicolumn{6}{l}{Panel C: $R^{\mathrm{Local}}_{\mathrm{chain}_{i},t}$ as the dependent variable}\\
    \addlinespace
    $\alpha_0$ 
      & -0.001   & 0.0001  & 0.002***  & 0.0003**  & -0.0002 \\
      & (-1.047) & (1.364)    & (3.373)   & (1.111)   & (-0.401)  \\

    $\alpha_1$
      & 0.670***  & 1.001***  & 1.054***   & 0.047***   & 0.723*** \\
      & (8.702)  & (526.856) & (1799.085) & (3.199)  & (11.955)  \\

    $\beta_{00}$
      & -0.022     & -0.002   & 0.002   & -0.024    & -0.064 \\
      & (-0.748)   & (-1.276) & (0.872)  & (-0.801)  & (-1.388)  \\

    $\beta_{01}$
      & -0.047    & 0.008   & 0.0007   & -0.139   & 0.133 \\
      & (-0.534)  & (0.699) & (0.089) & (-1.420) & (0.879)  \\

    $\beta_{02}$
      & -0.307    & 0.003    & 0.427   & 0.286  & 0.085 \\
      & (-0.304)   & (0.009) & (1.614)  & (1.204)  & (0.096)  \\

    $\beta_{10}$
      & -0.021    & 0.002***  & 0.096*    & 0.007* & -0.0007 \\
      & (-0.395)  & (2.704) & (1.919)  & (1.769)  & (-0.084)  \\

    $\beta_{11}$
      & -0.034   & 0.004   & -0.066   & -0.020   & -0.040 \\
      & (-0.426) & (0.662) & (-1.338) & (-0.987)  & (-0.854)  \\

    $\beta_{12}$
      & -0.432   & -0.042   & -3.106   & -0.958***   & 0.408 \\
      & (-0.793)  & (-0.624) & (-1.583) & (-3.013)  & (0.690)  \\

    $\beta_{20}$
      & 0.009   & -0.005*   & -0.032***   & 0.035***  & -0.015 \\
      & (0.524)   & (-1.799) & (-3.386) & (3.901)  & (-0.632)  \\

    $\beta_{21}$
      & -0.009    & -0.010    & -0.012    & -0.008   & 0.108 \\
      & (-0.110)  & (-0.906)  & (-0.390)  & (-0.281) & (0.737)  \\

    $\beta_{22}$
      & 3.871*  & 0.018     & 0.170   & -0.299***   & 0.190 \\
      & (1.915)   & (0.056) & (1.228) & (-2.764) & (0.769)  \\

    $\beta_{30}$
      & 0.025     & 0.079***  & -0.239***  & -0.002*   & 0.007 \\
      & (0.309)   & (6.000)   & (-4.710)  & (-1.625) & (1.521)  \\

    $\beta_{31}$
      & 0.042    & -0.145***  & -0.091   & 0.013    & 0.069*** \\
      & (0.196)  & (-3.681)   & (-0.923)  & (1.475)  & (3.268)  \\

    $\beta_{32}$
      & 0.425   & -0.153   & -0.381**  & -1.320**   & -1.272* \\
      & (0.943) & (-1.126) & (-2.482) & (-2.161) & (-1.751)  \\

    $\beta_{40}$
      & 0.052   & 0.040***    & 0.047** & -0.040**  & -0.280*** \\
      & (1.236) & (4.493)  & (2.058)  & (-2.110)  & (-4.121)  \\

    $\beta_{41}$
      & 0.041    & -0.005   & -0.070    &  0.045  &  0.116 \\
      & (0.450)  & (-0.199) & (-1.179)  & (1.351) & (0.502)  \\

    $\beta_{42}$
      & 1.020    & 0.177   & -0.455   & 0.198 & -1.368** \\
      & (1.204)  & (1.164) & (-0.835) & (0.426) & (-2.019)  \\

    $\theta_0$
      & 0.213   & -0.016   & -0.089   & -0.084    & -0.149 \\
      & (1.433) & (-0.430) & (-0.868) & (-1.345)  & (-1.338)  \\

    $\theta_1$
      & -0.016   & 0.002   &  0.022   & -0.0009 & 0.041 \\
      & (-0.226) & (0.055) & (0.752) & (-0.042)  & (0.716)  \\

    $\theta_2$
      & -0.221*  & -0.035   & 0.165   & 0.416*** &   -0.096 \\
      & (-1.955) & (-1.501) & (1.179) & (7.303)  & (-1.125)  \\

    $\theta_3$
      & 0.081***  & -0.009**   & 0.023   & 0.002    & -0.042* \\
      & (2.760)  & (-2.191)    & ( 0.592) & (0.165)    & (-1.913)  \\

    $\theta_4$
      & -0.036*** & -0.0008  & 0.042***    & -0.001    & -0.002 \\
      & (-2.667) & (-1.024)   & (3.238)  & (-0.407)  & (-0.254)  \\

    $\theta_5$
      & -0.009   & -0.002***  & 0.0002   & 0.001 & -0.014* \\
      & (-1.210) & (-2.767)  & (0.039) & (0.370)  & (-1.707)  \\

    \addlinespace

    p   & 2 & 3 & 2 & 1 & 1 \\
    o   & 2 & 0 & 2 & 0 & 0 \\
    q   & 3 & 3 & 2 & 1 & 1  \\
    \addlinespace

    Adjusted \(R^2\) & 0.041 & 0.976 & 0.995 & 0.179 & 0.380 \\

  \end{tabularx}

\end{ThreePartTable}
\end{singlespace}

% -------------------------------- table for non-limear models with extreme returns
\newpage

\begin{singlespace}
\begin{ThreePartTable}
  %──────────────────────────────────────────────────────────────────────────────
  % Notes, flush‐left and without indent
  \begin{TableNotes}[para,flushleft]
    \scriptsize
    Notes: This table shows estimates of the non-linear models with extreme return dummies, i.e., as illustrated by Equation (\ref{eq_nonlinear_extremeR_macro}), for the five chain portfolios in our study.  Panel A reports the estimated parameters of the models for all the five chains based on Equation Equation (\ref{eq_nonlinear_extremeR_macro}), which are as following:
      \begin{equation*}
      \begin{split}
      &R^{All}_{\mathrm{Ethereum},t}
        = \alpha_{0}
          + \alpha_{1}R^{CEX}_{\mathrm{Ethereum},t}
          + (\beta_{00}
             + \beta_{01}\,\mathrm{SR}_{\mathrm{Ethereum},t-1}
             + \beta_{02}\,R_{\$\mathrm{ETH},t-1})\,
             R^{All}_{\mathrm{Ethereum},t-1}
          + (\beta_{10}\\
           & + \beta_{11}\,\mathrm{SR}_{\mathrm{Solana},t}
             + \beta_{12}\,R_{\$\mathrm{SOL},t}
             + \beta_{13}\,D^{U}_{Solana,t}
             + \beta_{14}\,D^{L}_{Solana,t}
             )\,
             R^{All}_{\mathrm{Solana},t} 
          + (\beta_{20}
             + \beta_{21}\,\mathrm{SR}_{\mathrm{BSC},t}
             + \beta_{22}\,R_{\$\mathrm{BNB},t}\\
             & + \beta_{23}\,D^{U}_{BSC,t}
             + \beta_{24}\,D^{L}_{BSC,t}
             )\,       
             R^{All}_{\mathrm{BSC},t} 
          + (\beta_{30}
             + \beta_{31}\,\mathrm{SR}_{\mathrm{Arbitrum},t}
             + \beta_{32}\,R_{\$\mathrm{ARB},t}
             + \beta_{33}\,D^{U}_{Arbitrum,t}
             + \beta_{34}\,D^{L}_{Arbitrum,t}
             )\,\\
             & R^{All}_{\mathrm{Arbitrum},t}
          + (\beta_{40}
             + \beta_{41}\,\mathrm{SR}_{\mathrm{Avalanche},t}
             + \beta_{42}\,R_{\$\mathrm{AVAX},t}
             + \beta_{43}\,D^{U}_{Avalanche,t}
             + \beta_{44}\,D^{L}_{Avalanche,t}
             )\,
             R^{All}_{\mathrm{Avalanche},t}\\
          & + \theta_{0} \, \mathrm{FTSER}_{t}
          + \theta_{1} \, \mathrm{HSR}_{t}
          + \theta_{2} \, \mathrm{SPR}_{t}
          + \theta_{3} \, \mathrm{EURIBOR}_{t}
          + \theta_{4} \, \mathrm{HIBOR}_{t}
          + \theta_{5} \, \mathrm{TREA}_{t}
          + e_{t},
          \end{split}
      \end{equation*}
      \begin{equation*}
      \begin{split}
      &R^{All}_{\mathrm{Solana},t}
        = \alpha_{0}
          + \alpha_{1}R^{CEX}_{\mathrm{Solana},t}
          + (\beta_{00}
             + \beta_{01}\,\mathrm{SR}_{\mathrm{Solana},t-1}
             + \beta_{02}\,R_{\$\mathrm{SOL},t-1})\,
             R^{All}_{\mathrm{Solana},t-1}
          + (\beta_{10}
             + \beta_{11}\,\mathrm{SR}_{\mathrm{Ethereum},t}\\
             & + \beta_{12}\,R_{\$\mathrm{ETH},t}
             + \beta_{13}\,D^{U}_{Ethereum,t}
             + \beta_{14}\,D^{L}_{Ethereum,t}
             )\,
             R^{All}_{\mathrm{Ethereum},t} 
          + (\beta_{20}
             + \beta_{21}\,\mathrm{SR}_{\mathrm{BSC},t}
             + \beta_{22}\,R_{\$\mathrm{BNB},t}\\
             & + \beta_{23}\,D^{U}_{BSC,t}
             + \beta_{24}\,D^{L}_{BSC,t}
             )\,       
             R^{All}_{\mathrm{BSC},t} 
          + (\beta_{30}
             + \beta_{31}\,\mathrm{SR}_{\mathrm{Arbitrum},t}
             + \beta_{32}\,R_{\$\mathrm{ARB},t}
             + \beta_{33}\,D^{U}_{Arbitrum,t}\\
             & + \beta_{34}\,D^{L}_{Arbitrum,t}
             )\,
             R^{All}_{\mathrm{Arbitrum},t}
             + (\beta_{40}
             + \beta_{41}\,\mathrm{SR}_{\mathrm{Avalanche},t}
             + \beta_{42}\,R_{\$\mathrm{AVAX},t}
             + \beta_{43}\,D^{U}_{Avalanche,t}\\
             & + \beta_{44}\,D^{L}_{Avalanche,t}
             )\,
             R^{All}_{\mathrm{Avalanche},t} 
          + \theta_{0} \, \mathrm{FTSER}_{t}
          + \theta_{1} \, \mathrm{HSR}_{t}
          + \theta_{2} \, \mathrm{SPR}_{t}
          + \theta_{3} \, \mathrm{EURIBOR}_{t}
          + \theta_{4} \, \mathrm{HIBOR}_{t}
          + \theta_{5} \, \mathrm{TREA}_{t}\\
          & + e_{t},
      \end{split}
      \end{equation*}
      \begin{equation*}
      \begin{split}
      & R^{All}_{\mathrm{BSC},t}
         = \alpha_{0}
          + \alpha_{1}R^{CEX}_{\mathrm{BSC},t}
          + (\beta_{00}
             + \beta_{01}\,\mathrm{SR}_{\mathrm{BSC},t-1}
             + \beta_{02}\,R_{\$\mathrm{BSC},t-1})\,
             R^{All}_{\mathrm{BSC},t-1}
          + (\beta_{10}
             + \beta_{11}\,\mathrm{SR}_{\mathrm{Ethereum},t}\\
          &  + \beta_{12}\,R_{\$\mathrm{ETH},t}
             + \beta_{13}\,D^{U}_{Ethereum,t}
             + \beta_{14}\,D^{L}_{Ethereum,t}
             )\,
             R^{All}_{\mathrm{Ethereum},t} 
          + (\beta_{20}
             + \beta_{21}\,\mathrm{SR}_{\mathrm{Solana},t}
             + \beta_{22}\,R_{\$\mathrm{SOL},t}\\
          &  + \beta_{23}\,D^{U}_{Solana,t}
             + \beta_{24}\,D^{L}_{Solana,t}
             )\,
             R^{All}_{\mathrm{Solana},t} 
          + (\beta_{30}
             + \beta_{31}\,\mathrm{SR}_{\mathrm{Arbitrum},t}
             + \beta_{32}\,R_{\$\mathrm{ARB},t}
             + \beta_{33}\,D^{U}_{Arbitrum,t}\\
          &  + \beta_{34}\,D^{L}_{Arbitrum,t}
             )\,
             R^{All}_{\mathrm{Arbitrum},t}
          + (\beta_{40}
             + \beta_{41}\,\mathrm{SR}_{\mathrm{Avalanche},t}
             + \beta_{42}\,R_{\$\mathrm{AVAX},t}
             + \beta_{43}\,D^{U}_{Avalanche,t}\\
          &  + \beta_{44}\,D^{L}_{Avalanche,t}
             )\,
             R^{All}_{\mathrm{Avalanche},t} 
          + \theta_{0} \, \mathrm{FTSER}_{t}
          + \theta_{1} \, \mathrm{HSR}_{t}
          + \theta_{2} \, \mathrm{SPR}_{t}
          + \theta_{3} \, \mathrm{EURIBOR}_{t}
          + \theta_{4} \, \mathrm{HIBOR}_{t}
          + \theta_{5} \, \mathrm{TREA}_{t}\\
          & + e_{t},
      \end{split}
      \end{equation*}
      \begin{equation*}
      \begin{split}
      & R^{All}_{\mathrm{Arbitrum},t}
         = \alpha_{0}
          + \alpha_{1}R^{CEX}_{\mathrm{Arbitrum},t}
          + (\beta_{00}
             + \beta_{01}\,\mathrm{SR}_{\mathrm{Arbitrum},t-1}
             + \beta_{02}\,R_{\$\mathrm{ARB},t-1})\,
             R^{All}_{\mathrm{Arbitrum},t-1}
           + (\beta_{10}\\
           & + \beta_{11}\,\mathrm{SR}_{\mathrm{Ethereum},t}
             + \beta_{12}\,R_{\$\mathrm{ETH},t}
             + \beta_{13}\,D^{U}_{Ethereum,t}
             + \beta_{14}\,D^{L}_{Ethereum,t}
             )\,
             R^{All}_{\mathrm{Ethereum},t}
          + (\beta_{20}
             + \beta_{21}\,\mathrm{SR}_{\mathrm{Solana},t}\\
             & + \beta_{22}\,R_{\$\mathrm{SOL},t}
             + \beta_{23}\,D^{U}_{Solana,t}
             + \beta_{24}\,D^{L}_{Solana,t}
             )\,
             R^{All}_{\mathrm{Solana},t} 
          + (\beta_{30}
             + \beta_{31}\,\mathrm{SR}_{\mathrm{BSC},t}
             + \beta_{32}\,R_{\$\mathrm{BNB},t}
             + \beta_{33}\,D^{U}_{BSC,t}\\
             & + \beta_{34}\,D^{L}_{BSC,t}
             )\,       
             R^{All}_{\mathrm{BSC},t} 
          + (\beta_{40}
             + \beta_{41}\,\mathrm{SR}_{\mathrm{Avalanche},t}
             + \beta_{42}\,R_{\$\mathrm{AVAX},t}
             + \beta_{43}\,D^{U}_{Avalanche,t}\\
             & + \beta_{44}\,D^{L}_{Avalanche,t}
             )\,
             R^{All}_{\mathrm{Avalanche},t}
          + \theta_{0} \, \mathrm{FTSER}_{t}
          + \theta_{1} \, \mathrm{HSR}_{t}
          + \theta_{2} \, \mathrm{SPR}_{t}
          + \theta_{3} \, \mathrm{EURIBOR}_{t}
          + \theta_{4} \, \mathrm{HIBOR}_{t}
          + \theta_{5} \, \mathrm{TREA}_{t}\\
          & + e_{t},
      \end{split}
      \end{equation*}
      \begin{equation*}
      \begin{split}
      &R^{All}_{\mathrm{Avalanche},t}
         = \alpha_{0}
          + \alpha_{1}R^{CEX}_{\mathrm{Avalanche},t}
          + (\beta_{00}
             + \beta_{01}\,\mathrm{SR}_{\mathrm{Avalanche},t-1}
             + \beta_{02}\,R_{\$\mathrm{AVAX},t-1})\,
             R^{All}_{\mathrm{Avalanche},t-1}
           + (\beta_{10}\\
           & + \beta_{11}\,\mathrm{SR}_{\mathrm{Ethereum},t}
             + \beta_{12}\,R_{\$\mathrm{ETH},t}
             + \beta_{13}\,D^{U}_{Ethereum,t}
             + \beta_{14}\,D^{L}_{Ethereum,t}
             )\,
             R^{All}_{\mathrm{Ethereum},t}
          + (\beta_{20}
             + \beta_{21}\,\mathrm{SR}_{\mathrm{Solana},t}\\
             & + \beta_{22}\,R_{\$\mathrm{SOL},t}
             + \beta_{23}\,D^{U}_{Solana,t}
             + \beta_{24}\,D^{L}_{Solana,t}
             )\,
             R^{All}_{\mathrm{Solana},t} 
          + (\beta_{30}
             + \beta_{31}\,\mathrm{SR}_{\mathrm{BSC},t}
             + \beta_{32}\,R_{\$\mathrm{BNB},t}
             + \beta_{33}\,D^{U}_{BSC,t}\\
             & + \beta_{34}\,D^{L}_{BSC,t}
             )\,       
             R^{All}_{\mathrm{BSC},t} 
          + (\beta_{40}
             + \beta_{41}\,\mathrm{SR}_{\mathrm{Arbitrum},t}
             + \beta_{42}\,R_{\$\mathrm{ARB},t}
             + \beta_{43}\,D^{U}_{Arbitrum,t}
             + \beta_{44}\,D^{L}_{Arbitrum,t}
             )\,
             R^{All}_{\mathrm{Arbitrum},t} \\
          & + \theta_{0} \, \mathrm{FTSER}_{t}
          + \theta_{1} \, \mathrm{HSR}_{t}
          + \theta_{2} \, \mathrm{SPR}_{t}
          + \theta_{3} \, \mathrm{EURIBOR}_{t}
          + \theta_{4} \, \mathrm{HIBOR}_{t}
          + \theta_{5} \, \mathrm{TREA}_{t}
          + e_{t},
      \end{split}
      \end{equation*}
      while the series of residuals for all the modeled in the three panels are modeled by Glosten-Jagannathan-Runkle (GJR) asymmetric GARCH approach in Equation (\ref{garch_model})
    and the optimal combination of $p$, $o$, and $q$ for each model is reported in this table. 
    Panel B reports the estimated parameters of the models for all the five chains based on the specification with $R^{\mathrm{non-CEX}}_{\mathrm{chain}_{i},t}$ as the dependent variable and $R^{\mathrm{non-CEX}}_{\mathrm{chain}_{i},t-1}$ as the lagged return.
    Panel C reports the estimated parameters of the models for all the five chains based on the specification with $R^{\mathrm{Local}}_{\mathrm{chain}_{i},t}$ as the dependent variable and $R^{\mathrm{Local}}_{\mathrm{chain}_{i},t-1}$ as the lagged return. 
    Results in this table are based on unified sampling period from 17th Mar 2023 to 31st Mar 2025.
    t-statistics are in parentheses. 
      ***/**/* denote significance at 1\%/5\%/10\%.
  \end{TableNotes}
  %──────────────────────────────────────────────────────────────────────────────

  % Force the table to span the full text width
  \setlength\LTleft{0pt}
  \setlength\LTright{0pt}

  \begin{tabularx}{\textwidth}{@{}l*{5}{Y}@{}}
    \caption{Estimates of the non-linear models with global market and extreme return dummies.\label{tab:nonlinear_macro_extreme}}\\
    % insert notes at top
    \insertTableNotes\\
    \addlinespace
    \toprule
      Model: 
        & (1) & (2) & (3) & (4) & (5) \\
      Chain portfolios: 
        & $\mathrm{Ethereum}$ & $\mathrm{Solana}$ & $\mathrm{BSC}$ & $\mathrm{Arbitrum}$ & $\mathrm{Avalanche}$ \\
    \midrule
    \endfirsthead

    \caption[]{Estimates of the non-linear models with global market and extreme return dummies. (continued)}\\
    \toprule
      Model: 
        & (1) & (2) & (3) & (4) & (5) \\
      Chain portfolios: 
        & $\mathrm{Ethereum}$ & $\mathrm{Solana}$ & $\mathrm{BSC}$ & $\mathrm{Arbitrum}$ & $\mathrm{Avalanche}$ \\
    \midrule
    \endhead

    \midrule
    \multicolumn{6}{r}{\emph{Continued on next page}}
    \endfoot

    \bottomrule
    \endlastfoot

    %--------------- Panel A ----------------
    \multicolumn{6}{l}{Panel A: $R^{All}_{\mathrm{chain}_{i},t}$ as the dependent variable}\\
    \addlinespace

    $\alpha_0$ 
      & 0.0007  & -0.00007 & 0.0002  & 0.0003  & -0.00006 \\
      & (0.976) & (-0.592) & (0.220) & (1.218)  & (-0.342)  \\

    $\alpha_1$
      & 0.530***  &  0.987*** & 0.978***  & 0.411***  & 0.868*** \\
      & (11.433)  & (84.706) & (1253.073) & (25.984)  & (61.210)  \\

    $\beta_{00}$
      & -0.052**   & 0.002   & 0.001    & -0.004  & -0.001 \\
      & (-2.394)   & (0.779) & (0.300)  & (-0.190)  & (-0.107)  \\

    $\beta_{01}$
      &  -0.027    & -0.003  & -0.014   & -0.005   & 0.051 \\
      & (-0.379)  & (-0.301) & (-0.601) & (-0.143) & (0.495)  \\

    $\beta_{02}$
      & -0.086     & 0.049   & 0.976* & 0.308*    & -0.138 \\
      & (-0.125)   & (0.483) & (1.648)  & (1.992) & (-0.581)  \\

    $\beta_{10}$
      & 0.444*** & -0.038** & 0.021    & 0.066***  & -0.016 \\
      & (5.274)  & (-2.096) & (0.282)  & (4.135)  & (-0.736)  \\

    $\beta_{11}$
      & -0.0201   & -0.003   & -0.061**   & -0.004   & -0.008 \\
      & (-0.282)  & (-0.727) & (-2.401) & (-0.278)  & (-0.946)  \\

    $\beta_{12}$
      & -0.350   & -0.031   & -0.963   & -0.653***  & 0.033 \\
      & (-0.551) & (-0.412) & (-1.489) & (-2.892)  & (0.268)  \\

    $\beta_{13}$
      & -0.464*** & 0.036** & -0.014   & -0.062***  & 0.015 \\
      & (-4.378)  & (2.068) & (-0.181) & (-3.798)  & (0.660)  \\

    $\beta_{14}$
      & -0.473*** & 0.040**  & -0.013   & -0.071*** & 0.013 \\
      & (-3.814)  & (2.124) & (-0.169) & (-4.448)  & (0.593)  \\

    $\beta_{20}$
      & 0.150     & -0.051   & 0.151** & 0.096***  & -0.034* \\
      & (1.399)   & (-1.512) & (2.363) & (4.705)  & (-1.905)  \\

    $\beta_{21}$
      & -0.036    & -0.011   & -0.022   & -0.005   & 0.028 \\
      & (-1.432)  & (-1.397) & (-0.969)  & (-0.213) & (0.620)  \\

    $\beta_{22}$
      &  3.310*** & 0.555    & -0.068   & -0.170** & -0.045 \\
      & (2.891)   & (1.003) & (-0.865) & (-2.303) & (-0.707)  \\

    $\beta_{23}$
      & -0.147    & 0.051  & -0.152**   & -0.076*** & 0.035* \\
      & (-1.373) & (1.507) & (-2.333) & (-3.486)  & (1.876)  \\

    $\beta_{24}$
      & -0.091   & 0.038  & -0.171***   & -0.077*** & 0.020 \\
      & (-0.802) & (1.343) & (-2.729) & (-3.530)  & (0.925)  \\

    $\beta_{30}$
      & 0.575***   & 0.006   & -0.082    & 0.019   & -0.039 \\
      & (3.386)   & (0.378) & (-0.650) & (0.627) & (-1.690)  \\

    $\beta_{31}$
      & 0.270    & -0.061   & -0.281*   & 0.0005    & 0.016 \\
      & (1.588)  & (-1.578) & (-1.955)  & (0.080)  & (1.364)  \\

    $\beta_{32}$
      & 0.516    & 0.051   & -0.472   & -0.608*   & 0.047 \\
      & (1.576)  & (1.347) & (-1.367) & (-1.718) & (0.153)  \\

    $\beta_{33}$
      & -0.901*** & 0.021  & -0.060   & -0.021    & 0.042* \\
      & (-3.883)  & (0.915) & (-0.468) & (-0.689)  & (1.838)  \\

    $\beta_{34}$
      & -0.345**  & 0.052*  & -0.105   & -0.009    & 0.041* \\
      & (-2.128)  & (1.896) & (-0.793) & (-0.282)  & (1.743)  \\

    $\beta_{40}$
      & -0.195*** & 0.048*** & 0.090**  & -0.059***  & -0.056** \\
      & (-3.032)  & (2.681)  & (1.981)  & (-3.475)  & (-1.970)  \\

    $\beta_{41}$
      & 0.070    & 0.014   & -0.059   & 0.033    & 0.101 \\
      & (0.896)  & (0.593) & (-0.809)  & (1.255) & (1.317)  \\

    $\beta_{42}$
      & 0.192    & -0.043   & -0.077   & 0.152 & -0.065 \\
      & (0.331)  & (-0.085) & (-0.239) & (0.460) & (-0.802)  \\

    $\beta_{43}$
      & 0.259***  & -0.054   & -0.036   & 0.082***    & -0.009 \\
      & (3.390)   & (-0.622) & (-0.612) & (2.895)  & (-0.270)  \\

    $\beta_{44}$
      & 0.263***  & -0.006    & -0.083   & 0.0360    & -0.015 \\
      & (2.990)   & ( -0.292) & (-1.389) & (1.199)  & (-0.400)  \\

    $\theta_0$
      & 0.174*  & 0.047** &  -0.091   & -0.054  & -0.060** \\
      & (1.757) & (1.983) & (-1.451) & (-1.321)  & (-1.986)  \\

    $\theta_1$
      & -0.003   & -0.029   & 0.046   & -0.008  & 0.030** \\
      & (-0.052) & (-1.422) & (0.678) & (-0.554)  & (2.436)  \\

    $\theta_2$
      & -0.204*** & -0.003  & 0.201   & 0.232***  & -0.023 \\
      & (-2.583) & (-0.172) & (1.044) & (7.276)  & (-1.028)  \\

    $\theta_3$
      & 0.029   & -0.012*  & 0.028   & 0.008    & -0.008 \\
      & (1.396) & (-1.808) & (0.922) & (1.189)  & (-1.316)  \\

    $\theta_4$
      &  -0.035*** & -0.001*  & 0.0003   & -0.001  & 0.001 \\
      & (-3.769) & (-1.892) & (-0.080) & (-0.748)  & (0.843)  \\

    $\theta_5$
      & -0.001   & -0.0005 & -0.004  & 0.0008  & -0.0009 \\
      & (-0.237) & (-1.023) & (-1.245) & (0.373)  & (-0.981)  \\

    \addlinespace

    p   & 2 & 2 & 1 & 1 & 1 \\
    o   & 2 & 0 & 0 & 0 & 1 \\
    q   & 2 & 3 & 1 & 1 & 1  \\
    \addlinespace

    Adjusted \(R^2\) & 0.034 & 0.983 & 0.997 & 0.747 & 0.894 \\
    Observations     & 941   & 941   & 941   & 940  & 941 \\
  
  \midrule
  %--------------- Panel B ----------------
    \multicolumn{6}{l}{Panel B: $R^{\mathrm{non-CEX}}_{\mathrm{chain}_{i},t}$ as the dependent variable}\\
    \addlinespace
    $\alpha_0$ 
      & -0.002**  & -0.0003   & -0.0003  & 0.0005  & 0.0002 \\
      & (2.211) & (-0.302) & (-0.052) & (1.465)    & (0.180)  \\

    $\alpha_1$
      & 0.028    & 0.004  & -0.0001  & 0.019   & 0.438** \\
      & (0.202)  & (0.402) & (-0.002) & (1.249)  & (2.114)  \\

    $\beta_{00}$
      & -0.022     & 0.008   & 0.0002   & -0.014    & 0.004 \\
      & (-0.822)   & (0.437) & (0.364)  & (-0.444)  & (0.017)  \\

    $\beta_{01}$
      & 1.051    & -0.194   & 0.169   & -0.099   & 0.061 \\
      & (0.811)  & (-1.555) & (0.934) & (-0.996) & (0.014)  \\

    $\beta_{02}$
      & 0.077    & 0.975   & 0.113    & 0.805**  & -0.391 \\
      & (0.478)   & (0.581) & (0.019) & (2.407)   & (-0.090)  \\

    $\beta_{10}$
      & -0.042    & 0.250   & 0.076    & 0.102*** & -0.052 \\
      & (-0.224)  & (0.928) & (0.597)  & (4.184)  & (-0.429)  \\

    $\beta_{11}$
      & -0.157   & 0.156     & -0.079   & -0.013   & -0.066 \\
      & (-0.739)   & (0.899) & (-0.375) & (-0.643)  & (-0.170)  \\

    $\beta_{12}$
      & -0.107    & -0.664   & -0.072   & -0.984***   &  0.396 \\
      & (-0.238)  & (-0.824) & (-0.003) & (-3.437)  & (0.309)  \\

    $\beta_{13}$
      & 0.049    & -0.244   & -0.071   & -0.092***  & 0.050 \\
      & (0.227)  & (-0.923) & (-1.307) & (-3.680)  & (0.504)  \\

    $\beta_{14}$
      & 0.098    & -0.230  & -0.060   & -0.107*** & 0.045 \\
      & (0.403)  & (-0.880) & (-0.066) & (-4.421)  & (0.385)  \\

    $\beta_{20}$
      & 0.383     & 0.163   & 0.010 & 0.152***  & -0.084 \\
      & (1.600)   & (0.739) & (0.073) & (4.708)  & (-0.462)  \\

    $\beta_{21}$
      & 0.061    & 0.006   & -0.035   & -0.009   & 0.162 \\
      & (0.558)  & (0.136) & (-0.067)  & (-0.289) & (0.212)  \\

    $\beta_{22}$
      &  -0.054** & -2.138    & 0.068   & -0.243** & 0.162 \\
      & (-0.021)   & (-1.174) & (0.117) & (-1.961) & (0.212)  \\

    $\beta_{23}$
      & -0.380    & -0.167  & -0.017   & -0.125*** & 0.097\\
      & (-1.509) & (-0.748) & (-0.054) & (-3.597)  & (0.767)  \\

    $\beta_{24}$
      & -0.480   & -0.141  & -0.036   & -0.118*** & 0.007 \\
      & (-1.226) & (-0.637) & (-0.396) & (-3.467)  & (0.036)  \\

    $\beta_{30}$
      & 0.408   & 0.201   & -0.363   & 0.074   & -0.125 \\
      & (1.560)   & (1.026) & (-0.239) & (1.551) & (-0.324)  \\

    $\beta_{31}$
      & 1.487    & -0.060   & 0.067   & 0.007    & 0.017 \\
      & (0.491)  & (-0.103) & (0.319)  & (1.120)  & (0.083)  \\

    $\beta_{32}$
      & -0.750    & -0.331   & -0.904   & -1.238**   & 0.690 \\
      & (-1.183)  & (-0.433) & (-0.693) & (-2.120) & (0.560)  \\

    $\beta_{33}$
      & -1.203** & 0.424  & 0.169   & -0.077    & 0.133 \\
      & (-2.036)  & (0.850) & (0.079) & (-1.615)  & (0.324)  \\

    $\beta_{34}$
      & -0.013  & 0.520*  & 0.074   & -0.064    & 0.150 \\
      & (-0.035)  & (1.813) & (0.052) & (-1.322)  & (0.399)  \\

    $\beta_{40}$
      & -0.257    & 0.144* & -0.051  & -0.112***  & -0.307 \\
      & (-0.730)  & (1.824)  & (-0.267)  & (-4.231)  & (-0.632)  \\

    $\beta_{41}$
      & -0.526    & -0.061   & -0.086   & 0.063*    & 0.271 \\
      & (-0.422)  & (-0.494) & (-0.091)  & (1.714) & (0.138)  \\

    $\beta_{42}$
      & -2.672**    & -0.314   & -0.850   &  0.165 & -0.770 \\
      & (-2.070)  & (-0.241) & (-0.141) & (0.312) & (-0.526)  \\

    $\beta_{43}$
      & 0.262     & -0.289*   & 0.089   & 0.145***    & 0.088 \\
      & (0.607)   & (-1.668) & (0.304) & (3.399)  & (0.088)  \\

    $\beta_{44}$
      & 0.406    & -0.101    & 0.134   & 0.072    & -0.043 \\
      & (0.752)  & (-0.747) & (0.558) & (1.498)  & (-0.163)  \\

    $\theta_0$
      & 0.740***  & -0.156 &  -0.268   & -0.100  & -0.111 \\
      & (4.030) & (-0.792) & (-0.262) & (-1.569)  & (-0.694)  \\

    $\theta_1$
      & -0.009   & -0.002   & -0.133   & -0.002  & 0.015 \\
      & (-0.067) & (-0.020) & (-0.173) & (-0.079)  & (0.247)  \\

    $\theta_2$
      & -0.762   & 0.056  & 0.462   & 0.393***  & -0.169 \\
      & (-1.057) & (0.269) & (0.454) & (8.226)  & (-0.372)  \\

    $\theta_3$
      & -0.024   & -0.075*  & -0.011   & 0.004    & -0.013 \\
      & (-0.373) & (-1.656) & (-0.133) & (0.428)  & (-0.320)  \\

    $\theta_4$
      &  -0.009 & 0.003  & 0.0002   & -0.002  & -0.002 \\
      & (-0.738) & (0.233) & (0.128) & (-0.571)  & (-0.095)  \\

    $\theta_5$
      & 0.003   & -0.016 & 0.0005  & 0.0002  & -0.007 \\
      & (0.480) & (-0.574) & (0.037) & (0.061)  & (-1.063)  \\

    \addlinespace

    p   & 3 & 3 & 1 & 1 & 2 \\
    o   & 1 & 2 & 1 & 0 & 0 \\
    q   & 3 & 3 & 1 & 1 & 2  \\
    \addlinespace

    Adjusted \(R^2\) & -0.011 & 0.003 & 0.026 & 0.215 & 0.217 \\
    Observations     & 941 & 941 & 941 & 940 & 941 \\

  \midrule
  %--------------- Panel C ----------------
    \multicolumn{6}{l}{Panel C: $R^{\mathrm{Local}}_{\mathrm{chain}_{i},t}$ as the dependent variable}\\
    \addlinespace
    $\alpha_0$ 
      & 0.0007   & 0.00008  & 0.0009**  & 0.0004  & -0.0007*** \\
      & (0.661) & (0.192)    & (2.094)   & (1.247)   & (-28.415)  \\

    $\alpha_1$
      & 0.661***  & 1.001***  & 1.052***   & 0.038**   & 0.721*** \\
      & (8.412)  & (396.836) & (417.704) & (2.466)  & (9.790)  \\

    $\beta_{00}$
      & -0.053*   & -0.0008   & 0.002   & -0.032    & -0.069 \\
      & (-1.731)   & (-0.175) & (0.879)  & (-1.076)  & (-1.489)  \\

    $\beta_{01}$
      & -0.058    & 0.005   & 0.0003   & -0.130   & 0.118 \\
      & (-0.650)  & (0.286) & (0.028) & (-1.311) & (0.808)  \\

    $\beta_{02}$
      & -0.621    & -0.019    & 0.494**   & 0.235  & 0.160 \\
      & (-0.662)   & (-0.128) & (2.235)  & (1.040)  & (0.182)  \\

    $\beta_{10}$
      & 0.410*** & -0.034*  &  0.005    & 0.109***    & -0.004 \\
      & (2.656)  & (-1.820) & (0.181)  & (4.468)  & (-0.066)  \\

    $\beta_{11}$
      & -0.050   & 0.0009   & -0.066   & -0.012   & -0.034 \\
      & (-0.549) & (0.024) & (-1.358) & (-0.624)  & (-0.700)  \\

    $\beta_{12}$
      & -0.333    & -0.084   &  -3.127   & -0.942***   & 0.335 \\
      & (-0.412)  & (-0.667) & (-1.381) & (-3.105)  & (0.651)  \\

    $\beta_{13}$
      & -0.453***    & 0.036*   & 0.090   & -0.099***  & 0.013 \\
      & (-2.642)  & (1.900) & (1.062) & (-3.910)  & (0.213)  \\

    $\beta_{14}$
      & -0.473***    & 0.038**  & 0.084   & -0.114*** & 0.0008 \\
      & (-2.596)  & (2.037) & (0.954) & (-4.641)  & (0.013)  \\

    $\beta_{20}$
      & 0.178     & -0.094   & 0.010 & 0.163***  & -0.049 \\
      & (1.249)   & (-0.668) & (0.364) & (5.203)  & (-0.765)  \\

    $\beta_{21}$
      & -0.036    & -0.005   & -0.010   & -0.0005   & 0.104 \\
      & (-0.924)  & (-0.101) & (-0.330)  & (-0.016) & (0.840)  \\

    $\beta_{22}$
      &  3.311** & 0.132    & 0.081   & -0.242*** & -0.051 \\
      & (2.190)   & (0.127) & (0.703) & (-1.957) & (-0.262)  \\

    $\beta_{23}$
      & -0.175    & 0.092  & -0.031   & -0.137*** &  0.055\\
      & (-1.233) & (0.621) & (-1.047) & (-4.049)  & (0.801)  \\

    $\beta_{24}$
      & -0.101   & 0.081  & -0.053   & -0.128*** & 0.003 \\
      & (-0.635) & (0.500) & (-1.460) & (-3.839)  & (0.033)  \\

    $\beta_{30}$
      & 0.450**   & 0.022   & -0.154**   & 0.089*   & 0.015 \\
      & (2.076)   & (0.675) & (-2.249) & (1.914) & (0.150)  \\

    $\beta_{31}$
      & 0.286    & -0.146   & -0.112   & 0.006    & 0.073*** \\
      & (1.301)  & (-0.680) & (-1.212)  & (0.876)  & (3.280)  \\

    $\beta_{32}$
      & 0.766**    & -0.089   & -0.564*   & -1.300**   & -1.577** \\
      & (2.123)  & (-0.607) & (-1.697) & (-2.261) & (-2.108)  \\

    $\beta_{33}$
      & -0.888*** & 0.081  & -0.063   & -0.093**    & -0.009 \\
      & (-2.928)  & (0.567) & (-0.769) & (-1.991)  & (-0.096)  \\

    $\beta_{34}$
      & -0.290  & 0.093  & -0.147   & -0.081*    & 0.018 \\
      & (-1.375)  & (0.893) & (-1.148) & (-1.706)  & (0.178)  \\

    $\beta_{40}$
      & -0.158*    & 0.057*** & 0.093**  & -0.113***  & -0.438*** \\
      & (-1.730)  & (3.280)  & (2.190)  & (-4.372)  & (-3.963)  \\

    $\beta_{41}$
      & 0.038    & -0.008   & -0.067   &  0.064*    & 0.042 \\
      & (0.375)  & (-0.127) & (-1.268)  & (1.886) & (0.180)  \\

    $\beta_{42}$
      & 0.177    & 0.442   & -0.266   &  0.191 & -1.609* \\
      & (0.225)  & (0.501) & (-0.637) & (0.371) & (-1.898)  \\

    $\beta_{43}$
      &  0.306*** & 0.011   & -0.079   & 0.142***    & 0.304** \\
      & (2.820)   & (0.034) & (-1.582) & (3.357)  & (2.242)  \\

    $\beta_{44}$
      & 0.297**   & -0.058    &  -0.061   & 0.066    & 0.071 \\
      & (2.405)  & (-1.249) & (-1.615) & (1.370)  & (0.463)  \\

    $\theta_0$
      & 0.249  & -0.020 &  -0.072   & -0.085  & -0.162 \\
      & (1.565) & (-0.140) & (-0.869) & (-1.347)  & (-1.473)  \\

    $\theta_1$
      & -0.033   & -0.0008   & 0.017   & 0.0003  & 0.054 \\
      & (-0.431) & (-0.010) & (0.557) & (0.012)  & (0.979)  \\

    $\theta_2$
      & -0.295**   & -0.032  & 0.159***   & 0.393***  & -0.090 \\
      & (-2.569) & (-0.519) & (2.612) & (7.960)  & (-1.044)  \\

    $\theta_3$
      & 0.059**   & -0.008*  & 0.008   & 0.004    & -0.037* \\
      & (2.069) & (-1.892) & (0.088) & (0.391)  & (-1.746)  \\

    $\theta_4$
      &  -0.041*** & -0.002  & 0.013   & 0.0004  & -0.002 \\
      & (-3.279) & (-0.493) & (1.303) & (0.091)  & (-0.392)  \\

    $\theta_5$
      & -0.006   & -0.002* & -0.005  & 0.002  & -0.014* \\
      & (-0.790) & (-1.886) & (-1.173) & (0.449)  & (-1.812)  \\

    \addlinespace

    p   & 2 & 2 & 2 & 1 & 1 \\
    o   & 2 & 2 & 1 & 0 & 0 \\
    q   & 3 & 3 & 2 & 1 & 1  \\
    \addlinespace

    Adjusted \(R^2\) & 0.048 & 0.977 & 0.995 & 0.225 & 0.386 \\
    Observations     & 941 & 941 & 941 & 940 & 941 \\

  \end{tabularx}

\end{ThreePartTable}
\end{singlespace}

\newpage

\section*{Appendix}\label{appendex}

% -------------------------------- table for baseline models

\begin{singlespace}
\begin{ThreePartTable}
  %──────────────────────────────────────────────────────────────────────────────
  % Notes, flush‐left and without indent
  \begin{TableNotes}[para,flushleft]
    \scriptsize
    Notes: This table shows estimates of the baseline linear models, i.e., as illustrated by Equation (\ref{eq_baseline_all}), (\ref{eq_baseline_nonCEX}), and (\ref{eq_baseline_this_chain_only}), for the five chain portfolios in our study with data in sampling period from 17th Mar 2023 to 31st Mar 2025.  Panel A reports the estimated parameters of the models for all the five chains based on Equation (\ref{eq_baseline_all}), which are as following:
      \begin{equation*}
      R^{All}_{\mathrm{Ethereum},t}
        = \alpha_{0}
          + \alpha_{1}R^{All}_{\mathrm{Ethereum},t-1}
          + \beta_{0}R^{CEX}_{\mathrm{Ethereum},t}
          + \beta_{1}R^{All}_{\mathrm{Solana},t}
          + \beta_{2}R^{All}_{\mathrm{BSC},t}
          + \beta_{3}R^{All}_{\mathrm{Arbitrum},t}
          + \beta_{4}R^{All}_{\mathrm{Avalanche},t}
          + e_{t},
      \end{equation*}
      \begin{equation*}
      R^{All}_{\mathrm{Solana},t}
        = \alpha_{0}
          + \alpha_{1}R^{All}_{\mathrm{Solana},t-1}
          + \beta_{0}R^{CEX}_{\mathrm{Solana},t}
          + \beta_{1}R^{All}_{\mathrm{Ethereum},t}
          + \beta_{2}R^{All}_{\mathrm{BSC},t}
          + \beta_{3}R^{All}_{\mathrm{Arbitrum},t}
          + \beta_{4}R^{All}_{\mathrm{Avalanche},t}
          + e_{t},
      \end{equation*}
      \begin{equation*}
      R^{All}_{\mathrm{BSC},t}
        = \alpha_{0}
          + \alpha_{1}R^{All}_{\mathrm{BSC},t-1}
          + \beta_{0}R^{CEX}_{\mathrm{BSC},t}
          + \beta_{1}R^{All}_{\mathrm{Ethereum},t}
          + \beta_{2}R^{All}_{\mathrm{Solana},t}
          + \beta_{3}R^{All}_{\mathrm{Arbitrum},t}
          + \beta_{4}R^{All}_{\mathrm{Avalanche},t}
          + e_{t},
      \end{equation*}
      \begin{equation*}
      R^{All}_{\mathrm{Arbitrum},t}
        = \alpha_{0}
          + \alpha_{1}R^{All}_{\mathrm{Arbitrum},t-1}
          + \beta_{0}R^{CEX}_{\mathrm{Arbitrum},t}
          + \beta_{1}R^{All}_{\mathrm{Ethereum},t}
          + \beta_{2}R^{All}_{\mathrm{Solana},t}
          + \beta_{3}R^{All}_{\mathrm{BSC},t}
          + \beta_{4}R^{All}_{\mathrm{Avalanche},t}
          + e_{t},
      \end{equation*}
      \begin{equation*}
      R^{All}_{\mathrm{Avalanche},t}
        = \alpha_{0}
          + \alpha_{1}R^{All}_{\mathrm{Avalanche},t-1}
          + \beta_{0}R^{CEX}_{\mathrm{Avalanche},t}
          + \beta_{1}R^{All}_{\mathrm{Ethereum},t}
          + \beta_{2}R^{All}_{\mathrm{Solana},t}
          + \beta_{3}R^{All}_{\mathrm{BSC},t}
          + \beta_{4}R^{All}_{\mathrm{Arbitrum},t}
          + e_{t}.
      \end{equation*}
    Panel B reports the estimated parameters of the models for all the five chains based on Equation (\ref{eq_baseline_nonCEX}), which are as following:
      \begin{equation*}
      R^{non-CEX}_{\mathrm{Ethereum},t}
        = \alpha_{0}
          + \alpha_{1}R^{non-CEX}_{\mathrm{Ethereum},t-1}
          + \beta_{0}R^{CEX}_{\mathrm{Ethereum},t}
          + \beta_{1}R^{All}_{\mathrm{Solana},t}
          + \beta_{2}R^{All}_{\mathrm{BSC},t}
          + \beta_{3}R^{All}_{\mathrm{Arbitrum},t}
          + \beta_{4}R^{All}_{\mathrm{Avalanche},t}
          + e_{t},
      \end{equation*}
      \begin{equation*}
      R^{non-CEX}_{\mathrm{Solana},t}
        = \alpha_{0}
          + \alpha_{1}R^{non-CEX}_{\mathrm{Solana},t-1}
          + \beta_{0}R^{CEX}_{\mathrm{Solana},t}
          + \beta_{1}R^{All}_{\mathrm{Ethereum},t}
          + \beta_{2}R^{All}_{\mathrm{BSC},t}
          + \beta_{3}R^{All}_{\mathrm{Arbitrum},t}
          + \beta_{4}R^{All}_{\mathrm{Avalanche},t}
          + e_{t},
      \end{equation*}
      \begin{equation*}
      R^{non-CEX}_{\mathrm{BSC},t}
        = \alpha_{0}
          + \alpha_{1}R^{non-CEX}_{\mathrm{BSC},t-1}
          + \beta_{0}R^{CEX}_{\mathrm{BSC},t}
          + \beta_{1}R^{All}_{\mathrm{Ethereum},t}
          + \beta_{2}R^{All}_{\mathrm{Solana},t}
          + \beta_{3}R^{All}_{\mathrm{Arbitrum},t}
          + \beta_{4}R^{All}_{\mathrm{Avalanche},t}
          + e_{t},
      \end{equation*}
      \begin{equation*}
      R^{non-CEX}_{\mathrm{Arbitrum},t}
        = \alpha_{0}
          + \alpha_{1}R^{non-CEX}_{\mathrm{Arbitrum},t-1}
          + \beta_{0}R^{CEX}_{\mathrm{Arbitrum},t}
          + \beta_{1}R^{All}_{\mathrm{Ethereum},t}
          + \beta_{2}R^{All}_{\mathrm{Solana},t}
          + \beta_{3}R^{All}_{\mathrm{BSC},t}
          + \beta_{4}R^{All}_{\mathrm{Avalanche},t}
          + e_{t},
      \end{equation*}
      \begin{equation*}
      R^{non-CEX}_{\mathrm{Avalanche},t}
        = \alpha_{0}
          + \alpha_{1}R^{non-CEX}_{\mathrm{Avalanche},t-1}
          + \beta_{0}R^{CEX}_{\mathrm{Avalanche},t}
          + \beta_{1}R^{All}_{\mathrm{Ethereum},t}
          + \beta_{2}R^{All}_{\mathrm{Solana},t}
          + \beta_{3}R^{All}_{\mathrm{BSC},t}
          + \beta_{4}R^{All}_{\mathrm{Arbitrum},t}
          + e_{t}.
      \end{equation*}
    Panel C reports the estimated parameters of the models for all the five chains based on Equation (\ref{eq_baseline_this_chain_only}), which are as following:
      \begin{equation*}
      R^{Local}_{\mathrm{Ethereum},t}
        = \alpha_{0}
          + \alpha_{1}R^{Local}_{\mathrm{Ethereum},t-1}
          + \beta_{0}R^{CEX}_{\mathrm{Ethereum},t}
          + \beta_{1}R^{All}_{\mathrm{Solana},t}
          + \beta_{2}R^{All}_{\mathrm{BSC},t}
          + \beta_{3}R^{All}_{\mathrm{Arbitrum},t}
          + \beta_{4}R^{All}_{\mathrm{Avalanche},t}
          + e_{t},
      \end{equation*}
      \begin{equation*}
      R^{Local}_{\mathrm{Solana},t}
        = \alpha_{0}
          + \alpha_{1}R^{Local}_{\mathrm{Solana},t-1}
          + \beta_{0}R^{CEX}_{\mathrm{Solana},t}
          + \beta_{1}R^{All}_{\mathrm{Ethereum},t}
          + \beta_{2}R^{All}_{\mathrm{BSC},t}
          + \beta_{3}R^{All}_{\mathrm{Arbitrum},t}
          + \beta_{4}R^{All}_{\mathrm{Avalanche},t}
          + e_{t},
      \end{equation*}
      \begin{equation*}
      R^{Local}_{\mathrm{BSC},t}
        = \alpha_{0}
          + \alpha_{1}R^{Local}_{\mathrm{BSC},t-1}
          + \beta_{0}R^{CEX}_{\mathrm{BSC},t}
          + \beta_{1}R^{All}_{\mathrm{Ethereum},t}
          + \beta_{2}R^{All}_{\mathrm{Solana},t}
          + \beta_{3}R^{All}_{\mathrm{Arbitrum},t}
          + \beta_{4}R^{All}_{\mathrm{Avalanche},t}
          + e_{t},
      \end{equation*}
      \begin{equation*}
      R^{Local}_{\mathrm{Arbitrum},t}
        = \alpha_{0}
          + \alpha_{1}R^{Local}_{\mathrm{Arbitrum},t-1}
          + \beta_{0}R^{CEX}_{\mathrm{Arbitrum},t}
          + \beta_{1}R^{All}_{\mathrm{Ethereum},t}
          + \beta_{2}R^{All}_{\mathrm{Solana},t}
          + \beta_{3}R^{All}_{\mathrm{BSC},t}
          + \beta_{4}R^{All}_{\mathrm{Avalanche},t}
          + e_{t},
      \end{equation*}
      \begin{equation*}
      R^{Local}_{\mathrm{Avalanche},t}
        = \alpha_{0}
          + \alpha_{1}R^{Local}_{\mathrm{Avalanche},t-1}
          + \beta_{0}R^{CEX}_{\mathrm{Avalanche},t}
          + \beta_{1}R^{All}_{\mathrm{Ethereum},t}
          + \beta_{2}R^{All}_{\mathrm{Solana},t}
          + \beta_{3}R^{All}_{\mathrm{BSC},t}
          + \beta_{4}R^{All}_{\mathrm{Arbitrum},t}
          + e_{t},
      \end{equation*}
      while the series of residuals for all the modeled in the three panels are modeled by Glosten-Jagannathan-Runkle (GJR) asymmetric GARCH approach:
    \begin{equation}
    \begin{split}
        e_{t} \sim \mathcal{N}(0, \sigma_t^2), \;
        \sigma_t^2 = \omega + \sum_{i=1}^p \alpha_i e_{t-i}^2 + \sum_{j=1}^o \gamma_j e_{t-j}^2 \cdot \mathbb{I}_{\{e_{t-j} < 0\}} + \sum_{k=1}^q \beta_k \sigma_{t-k}^2,
    \end{split}
    \end{equation}
    and the optimal combination of $p$, $o$, and $q$ for each model is reported in this table. \\
    Results in this table are based on unified sampling period from 17th Mar 2023 to 31st Mar 2025.\\
    t-statistics are in parentheses. \\
      ***/**/* denote significance at 1\%/5\%/10\%.
  \end{TableNotes}
  %──────────────────────────────────────────────────────────────────────────────

  % Force the table to span the full text width
  \setlength\LTleft{0pt}
  \setlength\LTright{0pt}

  \begin{tabularx}{\textwidth}{@{}l*{5}{Y}@{}}
    \caption{Estimates of the baseline linear models with data from 17th Mar 2023 to 31st Mar 2025.\label{tab:baseline_truncated}}\\
    % insert notes at top
    \insertTableNotes\\
    \addlinespace
    \toprule
      Model: 
        & (1) & (2) & (3) & (4) & (5) \\
      Chain portfolios: 
        & $\mathrm{Ethereum}$ & $\mathrm{Solana}$ & $\mathrm{BSC}$ & $\mathrm{Arbitrum}$ & $\mathrm{Avalanche}$ \\
    \midrule
    \endfirsthead

    \caption[]{Estimates of the baseline linear models with data from 17th Mar 2023 to 31st Mar 2025. (continued)}\\
    \toprule
      Model: 
        & (1) & (2) & (3) & (4) & (5) \\
      Chain portfolios: 
        & $\mathrm{Ethereum}$ & $\mathrm{Solana}$ & $\mathrm{BSC}$ & $\mathrm{Arbitrum}$ & $\mathrm{Avalanche}$ \\
    \midrule
    \endhead

    \midrule
    \multicolumn{6}{r}{\emph{Continued on next page}}
    \endfoot

    \bottomrule
    \endlastfoot

    %--------------- Panel A ----------------
    \multicolumn{6}{l}{Panel A: $R^{All}_{\mathrm{chain}_{i},t}$ as the dependent variable}\\
    \addlinespace

    $\alpha_0$ 
      & -0.0008  & 0.00006** & 0.0004 & 0.0002*  & 0.00007 \\
      & (-1.607) & (0.623) & (0.419)  & (1.801)  & (0.801)  \\

    $\alpha_1$
      & -0.008    & 0.001 & -0.00008 & -0.007  & -0.00005 \\
      & (-0.329)  & (0.330) & (-0.608)  & (-0.433)  & (-0.007)  \\

    $\beta_0$
      & 0.636***  & 0.986***  & 0.978***   & 0.433***  & 0.837*** \\
      & (16.081)  & (344.907) & (877.765)  & (40.153)  & (66.065)  \\

    $\beta_1$
      & -0.046    & -0.0007* & -0.017**   & 0.0003  & -0.001* \\
      & (-1.275)  & (-1.854) & (-2.474)  & (0.254)  & (-1.812)  \\

    $\beta_2$
      & -0.001** & 0.0004  & -0.008***  & 0.017*** & -0.007 \\
      & (-2.210) & (0.265) & (-3.993) & (3.159) & (-0.010)  \\

    $\beta_3$
      & 0.086**  & 0.021** & -0.117   & 0.00009  & -0.0007 \\
      & (2.400)  & (2.345) & (-1.028) & (1.417)  & (-1.568)  \\

    $\beta_4$
      & -0.014    & 0.027* & 0.015   & -0.009  & -0.073*** \\
      & (-1.114) & (1.933) & (1.223) & (-1.189)  & (-6.795)  \\

    \addlinespace

    p   & 2 & 2 & 2 & 1 & 3 \\
    o   & 2 & 2 & 2 & 0 & 0 \\
    q   & 3 & 3 & 3 & 1 & 2  \\
    \addlinespace

    \(R^2\) & 0.017 & 0.981 & 0.995 & 0.731 & 0.890 \\
  
  \midrule
  %--------------- Panel B ----------------
    \multicolumn{6}{l}{Panel B: $R^{\mathrm{non-CEX}}_{\mathrm{chain}_{i},t}$ as the dependent variable}\\
    \addlinespace
    $\alpha_0$ 
      & -0.002    & -0.001   & 0.0008 & 0.0002  & 0.0008** \\
      & (-1.301)  & (-1.630) & (0.758)  & (1.393)  & (2.109)  \\

    $\alpha_1$
      & -0.012**  & -0.064    & -0.015    & 0.005  & -0.031 \\
      & (-2.567)   & (-1.505) & (-0.062)  & (0.151)  & (-1.162)  \\

    $\beta_0$
      & 0.186**  & 0.018   & 0.0006   & 0.052***  & 0.292*** \\
      & (2.026)  & (0.991) & (1.172)  & (4.298)  & (6.796)  \\

    $\beta_1$
      & -0.029   & -0.007   & -0.041    & 0.002  & -0.009*** \\
      & (-0.629) & (-0.598) & (-1.072)  & (0.986)  & (-4.018)  \\

    $\beta_2$
      & -0.002   & 0.002*** & -0.002   & 0.026*** & -0.002 \\
      & (-0.463) & (8.496) & (-0.308)  & (3.327) & (-0.061)  \\

    $\beta_3$
      & -0.159  & 0.464**  & -0.734***  & 0.00004  & 0.002 \\
      & (-1.074) & (2.520) & (-2.691)   & (0.195)  & (0.153)  \\

    $\beta_4$
      & -0.259   & 0.107   & 0.013   & -0.019  & -0.405*** \\
      & (-1.305) & (1.161) & (0.383) & (-1.505)  & (-6.341)  \\

    \addlinespace

    p   & 3 & 1 & 1 & 1 & 2 \\
    o   & 2 & 0 & 1 & 0 & 0 \\
    q   & 1 & 1 & 2 & 1 & 2  \\
    \addlinespace

    \(R^2\) & -0.004 & 0.021 & -0.025 & 0.053 & 0.148 \\

  \midrule
  %--------------- Panel C ----------------
    \multicolumn{6}{l}{Panel C: $R^{\mathrm{Local}}_{\mathrm{chain}_{i},t}$ as the dependent variable}\\
    \addlinespace
    $\alpha_0$ 
      & -0.0007  & 0.0004*** & 0.001 & 0.0002***  & -0.0005 \\
      & (-1.108)  & (4.696) & (1.464)  & (2.871)  & (1.147)  \\

    $\alpha_1$
      & -0.0112 & -0.002** & -0.0002*** & -0.007  & -0.063** \\
      & (-0.363) & (-2.299) & (-3.034)  & (-0.247)  & (-3.255)  \\

    $\beta_0$
      & 0.785*** & 0.999*** & 1.053***  & 0.073***  & 0.789*** \\
      & (12.512)  & (648.472) & (495.490)  & (6.014)  & (7.108)  \\

    $\beta_1$
      & -0.051   & -0.0006*** & -0.045   & 0.002  & -0.007** \\
      & (-0.906) & (-2.637) & (-1.544)  & (1.477)  & (-2.371)  \\

    $\beta_2$
      & -0.001   & -0.0001  & -0.012**   & 0.026*** & 0.004 \\
      & (-1.411) & (-0.215) & (-1.994) & (3.369) & (0.259)  \\

    $\beta_3$
      & -0.022  & 0.071*** & -0.258***   & 0.0002  & -0.0007 \\
      & (-0.403) & (4.234) & (-5.073) & (1.094)  & (-0.806)  \\

    $\beta_4$
      & 0.031    & 0.034** & 0.025   & -0.019  & -0.242*** \\
      & (0.161) & (2.327) & (1.514) & (-1.527)  & (-3.803)  \\

    \addlinespace

    p   & 2 & 2 & 2 & 1 & 1 \\
    o   & 2 & 0 & 2 & 0 & 0 \\
    q   & 3 & 3 & 3 & 1 & 1  \\
    \addlinespace

    \(R^2\) & 0.025 & 0.975 & 0.990 & 0.049 & 0.392 \\

  \end{tabularx}

\end{ThreePartTable}
\end{singlespace}
% --------------------------------------------------

% -------------------------------------- bib

\printbibliography

\end{document}